\documentclass[12pt]{article}
\usepackage{graphicx}

\begin{document}

\title{The intercept/resend and translucent attacks on the quantum key distribution protocol based on the pre- and post-selection effect}

\author{Hiroo Azuma${}^{1,}$\thanks{On leave from
Advanced Algorithm \& Systems Co., Ltd.,
7F Ebisu-IS Building, 1-13-6 Ebisu, Shibuya-ku, Tokyo 150-0013, Japan.
Email: hiroo.azuma@m3.dion.ne.jp}
\ \ 
and
\ \ 
Masashi Ban${}^{2,}$\thanks{Email: m.ban@phys.ocha.ac.jp}
\\
\\
{\small ${}^{1}$Nisshin-scientia Co., Ltd.,}\\
{\small 8F Omori Belport B, 6-26-2 MinamiOhi, Shinagawa-ku, Tokyo 140-0013, Japan}\\
{\small ${}^{2}$Graduate School of Humanities and Sciences, Ochanomizu University,}\\
{\small 2-1-1 Ohtsuka, Bunkyo-ku, Tokyo 112-8610, Japan}
}

\date{\today}

\maketitle

\begin{abstract}
We investigate the security against the intercept/resend and translucent attacks
on the quantum key distribution protocol based on the pre- and post-selection effect.
In 2001, Bub proposed the quantum cryptography scheme,
which was an application of the so-called mean king's problem.
We evaluate a probability that legitimate users cannot detect eavesdropper's malicious acts for Bub's protocol.
We also estimate a probability that the eavesdropper guesses right at the random secret key one of the legitimate users tries to share with the other one.
From rigorous mathematical and numerical analyses,
we conclude that Bub's protocol is weaker than the Bennett-Brassard protocol of 1984 (BB84) against both the intercept/resend and translucent attacks.
Because Bub's protocol uses a two-way quantum channel,
the analyses of its security are tough to accomplish.
We refer to their technical points accurately in the current paper.
For example,
we impose some constraints upon the eavesdropper's strategies in order to let their degrees of freedom be small.
\end{abstract}

\section{\label{section-introduction}Introduction}
The quantum key distribution is one of the practical goals that researchers in the field of quantum information attempt to achieve
from both theoretical and experimental points of view.
Since the Bennett-Brassard protocol of 1984 (BB84) and the Ekert protocol of 1991 (E91) were proposed,
the security of the quantum key distribution protocols has been studied eagerly \cite{Bennett1984,Ekert1991,Bennett1992a}.

The quantum key distribution is aimed at establishing a secure random secret key between two parties, Alice and Bob.
For example, Alice and Bob can use this key for the one-time pad cipher.
To examine the security of the quantum key distribution,
we assume that the eavesdropper Eve can interact with the quantum channel through which Alice and Bob send and receive signals.
For evaluating the security of the quantum cryptography in concrete terms,
we study some typical strategies Eve pursues.

One of the simplest strategies that Eve follows is the intercept/resend attack \cite{Bennett1992b}.
In this attack,
Eve makes a strong projective measurement on a signal emitted by Alice in an arbitrary basis
and resends another new one to Bob
depending on the result Eve obtains.
If we construct quantum bits (qubits) from single photons,
Eve can perform the intercept/resend attack only with passive linear optics,
that is to say,
beamsplitters, waveplates, photodetectors, and a single photon source.
Eve does not need to prepare two-qubit gates that generate entanglement.

Another basic strategy that we consider is the translucent attack \cite{Ekert1994,Biham1997,Enzer2002,Cirac1997}.
In this strategy,
Eve produces entanglement between the single qubit the legitimate user sends and her auxiliary quantum system,
and waits until she obtains the public discussion between Alice and Bob to make a measurement on her auxiliary system.
To make this attack on the single qubit Alice and Bob transmit,
Eve has to use quantum circuits,
in other words,
a quantum computer,
to generate entanglement between the signal and her own auxiliary system.
Thus, the translucent attack is more difficult and sophisticated than the intercept/resend attack.

In 2001, Bub proposed a unique protocol for the quantum key distribution \cite{Bub2001}.
It is based on the pre- and post-selection effect,
which Aharonov, Bergmann, and Lebowitz discovered \cite{Aharonov1964,Aharonov2008}.
In Ref.~\cite{Aharonov1964},
they discussed the measurement of a quantum system at time $t$ between two other measurements performed at times $t_{1}$ and $t_{2}$,
where $t_{1}<t<t_{2}$,
in the following situation.

We assume that the measurement at the time $t_{1}$ lets the quantum system be in the state $|\psi_{1}(t_{1})\rangle$.
This state has the standard time evolution,
$|\psi_{1}(t)\rangle=U(t_{1},t)|\psi_{1}(t_{1})\rangle$.
Simultaneously, we assume that the measurement at the time $t_{2}$ generates the state $|\psi_{2}(t_{2})\rangle$
for the quantum system.
Its backward time evolution is given by
$\langle\psi_{2}(t)|=\langle\psi_{2}(t_{2})|U(t,t_{2})$.
Then, the measurement at the time $t$ of a variable $C$ is obtained as
\begin{equation}
\mbox{prob}(C=c_{n})
=
\frac{|\langle\psi_{2}(t)|\hat{P}(C=c_{n})|\psi_{1}(t)\rangle|^{2}}{\sum_{j}|\langle\psi_{2}(t)|\hat{P}(C=c_{j})|\psi_{1}(t)\rangle|^{2}},
\label{ABL-rule-01}
\end{equation}
where $\hat{P}(C=c_{j})$ is a projection operator made of an eigenvector with an eigenvalue $c_{j}$.
We put a hat on the symbols of the projection operators hereafter to draw your attention on them.
Equation~(\ref{ABL-rule-01}) is called the Aharonov-Bergmann-Lebowitz rule (ABL-rule).
The ABL-rule is regarded as a new concept that gives a complete description of a quantum system in the time interval between two measurements,
in other words,
information about the system both from the past and from the future \cite{Aharonov1991}.

In Ref.~\cite{Vaidman1987},
according to the ABL-rule,
Vaidman, Aharonov, and Albert found a process,
in which the results of measurements of $\sigma_{x}$, $\sigma_{y}$, and $\sigma_{z}$ were ascertained with a probability of unity.
In their process, first, we prepare the maximally entangled initial state of two spin-$1/2$ particles.
Second, we perform the spin measurement of $\sigma_{x}$, $\sigma_{y}$, or $\sigma_{z}$
on a single qubit that belongs to the initial entangled state.
Third, we take a measurement on the composite system with an operator whose eigenvectors are entangled.
Then, we obtain results of measurements of $\sigma_{x}$, $\sigma_{y}$, and $\sigma_{z}$ with a probability of unity
although these operators do not commute.
This counter-intuitive phenomenon is regarded as one of the pre- and post-selection effects.

The above process is known as a solution of the mean king's problem \cite{Aharonov2001}.
Bub's quantum key distribution protocol is a natural application of the result obtained in Ref.~\cite{Vaidman1987}.
First, Alice prepares the initial two-qubit entangled state,
and second,
Bob performs the measurement of $\sigma_{x}$ or $\sigma_{z}$ at random on the single qubit
owned by the initial two-qubit state.
Third, Alice carries out the final measurement on the whole system with the entangled basis.
Because the results of the measurements of $\sigma_{x}$ and $\sigma_{z}$ are ascertained with a probability of unity,
Alice can share a random key with Bob.

In the present paper, first, we examine the security against the intercept/resend attack on Bub's quantum key distribution protocol
by evaluating a probability $P_{\mbox{\scriptsize AB}}$ that Alice and Bob cannot detect Eve's illegal acts and a probability $P_{\mbox{\scriptsize E}}$ that Eve guesses right
at the random secret key Alice obtains.
We show that $P_{\mbox{\scriptsize AB}}$ and $P_{\mbox{\scriptsize E}}$ are given by $5/6\simeq 0.833$ and $(5+3\sqrt{2})/10\simeq 0.924$ respectively
if Eve uses the Breidbart basis.
From these results,
we can conclude that Bub's protocol is more vulnerable than the BB84 scheme under the intercept/resend attack.

Second, we study the security against the translucent attack on Bub's protocol by estimating $P_{\mbox{\scriptsize AB}}$ and $P_{\mbox{\scriptsize E}}$.
For the translucent attack,
Eve lets her probe interact with the single qubit flying between Alice and Bob.
Because the unitary transformation applied to Eve's probe and the single qubit has many real parameters,
it is very difficult to find Eve's optimum strategy.
Thus, imposing some constraints upon Eve's unitary transformation,
we make its degrees of freedom small
and optimize her strategies.
From these analyses, we can conclude that Bub's protocol is not safer than the BB84 scheme under the translucent attacks, as well.

So far, the security against each specified attack on Bub's quantum key distribution protocol has not been studied in a systematic manner.
How about the security situation of this protocol is as follows.
It has not been proved to be secure under the translucent attack or the coherent attack.
Even its security against the intercept/resend attack has not been examined precisely yet.
We have to say that the study of security of Bub's protocol is still at a very early stage of development.
This point is the motivation for the current paper.

In Sec.~\ref{section-protocol}, we are going to explain that
Bub's protocol has a two-way quantum channel, that is to say, a quantum transmission from Alice to Bob and that from Bob to Alice.
Because of this circumstance,
we can divide Eve's strategies into two groups, one-way and two-way attacks.
In the present paper, we investigate the following four attacks:
a one-way intercept/resend attack,
a two-way intercept/resend attack,
a one-way translucent attack,
and a two-way translucent attack.

Here, we pay attention to the fact that there are two different two-way
intercept/resend attacks rigorously.
The first one is that Eve eavesdrops on each of quantum channels of the opposite directions between Alice and Bob independently.
In this strategy, Eve can utilize a classical correlation between two measurements obtained on the transmitted qubit's travels
from Alice to Bob and back again.
To make this attack,
Eve only has to perform strong projective measurements twice on the transmitted qubit and does not need to prepare her own auxiliary qubits.

The second one is taking an entangled joint attack,
specifically two interactive measurements on the qubit flying there and back.
In this strategy,
Eve can employ a quantum correlation between the two outputs of observations.
To carry out this attack,
Eve has to make her auxiliary qubits ready for interaction with the two-way quantum channel.
However,
this process is equivalent to the two-way translucent attack.
Hence, for the two-way intercept/resend attack,
we consider only the classical correlation between the two projective measurements of the qubit flying there and back again.

The most sophisticated strategy Eve can make is the coherent attack.
In this attack,
Eve lets her quantum probe interact with multiple qubits the legitimate users send and receive,
keeps her probe until she learns Alice and Bob's public announcement about the error correction and the privacy amplification method,
and finally observes her probe according to the information disclosed by the legitimate users.
In the present paper,
we do not consider the coherent attack.
This topic has to be investigated in the future.

Here, we mention some previous studies.
Yoshida {\it et al}. derived trade-off inequalities indicating that an increase of Eve's information gain caused
a rise of Alice and Bob's detection rate for Eve's illegal acts
during slightly modified Bub's quantum key distribution protocol \cite{Yoshida2010}.
In the protocol, they did not divide the sequence of the transmissions into particular subsequences as Bub's original protocol does,
but chose transmitted qubits at random for monitoring Eve's malicious acts.

In Ref.~\cite{Yoshida2012},
Yoshida {\it et al}. proposed simplified quantum key distribution protocols that were applications of the mean king's problem.
Werner {\it et al}. also proposed another version of the quantum key distribution protocol based on the mean king's problem
and examined its security against coherent attacks \cite{Werner2009}.

The present paper is organized as follows.
In Sec.~\ref{section-protocol},
we explain Bub's quantum key distribution protocol step by step.
In Sec.~\ref{section-previous-works},
we explain differences between early works and our results.
In Sec.~\ref{section-one-way-intercept-resend-attack},
we formulate the one-way intercept/resend attack.
In Sec.~\ref{section-symmetric-strategy-Eve},
we consider cases where specific relations hold for the one-way intercept/resend attack.
We show that the Breidbart basis is optimum for Eve's attack.
In Sec.~\ref{section-two-way-intercept-resend-attack},
we examine the two-way intercept/resend attack numerically.
In Sec.~\ref{section-one-way-translucent-attack},
we investigate the one-way translucent attack numerically.
In Sec.~\ref{section-two-way-translucent-attack},
we evaluate the security against the two-way translucent attack numerically.
In Sec.~\ref{section-Eve-careful-attack},
we show that Eve's one-way translucent attack gives her exactly zero information if she is restricted to add no noise.
In Sec.~\ref{section-discussion},
we give brief discussion.
In Appendix~\ref{section-appendix-A},
we list some useful functions for Secs.~\ref{section-one-way-intercept-resend-attack} and \ref{section-symmetric-strategy-Eve}.
In Appendix~\ref{section-appendix-B},
we list some useful functions for Sec.~\ref{section-two-way-intercept-resend-attack}.
In Appendix~\ref{section-appendix-C},
we give explicit forms of states of Eve's probe that are utilized in Sec.~\ref{section-one-way-translucent-attack}.
In Appendix~\ref{section-appendix-D},
we give explicit forms of states of Eve's probes that are made use of in Sec.~\ref{section-two-way-translucent-attack}.

\section{\label{section-protocol}The quantum key distribution protocol based on the pre- and post-selection effect}
In 2001, Bub proposed a quantum key distribution protocol based on the pre- and post-selection effect \cite{Bub2001}.
This protocol is a natural application of the result obtained in Ref.~\cite{Vaidman1987}.
We explain the protocol step by step in this section.
This section is a brief review of Refs.~\cite{Bub2001} and \cite{Vaidman1987}.

The purpose of the scheme is for the legitimate users, Alice and Bob,
to exchange a random secret key.
We assume that Alice and Bob can utilize both classical and quantum channels.
On the one hand, through the classical channel,
signals are disclosed publicly and the eavesdropper Eve can learn the entire contents of the classical messages.
On the other hand, via the quantum channel, Alice and Bob can exchange qubits with each other.
We assume that Eve can interact with the quantum channel, but she cannot do this without disturbing the qubits in general.

First, Alice prepares the following maximally entangled initial state,
\begin{equation}
|\psi\rangle_{\mbox{\scriptsize AC}}
=
\frac{1}{\sqrt{2}}
(|0\rangle_{\mbox{\scriptsize A}}|0\rangle_{\mbox{\scriptsize C}}
+
|1\rangle_{\mbox{\scriptsize A}}|1\rangle_{\mbox{\scriptsize C}}),
\label{initial-state}
\end{equation}
where the subscripts A and C represent the auxiliary and channel qubits, respectively.
The basis $\{|0\rangle, |1\rangle \}$ denotes a pair of the eigenstates of $\sigma_{z}$.
Alice keeps the auxiliary qubit close at hand.
The channel qubit is transmitted between Alice and Bob through the quantum channel.
Second, Alice sends the channel qubit to Bob.
Third, receiving the channel qubit from Alice,
Bob observes either $\sigma_{x}$ or $\sigma_{z}$ at random on it.
This observation causes a collapse of the state of the channel qubit depending on Bob's choice of the observables
and its outcome.
After the measurement, Bob returns the channel qubit to Alice.
Fourth, receiving the channel qubit,
Alice measures an observable $R$ on the pair of the auxiliary and channel qubits.
The operator $R$
has the following four eigenstates:
\begin{eqnarray}
|r_{1}\rangle_{\mbox{\scriptsize AC}}
&=&
\frac{1}{\sqrt{2}}|0\rangle_{\mbox{\scriptsize A}}|0\rangle_{\mbox{\scriptsize C}}
+
\frac{1}{2}
(e^{i\pi/4}|0\rangle_{\mbox{\scriptsize A}}|1\rangle_{\mbox{\scriptsize C}}
+
e^{-i\pi/4}|1\rangle_{\mbox{\scriptsize A}}|0\rangle_{\mbox{\scriptsize C}}) \nonumber \\
&=&
\frac{1}{\sqrt{2}}|+\rangle_{\mbox{\scriptsize A}}|+\rangle_{\mbox{\scriptsize C}}
+
\frac{1}{2}
(e^{-i\pi/4}|+\rangle_{\mbox{\scriptsize A}}|-\rangle_{\mbox{\scriptsize C}}
+
e^{i\pi/4}|-\rangle_{\mbox{\scriptsize A}}|+\rangle_{\mbox{\scriptsize C}}), \nonumber \\
|r_{2}\rangle_{\mbox{\scriptsize AC}}
&=&
\frac{1}{\sqrt{2}}|0\rangle_{\mbox{\scriptsize A}}|0\rangle_{\mbox{\scriptsize C}}
-
\frac{1}{2}
(e^{i\pi/4}|0\rangle_{\mbox{\scriptsize A}}|1\rangle_{\mbox{\scriptsize C}}
+
e^{-i\pi/4}|1\rangle_{\mbox{\scriptsize A}}|0\rangle_{\mbox{\scriptsize C}}) \nonumber \\
&=&
\frac{1}{\sqrt{2}}|-\rangle_{\mbox{\scriptsize A}}|-\rangle_{\mbox{\scriptsize C}}
+
\frac{1}{2}
(e^{i\pi/4}|+\rangle_{\mbox{\scriptsize A}}|-\rangle_{\mbox{\scriptsize C}}
+
e^{-i\pi/4}|-\rangle_{\mbox{\scriptsize A}}|+\rangle_{\mbox{\scriptsize C}}), \nonumber \\
|r_{3}\rangle_{\mbox{\scriptsize AC}}
&=&
\frac{1}{\sqrt{2}}|1\rangle_{\mbox{\scriptsize A}}|1\rangle_{\mbox{\scriptsize C}}
+
\frac{1}{2}
(e^{-i\pi/4}|0\rangle_{\mbox{\scriptsize A}}|1\rangle_{\mbox{\scriptsize C}}
+
e^{i\pi/4}|1\rangle_{\mbox{\scriptsize A}}|0\rangle_{\mbox{\scriptsize C}}) \nonumber \\
&=&
\frac{1}{\sqrt{2}}|+\rangle_{\mbox{\scriptsize A}}|+\rangle_{\mbox{\scriptsize C}}
-
\frac{1}{2}
(e^{-i\pi/4}|+\rangle_{\mbox{\scriptsize A}}|-\rangle_{\mbox{\scriptsize C}}
+
e^{i\pi/4}|-\rangle_{\mbox{\scriptsize A}}|+\rangle_{\mbox{\scriptsize C}}), \nonumber \\
|r_{4}\rangle_{\mbox{\scriptsize AC}}
&=&
\frac{1}{\sqrt{2}}|1\rangle_{\mbox{\scriptsize A}}|1\rangle_{\mbox{\scriptsize C}}
-
\frac{1}{2}
(e^{-i\pi/4}|0\rangle_{\mbox{\scriptsize A}}|1\rangle_{\mbox{\scriptsize C}}
+
e^{i\pi/4}|1\rangle_{\mbox{\scriptsize A}}|0\rangle_{\mbox{\scriptsize C}}) \nonumber \\
&=&
\frac{1}{\sqrt{2}}|-\rangle_{\mbox{\scriptsize A}}|-\rangle_{\mbox{\scriptsize C}}
-
\frac{1}{2}
(e^{-i\pi/4}|+\rangle_{\mbox{\scriptsize A}}|-\rangle_{\mbox{\scriptsize C}}
+
e^{i\pi/4}|-\rangle_{\mbox{\scriptsize A}}|+\rangle_{\mbox{\scriptsize C}}),
\label{Alice-measurement-basis}
\end{eqnarray}
where $\{|+\rangle,|-\rangle\}$ are eigenvectors of $\sigma_{x}$.
We pay attention to the facts that $\{|r_{i}\rangle_{\mbox{\scriptsize AC}}: i=1,2,3,4\}$ forms an orthonormal basis and
$\sum_{i=1}^{4}|r_{i}\rangle_{\mbox{\scriptsize AC}}{}_{\mbox{\scriptsize AC}}\langle r_{i}|=\mbox{\boldmath $I$}_{\mbox{\scriptsize AC}}$.
Alice and Bob repeat these four steps many times.

Here, for example, we calculate a probability that Bob obtains an outcome `$+1$' for $\sigma_{x}$
on condition that Alice detects $|r_{1}\rangle_{\mbox{\scriptsize AC}}$.
We write projection operators of the eigenvalues `$+1$' and `$-1$' for $\sigma_{x}$
as $\hat{P}(\sigma_{x}=1)$ and $\hat{P}(\sigma_{x}=-1)$, respectively.
We pay attention to the fact that $\hat{P}(\sigma_{x}=\pm 1)$ act on the channel qubit.
We obtain
\begin{eqnarray}
{}_{\mbox{\scriptsize AC}}\langle r_{1}|\hat{P}(\sigma_{x}=1)|\psi\rangle_{\mbox{\scriptsize AC}}=1/2, \nonumber \\
{}_{\mbox{\scriptsize AC}}\langle r_{1}|\hat{P}(\sigma_{x}=-1)|\psi\rangle_{\mbox{\scriptsize AC}}=0,
\end{eqnarray}
with ease.
Thus, describing the probability that Bob's outcome is `$+1$' for the measurement of $\sigma_{x}$
on condition that Alice detects $|r_{1}\rangle_{\mbox{\scriptsize AC}}$
as $\mbox{prob}(\sigma_{x}=1,r_{1})$,
we achieve
\begin{eqnarray}
\mbox{prob}(\sigma_{x}=1,r_{1})
&=&
\frac{|_{\mbox{\scriptsize AC}}\langle r_{1}|\hat{P}(\sigma_{x}=1)|\psi\rangle_{\mbox{\scriptsize AC}}|^{2}}
{\sum_{i\in\{1,-1\}}|_{\mbox{\scriptsize AC}}\langle r_{1}|\hat{P}(\sigma_{x}=i)|\psi\rangle_{\mbox{\scriptsize AC}}|^{2}} \nonumber \\
&=&
1,
\label{prob_sigma_x_plus_r1}
\end{eqnarray}
according to the ABL-rule, that is to say Eq.~(\ref{ABL-rule-01}).
Similarly, we arrive at
\begin{eqnarray}
\mbox{prob}(\sigma_{x}=-1,r_{1})
&=&
\frac{|_{\mbox{\scriptsize AC}}\langle r_{1}|\hat{P}(\sigma_{x}=-1)|\psi\rangle_{\mbox{\scriptsize AC}}|^{2}}
{\sum_{i\in\{1,-1\}}|_{\mbox{\scriptsize AC}}\langle r_{1}|\hat{P}(\sigma_{x}=i)|\psi\rangle_{\mbox{\scriptsize AC}}|^{2}} \nonumber \\
&=&
0,
\label{prob_sigma_x_minus_r1}
\end{eqnarray}
as well.

Equations~(\ref{prob_sigma_x_plus_r1}) and (\ref{prob_sigma_x_minus_r1}) imply that Bob obtains the outcome `$+1$' for the observation of $\sigma_{x}$
with a probability of unity
if Alice detects $|r_{1}\rangle_{\mbox{\scriptsize AC}}$.
Moreover, carrying out other similar calculations,
we obtain
\begin{eqnarray}
&&
\mbox{prob}(\sigma_{y}=1,r_{1})
=
1, \quad
\mbox{prob}(\sigma_{y}=-1,r_{1})
=
0, \nonumber \\
&&
\mbox{prob}(\sigma_{z}=1,r_{1})
=
1, \quad
\mbox{prob}(\sigma_{z}=-1,r_{1})
=
0,
\end{eqnarray}
and we can show that Bob obtains the outcome `$+1$' for the observations of $\sigma_{y}$ and $\sigma_{z}$
with a probability of unity if Alice finds $|r_{1}\rangle_{\mbox{\scriptsize AC}}$.

Alice's detections of $|r_{2}\rangle_{\mbox{\scriptsize AC}}$, $|r_{3}\rangle_{\mbox{\scriptsize AC}}$, and $|r_{4}\rangle_{\mbox{\scriptsize AC}}$
also lead to Bob's determined outcomes of measurements
for $\sigma_{x}$, $\sigma_{y}$, and $\sigma_{z}$.
We summarize these results in Table~\ref{Table01}.
Then, we have reached an incredible result.
Although the operators $\sigma_{x}$, $\sigma_{y}$, and $\sigma_{z}$ do not commute with each other,
their outcomes of measurements are ascertained with a probability of unity
in the case where Alice detects $\{|r_{i}\rangle_{\mbox{\scriptsize AC}}:i=1,2,3,4\}$.
This phenomenon is regarded as one of the pre- and post-selection effects.
Bub's protocol utilizes this counter-intuitive fact for $\sigma_{x}$ and $\sigma_{z}$.

\begin{table}
\caption{Relations of Alice's detection for the observable $R$ and Bob's outcomes of measurements for $\sigma_{x}$, $\sigma_{y}$, and $\sigma_{z}$.
These relations are realized with a probability of unity.}
\label{Table01}
\begin{center}
\begin{tabular}{|c|ccc|}
\hline
$R$ & $\sigma_{x}$ & $\sigma_{y}$ & $\sigma_{z}$ \\
\hline
$|r_{1}\rangle_{\mbox{\scriptsize AC}}$ & $1$  & $1$  & $1$ \\
$|r_{2}\rangle_{\mbox{\scriptsize AC}}$ & $-1$ & $-1$ & $1$ \\
$|r_{3}\rangle_{\mbox{\scriptsize AC}}$ & $1$  & $-1$ & $-1$ \\
$|r_{4}\rangle_{\mbox{\scriptsize AC}}$ & $-1$ & $1$  & $-1$ \\
\hline
\end{tabular}
\end{center}
\end{table}

Here, according to Bub's protocol,
we divide the sequence of communications between Alice and Bob into two subsequences.
The subsequence $S_{14}$ consists of transmissions for which Alice detects $|r_{1}\rangle_{\mbox{\scriptsize AC}}$ or $|r_{4}\rangle_{\mbox{\scriptsize AC}}$.
The subsequence $S_{23}$ consists of transmissions for which she finds $|r_{2}\rangle_{\mbox{\scriptsize AC}}$ or $|r_{3}\rangle_{\mbox{\scriptsize AC}}$.
On the one hand, for $S_{14}$,
if Alice obtains $|r_{1}\rangle_{\mbox{\scriptsize AC}}$,
Bob's outcome has to be `$+1$' for both $\sigma_{x}$ and $\sigma_{z}$.
On the other hand, for $S_{14}$,
if Alice obtains $|r_{4}\rangle_{\mbox{\scriptsize AC}}$,
Bob's outcome has to be `$-1$'
for both $\sigma_{x}$ and $\sigma_{z}$.
Contrastingly, on the one hand, for $S_{23}$,
if Alice detects $|r_{2}\rangle_{\mbox{\scriptsize AC}}$,
Bob obtains the outcome `$-1$' for $\sigma_{x}$ and the outcome `$+1$' for $\sigma_{z}$.
On the other hand, for $S_{23}$,
if Alice detects $|r_{3}\rangle_{\mbox{\scriptsize AC}}$,
Bob obtains the outcome `$+1$' for $\sigma_{x}$ and the outcome `$-1$' for $\sigma_{z}$.

Alice and Bob utilize the subsequence $S_{23}$
for checking whether or not the channel qubits are monitored by Eve.
By contrast, they use the subsequence $S_{14}$ to establish a shared random secret key.
First of all, Alice publicly announces the indices of the subsequence $S_{23}$
via the classical channel.
At the same time, Alice discloses whether she detects $|r_{2}\rangle_{\mbox{\scriptsize AC}}$ or $|r_{3}\rangle_{\mbox{\scriptsize AC}}$
in each transmission of $S_{23}$.
Because Alice uses the classical channel for making the public announcements,
Eve knows these pieces of information.
Receiving these public notices,
Bob examines whether or not his outcomes for $\sigma_{x}$ and $\sigma_{z}$ are consistent with Alice's announcements.
If he finds even a single discrepancy between his measurements and Alice's disclosed statements,
he concludes that Eve eavesdrops on their transmissions.
By contrast, if Bob cannot find any contradictions,
he believes that there is no illegal act executed by Eve.
If Bob judges that the transmissions are not interfered in by Eve,
Alice obtains a series of `$+1$' and `$-1$' according to detections of $|r_{1}\rangle_{\mbox{\scriptsize AC}}$ and $|r_{4}\rangle_{\mbox{\scriptsize AC}}$
and Bob obtains that by the outcomes of measurements of $\sigma_{x}$ and $\sigma_{z}$.

The probabilities that Alice obtains $|r_{i}\rangle_{\mbox{\scriptsize AC}}$ for $i=1,2,3,4$
on average of Bob's outcomes are the same and given by $1/4$
if Alice and Bob follow the protocol correctly without Eve's disturbance.

As explained in this section,
Bub's protocol requires the observation of two-qubit entangled states.
This requirement is severer compared with the BB84 and E91 schemes.
However,
Bub's protocol is an application of the ABL-rule,
the measurement problem in quantum mechanics,
so that it is very unique and interesting among various quantum cryptography schemes.
This is why we study Bub's protocol.

\section{\label{section-previous-works}Differences between early works and our results}
In this section, we explain differences between previous works,
Refs.~\cite{Bub2001,Yoshida2010,Yoshida2012,Werner2009} and \cite{Reimpell2007},
and our results.
In the security analyses of these references,
error-free transmissions were assumed, so that robustness proofs (and not security proofs) were provided.
First of all,
we give a short review of Refs.~\cite{Werner2009} and \cite{Reimpell2007}.
Reference~\cite{Werner2009} is a sequel to Ref.~\cite{Reimpell2007}.

The mean king's problem is the following game
played by Alice and the king.
First, Alice prepares a maximally entangled state on ${\cal H}_{1}\otimes{\cal H}_{2}$
where ${\cal H}_{1}$ and ${\cal H}_{2}$ are Hilbert spaces and $\mbox{dim}{\cal H}_{1}=\mbox{dim}{\cal H}_{2}=d$.
For example, we can suppose that $d=2$ and ${\cal H}_{1}$ and ${\cal H}_{2}$ are made of single qubits.

Second, Alice sends a particle of ${\cal H}_{2}$ to the king.
Third, the king gets $(d+1)$ orthonormal bases on ${\cal H}_{2}$ ready.
Werner {\it et al}.  described these bases as
$\{|\Phi_{b}(i)\rangle:i=1, ..., d, b=1, ..., d+1\}$.
If ${\cal H}_{1}$ and ${\cal H}_{2}$ are constructed with two single qubits,
we can prepare
$|\Phi_{z}(1)\rangle=|0\rangle$,
$|\Phi_{z}(2)\rangle=|1\rangle$,
$|\Phi_{x}(1)\rangle=|+\rangle$,
$|\Phi_{x}(2)\rangle=|-\rangle$,
$|\Phi_{y}(1)\rangle=|\tilde{0}\rangle$,
and
$|\Phi_{y}(2)\rangle=|\tilde{1}\rangle$,
where
$\{|\tilde{0}\rangle,|\tilde{1}\rangle\}$
are eigenvectors of $\sigma_{y}$.

Choosing one basis $b\in\{1, ..., d+1\}$ on ${\cal H}_{2}$ at random and applying it to the particle of the maximally entangled state sent by Alice
for performing the von Neumann measurement,
the king obtains an outcome of the observation $i\in\{1, ..., d\}$.
The king keeps $b$ and $i$ secret and returns the particle that he has observed on ${\cal H}_{2}$ to Alice.
Fourth, Alice executes a measurement of the whole system defined on ${\cal H}_{1}\otimes{\cal H}_{2}$
with positive operators $\{F_{x}\}$ and obtains an output $x$.
Fifth, the king discloses $b$ to Alice.
If Alice correctly names $i$ from $x$ and $b$ with a probability of unity,
Alice wins the game.

In Ref.~\cite{Reimpell2007},
the following was shown.
Reimpell and Werner defined ${\cal R}$ as a space spanned by Hermitian operators $\{|\Phi_{b}(i)\rangle\langle\Phi_{b}(i)|: i=1, ...,d, b=1, ..., d+1\}$.
If $\mbox{dim}{\cal R}=d^{2}$ holds, that is to say the chosen basis set is non-degenerate,
and $\{|\Phi_{b}(i)\rangle\}$ admit a classical model,
there exists a safe strategy for Alice to win the game.
That the $(d+1)$ bases $\{|\Phi_{b}(i)\rangle\}$ admit a classical model implies the following.
There exists a probability distribution of $(d+1)$ variables,
each of which takes $d$ values,
and its marginals are equal to the probability distributions of the joint probabilities
$p_{ab}(i,j)=(1/d)|\langle\Phi_{a}(i)|\Phi_{b}(j)\rangle|^{2}$
for all pairs of bases.
If Alice chooses a measurement $\{F_{x}\}$ that incorporates all of the projectors
$p(x)|\eta_{x}\rangle\langle\eta_{x}|$ for $p(x)\neq 0$,
her strategy is regarded as maximal,
where $|\eta_{x}\rangle$ was defined in Ref.~\cite{Reimpell2007} and named the safe vector.

In the case where we choose eigenvectors of $\sigma_{x}$, $\sigma_{y}$, and $\sigma_{z}$ as $\{|\Phi_{b}(i)\rangle\}$,
${\cal R}$ is a space spanned by $2\times 2$ Hermitian matrices and the dimension of ${\cal R}$ is given by four,
so that the chosen basis set is non-degenerate.
Here, we explain this circumstance in detail.
We take an arbitrary density operator $\rho\in{\cal R}$.
Then, the number of real parameters of $\rho$ obtained from
$\mbox{tr}[\rho|\Phi_{b}(i)\rangle\langle\Phi_{b}(i)|]$ for $i=1,2$
is equal to one because the relation
$\sum_{i}\mbox{tr}[\rho|\Phi_{b}(i)\rangle\langle\Phi_{b}(i)|]=\mbox{tr}\rho$
reduces the degree of freedom of the real parameters.
The number of bases is given by three,
that is to say,
for $\sigma_{x}$, $\sigma_{y}$, and $\sigma_{z}$,
and we obtain three parameters in total.
Finally, adding one real parameter $\mbox{tr}\rho$ to them,
we obtain four real parameters which specify a $2\times 2$ Hermitian matrix perfectly.
This situation is expressed as the word ``non-degenerate".

In Ref.~\cite{Werner2009},
the following quantum key distribution protocol was proposed as an application of the mean king problem.
Here, it is assumed that Alice has the maximal and successful strategy,
where the word ``successful" means that Alice makes a wrong guess with a probability of zero.
\begin{enumerate}
\item Alice and Bob share $n$ maximally entangled states with each other.
\item {Bob chooses $n$ bases for observations as $\mbox{\boldmath $b$}=(b_{1}, ..., b_{n})$.
       He performs a projective measurement upon the $k$th particle with the basis $b_{k}$ and obtains an outcome $i_{k}$.
       Then, Bob's states reduce to $|\Phi_{\mbox{\scriptsize \boldmath $b$}}(\mbox{\boldmath $i$})\rangle=\otimes_{k=1}^{n}|\Phi_{b_{k}}(i_{k})\rangle$.
       Bob keeps $\mbox{\boldmath $b$}$ and $\mbox{\boldmath $i$}$ secret and returns $|\Phi_{\mbox{\scriptsize \boldmath $b$}}(\mbox{\boldmath $i$})\rangle$ to Alice.}
\item {After Alice executes the measurement $\{F_{x}\}$ on the particles sent by Bob and the ones she keeps close at hand,
       she obtains guess functions $\mbox{\boldmath $x$}$ as outputs of the observation.}
\item {Alice tells Bob that she has completed the measurement.
       Bob discloses $\mbox{\boldmath $b$}$ to Alice.
       Then, Alice obtains $i'_{k}=x_{k}(b_{k})$ for $k=1, ..., n$.
       If Eve does not disturb the transmissions between Alice and Bob,
       $i'_{k}=i_{k}$ holds for $i=1, ..., n$ and Alice and Bob share a random secret string of bits.
       At this stage, Alice and Bob can detect Eve's interference by selecting the $k$th particle at random
       and confirming whether or not $i_{k}$ and $i'_{k}$ correspond to each other through the classical channel.}
\end{enumerate}

In Ref.~\cite{Werner2009},
in order to provide robustness analysis (and not security analysis),
it is assumed that there are no transmission errors during the protocol.
In step one, Werner {\it et al}. assumed that Eve could replace Alice and Bob's initial states with states Eve preferred,
so that Eve's particles entangled themselves with Alice and Bob's particles.
Moreover, in step two, they assumed that Eve could make a coherent attack on the quantum channel
through which Bob returned the particles having been observed with $\{|\Phi_{b}(i)\rangle\}$ to Alice.
Under these attacks of Eve's, the following was proved in Ref.~\cite{Werner2009}:
If Alice and Bob eventually share the same random key string,
Eve cannot learn anything about it.

There are differences between the quantum key distribution scheme proposed by Werner {\it et al}. in Ref.~\cite{Werner2009}
and Bub's protocol.
In the scheme of Werner {\it et al}.,
Alice and Bob select the $k$th particle at random and disclose $i_{k}$ and $i'_{k}$ through the classical channel to detect Eve's interference.
Eve can change methods for observing the particles she keeps close at hand according to $\mbox{\boldmath $b$}$ disclosed by Bob,
so that she can enlarge the amount of information obtained by eavesdropping.
By contrast, in Bub's protocol,
Alice and Bob detect Eve's malicious acts from the subsequence $S_{23}$ and establish a random bit string from the subsequence $S_{14}$.
Eve observes the particles that she keeps close at hand for eavesdropping on Alice and Bob's bit string in the subsequence of $S_{14}$.

In Ref.~\cite{Werner2009},
Werner {\it et al}. showed the following.
Even if Eve steals only a little bit of information from eavesdropping on Alice and Bob's transmission,
there must exist discrepancies between Alice and Bob's random bit strings,
so that Alice and Bob can detect Eve's interference in a probabilistic manner.

However, Ref.~\cite{Werner2009} did not estimate the probability $P_{\mbox{\scriptsize E}}$ that Eve guessed right at the random secret bit Alice and Bob established
or the probability $P_{\mbox{\scriptsize AB}}$ that Alice and Bob did not notice Eve's disturbance analytically or numerically.
Contrastingly, in the current paper,
we calculate both these probabilities for the intercept/resend and translucent attacks rigorously.
Although we impose some constraints upon Eve's strategies,
we evaluate effects caused by Eve's attacks in concrete terms.
These results are differences between Ref.~\cite{Werner2009} and the present paper.
Furthermore, they are main conclusions of the current paper.

Next, we explain differences between Refs.~\cite{Yoshida2010} and \cite{Yoshida2012} and our work.
In Ref.~\cite{Yoshida2010},
Yoshida {\it et al}. examined the security of slightly modified Bub's protocol,
in which the legitimate users repeated tasks of the transmission $2n$ times,
chose $n$ bits at random through the public channel,
estimated an error rate,
and detected Eve's eavesdropping according to the error rate.
They derived trade-off inequalities between the information Eve gained and the error probability Alice and Bob calculated for two attack scenarios.
Yoshida {\it et al}. considered the following scenarios of Eve's.
The first one was a one-way translucent attack on the quantum channel
where the qubit was flying from Alice to Bob.
The second one was a one-way translucent attack on the quantum channel where the qubit was travelling from Bob to Alice.
Thus, in Ref.~\cite{Yoshida2010},
the two-way translucent attack was not studied.

In Ref.~\cite{Yoshida2012}, Yoshida {\it et al}. proposed three protocols in which Alice used simplified observables to solve the mean king's problem.
Bub's original protocol employs a projective measurement with an entangled orthogonal basis for Alice's two-qubit observation.
By contrast, the proposals of Yoshida {\it et al}. used unentangled observables for two-qubit measurements.
They examined the security of their protocols against the following three attacks of Eve's.
The first one was a one-way translucent attack on the qubit going on the way.
The second one was a one-way translucent attack on the qubit going the way back.
The third one was slightly not usual.
In this case, two eavesdroppers Eve1 and Eve2 appeared.
Eve1 and Eve2 eavesdropped on the qubit travelling from Alice to Bob and one flying from Bob to Alice, respectively and independently.
Thus, the third scenario can be regarded as a one-way translucent attack, as well.

Therefore, in Refs.~\cite{Yoshida2010} and \cite{Yoshida2012},
the two-way translucent attack was not investigated.
This is the difference between the works of Yoshida {\it et al}. and the current paper.

Finally, we explain the relation between the security analysis of Ref.~\cite{Bub2001} and our results.
In Ref.~\cite{Bub2001}, Bub wrote that the probability that the legitimate users detected Eve's intercept/resend attack
with observables $\sigma_{x}$, $\sigma_{y}$, and $\sigma_{z}$ was equal to $3/8$.
In the current paper, we examine more general intercept/resend attacks,
and this point is one of our aims for the present work.

\section{\label{section-one-way-intercept-resend-attack}The one-way intercept/resend attack}
We define the one-way intercept/resend attack as follows.
Eve measures the single qubit Alice sends with an orthonormal basis $\{|\xi_{0}\rangle,|\xi_{1}\rangle\}$
and resends an alternative one to Bob with the same basis $\{|\xi_{0}\rangle,|\xi_{1}\rangle\}$
according to the result of Eve's observation.
This process is equal to the situation where Eve performs the measurement on the single qubit Alice sends with the projection operators
$\hat{P}(\sigma_{\xi}=1)$ and $\hat{P}(\sigma_{\xi}=-1)$.

In this section, we consider the one-way attack where Eve measures $\sigma_{\xi}$ on the channel qubit sent by Alice and resends it to Bob.
We assume that $\sigma_{\xi}$ represents an observable of the spin along an arbitrary direction.

We start deriving an explicit form of the projection operators of $\sigma_{\xi}$.
We prepare Euler's rotation matrix for SU(2) as follows \cite{Sakurai}:
\begin{eqnarray}
U(\alpha,\beta,\gamma)
&=&
\exp\Biggl(-\frac{i}{2}\alpha\sigma_{z}\Biggr)
\exp\Biggl(-\frac{i}{2}\beta\sigma_{y}\Biggr)
\exp\Biggl(-\frac{i}{2}\gamma\sigma_{z}\Biggr) \nonumber \\
&=&
\left(
\begin{array}{cc}
e^{-i(\alpha+\gamma)/2} \cos(\beta/2) & -e^{-i(\alpha-\gamma)/2} \sin(\beta/2) \\
e^{i(\alpha-\gamma)/2} \sin(\beta/2) & e^{i(\alpha+\gamma)/2} \cos(\beta/2) \\
\end{array}
\right),
\end{eqnarray}
where $0\leq\alpha<4\pi$, $0\leq\beta<4\pi$, and $0\leq\gamma<4\pi$.
Using $U(\alpha,\beta,\gamma)$, we can write down the projection operators of $\sigma_{\xi}$ as
\begin{eqnarray}
\hat{P}(\sigma_{\xi}=1)
&=&
U(\alpha,\beta,\gamma)
\left(
\begin{array}{cc}
1 & 0 \\
0 & 0 \\
\end{array}
\right)
U^{\dagger}(\alpha,\beta,\gamma) \nonumber \\
&=&
\left(
\begin{array}{cc}
\cos^{2}(\beta/2) & e^{-i\alpha}\sin(\beta/2)\cos(\beta/2) \\
e^{i\alpha}\sin(\beta/2)\cos(\beta/2) & \sin^{2}(\beta/2) \\
\end{array}
\right),
\label{projection-sigma-plus}
\end{eqnarray}
\begin{eqnarray}
\hat{P}(\sigma_{\xi}=-1)
&=&
U(\alpha,\beta,\gamma)
\left(
\begin{array}{cc}
0 & 0 \\
0 & 1 \\
\end{array}
\right)
U^{\dagger}(\alpha,\beta,\gamma) \nonumber \\
&=&
\left(
\begin{array}{cc}
\sin^{2}(\beta/2) & -e^{-i\alpha}\sin(\beta/2)\cos(\beta/2) \\
-e^{i\alpha}\sin(\beta/2)\cos(\beta/2) & \cos^{2}(\beta/2) \\
\end{array}
\right).
\label{projection-sigma-minus}
\end{eqnarray}
For example,
putting $(\alpha,\beta)=(0,0)$, $(0,\pi/2)$, and $(\pi/2,\pi/2)$,
we obtain $\sigma_{\xi}=\sigma_{z}$, $\sigma_{x}$, and $\sigma_{y}$, respectively.

Here, we define the probability that is useful for discussion in the rest of this section and the next section.
For example, we describe the probability that Bob has $\sigma_{x}=i$ and Eve obtains $\sigma_{\xi}=j$
on condition that Alice detects $|r_{2}\rangle_{\mbox{\scriptsize AC}}$ as
\begin{eqnarray}
\mbox{prob}(\sigma_{x}=i,\sigma_{\xi}=j,r_{2})
&=&
\frac{|_{\mbox{\scriptsize AC}}\langle r_{2}|\hat{P}(\sigma_{x}=i)\hat{P}(\sigma_{\xi}=j)|\psi\rangle_{\mbox{\scriptsize AC}}|^{2}}
{\sum_{k,l\in\{1,-1\}}|_{\mbox{\scriptsize AC}}\langle r_{2}|\hat{P}(\sigma_{x}=k)\hat{P}(\sigma_{\xi}=l)|\psi\rangle_{\mbox{\scriptsize AC}}|^{2}} \nonumber \\
&&\quad
\mbox{for $i,j\in\{1,-1\}$}.
\label{prob-sigma-x-r2-one-way-intercept-resend-attak}
\end{eqnarray}

In the following paragraphs,
we consider Eve's strategy.
First, Eve needs to let Alice and Bob not notice her illegal acts.
To think in concrete terms,
we assume a case where Bob observes $\sigma_{x}$ and Alice detects $|r_{2}\rangle_{\mbox{\scriptsize AC}}$.
In this case, Eve had better make
$\sum_{j\in\{1,-1\}}\mbox{prob}(\sigma_{x}=-1,\sigma_{\xi}=j,r_{2})$
greater in value and let Alice and Bob not find evidence of her eavesdropping.
Similarly,
if Bob observes $\sigma_{z}$ and Alice detects $|r_{2}\rangle_{\mbox{\scriptsize AC}}$,
Eve has to make
$\sum_{j\in\{1,-1\}}\mbox{prob}(\sigma_{z}=1,\sigma_{\xi}=j,r_{2})$
greater in value.
If Bob measures $\sigma_{x}$ and Alice finds $|r_{3}\rangle_{\mbox{\scriptsize AC}}$,
Eve had better let
$\sum_{j\in\{1,-1\}}\mbox{prob}(\sigma_{x}=1,\sigma_{\xi}=j,r_{3})$
be larger.
If Bob measures $\sigma_{z}$ and Alice finds $|r_{3}\rangle_{\mbox{\scriptsize AC}}$,
Eve should have
$\sum_{j\in\{1,-1\}}\mbox{prob}(\sigma_{z}=-1,\sigma_{\xi}=j,r_{3})$
larger in value.

Second, Eve has to guess right at the random bit of the secret key that Alice obtains.
To put the discussion more concretely,
we consider a case where Bob observes $\sigma_{x}$ and Alice detects $|r_{1}\rangle_{\mbox{\scriptsize AC}}$.
In this case, Eve had better make
$\sum_{i\in\{1,-1\}}\mbox{prob}(\sigma_{x}=i,\sigma_{\xi}=1,r_{1})$
greater in value.
If Eve wants to guess right at the random secret bit that Bob obtains,
she has to let
$\sum_{i\in\{1,-1\}}\mbox{prob}(\sigma_{x}=i,\sigma_{\xi}=i,r_{1})$
be larger.
However, in the current paper,
we do not examine this strategy.
If Bob measures $\sigma_{z}$ and Alice detects $|r_{1}\rangle_{\mbox{\scriptsize AC}}$,
Eve has to enlarge
$\sum_{i\in\{1,-1\}}\mbox{prob}(\sigma_{z}=i,\sigma_{\xi}=1,r_{1})$.
If Bob observes $\sigma_{x}$ and Alice detects $|r_{4}\rangle_{\mbox{\scriptsize AC}}$,
Eve should have
$\sum_{i\in\{1,-1\}}\mbox{prob}(\sigma_{x}=i,\sigma_{\xi}=-1,r_{4})$
larger.
If Bob measures $\sigma_{z}$ and Alice finds $|r_{4}\rangle_{\mbox{\scriptsize AC}}$,
Eve had better enlarge
$\sum_{i\in\{1,-1\}}\mbox{prob}(\sigma_{z}=i,\sigma_{\xi}=-1,r_{4})$.

To evaluate the probabilities that Alice and Bob do not notice Eve's malicious acts
in the subsequence $S_{23}$,
we prepare eight functions $f_{k}(\alpha,\beta)$ and $g_{k}(\alpha,\beta)$ for $k=1, 2, 3, 4$
in Eqs.~(\ref{f1g1-alpha-beta-01}), (\ref{f2g2-alpha-beta-02}), (\ref{f3g3-alpha-beta-03}), and (\ref{f4g4-alpha-beta-04})
in Appendix~\ref{section-appendix-A}.
Then, the following relations hold between the probabilities and the eight functions:
\begin{eqnarray}
\sum_{j\in\{1,-1\}}\mbox{prob}(\sigma_{x}=1,\sigma_{\xi}=j,r_{2})
&\propto&
f_{1}(\alpha,\beta), \nonumber \\
\sum_{j\in\{1,-1\}}\mbox{prob}(\sigma_{x}=-1,\sigma_{\xi}=j,r_{2})
&\propto&
g_{1}(\alpha,\beta), \nonumber \\
\sum_{j\in\{1,-1\}}\mbox{prob}(\sigma_{z}=1,\sigma_{\xi}=j,r_{2})
&\propto&
f_{2}(\alpha,\beta), \nonumber \\
\sum_{j\in\{1,-1\}}\mbox{prob}(\sigma_{z}=-1,\sigma_{\xi}=j,r_{2})
&\propto&
g_{2}(\alpha,\beta), \nonumber \\
\sum_{j\in\{1,-1\}}\mbox{prob}(\sigma_{x}=1,\sigma_{\xi}=j,r_{3})
&\propto&
f_{3}(\alpha,\beta), \nonumber \\
\sum_{j\in\{1,-1\}}\mbox{prob}(\sigma_{x}=-1,\sigma_{\xi}=j,r_{3})
&\propto&
g_{3}(\alpha,\beta), \nonumber \\
\sum_{j\in\{1,-1\}}\mbox{prob}(\sigma_{z}=1,\sigma_{\xi}=j,r_{3})
&\propto&
f_{4}(\alpha,\beta), \nonumber \\
\sum_{j\in\{1,-1\}}\mbox{prob}(\sigma_{z}=-1,\sigma_{\xi}=j,r_{3})
&\propto&
g_{4}(\alpha,\beta).
\label{probabilities-fg}
\end{eqnarray}

Using these functions,
we can evaluate the probability that Alice and Bob cannot notice evidence of Eve's illegal acts as follows.
Alice and Bob do not become aware of Eve's interference if results of their measurements are consistent with the relations shown in Table~\ref{Table01}.
The probability that Bob obtains $\sigma_{x}=-1$ in the case where Alice detects $|r_{2}\rangle_{\mbox{\scriptsize AC}}$ is given by
\begin{equation}
\sum_{j\in\{1,-1\}}\mbox{prob}(\sigma_{x}=-1,\sigma_{\xi}=j,r_{2})
=
g_{1}(\alpha,\beta)/[f_{1}(\alpha,\beta)+g_{1}(\alpha,\beta)].
\end{equation}
The probability that Bob has $\sigma_{z}=1$ in the case where Alice finds $|r_{2}\rangle_{\mbox{\scriptsize AC}}$ is given by
\begin{equation}
\sum_{j\in\{1,-1\}}\mbox{prob}(\sigma_{z}=1,\sigma_{\xi}=j,r_{2})
=
f_{2}(\alpha,\beta)/[f_{2}(\alpha,\beta)+g_{2}(\alpha,\beta)].
\end{equation}
The probability that Bob obtains $\sigma_{x}=1$ on condition that Alice detects $|r_{3}\rangle_{\mbox{\scriptsize AC}}$ is given by
\begin{equation}
\sum_{j\in\{1,-1\}}\mbox{prob}(\sigma_{x}=1,\sigma_{\xi}=j,r_{3})
=
f_{3}(\alpha,\beta)/[f_{3}(\alpha,\beta)+g_{3}(\alpha,\beta)].
\end{equation}
The probability that Bob has $\sigma_{z}=-1$ on condition that Alice finds $|r_{3}\rangle_{\mbox{\scriptsize AC}}$ is given by
\begin{equation}
\sum_{j\in\{1,-1\}}\mbox{prob}(\sigma_{z}=-1,\sigma_{\xi}=j,r_{3})
=
g_{4}(\alpha,\beta)/[f_{4}(\alpha,\beta)+g_{4}(\alpha,\beta)].
\end{equation}

To let Alice and Bob not find a sign of Eve's eavesdropping,
Eve has to make $g_{1}(\alpha,\beta)$, $f_{2}(\alpha,\beta)$, $f_{3}(\alpha,\beta)$, and $g_{4}(\alpha,\beta)$ larger
and
$f_{1}(\alpha,\beta)$, $g_{2}(\alpha,\beta)$, $g_{3}(\alpha,\beta)$, and $f_{4}(\alpha,\beta)$ smaller in value.
However, only from this principle,
it is difficult for us to obtain optimum $\alpha$ and $\beta$ for Eve's attack.

Thus, to let the problem be simple,
we make it a condition that the following relation holds:
\begin{equation}
f_{1}(\alpha,\beta)
=
g_{2}(\alpha,\beta)
=
g_{3}(\alpha,\beta)
=
f_{4}(\alpha,\beta).
\label{Eve-strategy-symmetric-condition-01}
\end{equation}
In other words, Eve sets a plan in which the parameters $\alpha$ and $\beta$ satisfy Eq.~(\ref{Eve-strategy-symmetric-condition-01}).
In Sec.~\ref{section-symmetric-strategy-Eve}, we analyse this plan of Eve's in detail.

Here, we evaluate the probabilities that Eve guesses right at the random secret bit Alice obtains in the subsequence $S_{14}$.
To perform this evaluation, we prepare eight functions $u_{k}(\alpha,\beta)$ and $v_{k}(\alpha,\beta)$ for $k=1, 2, 3, 4$
in Eqs.~(\ref{u1v1-alpha-beta-01}), (\ref{u2v2-alpha-beta-02}), (\ref{u3v3-alpha-beta-03}), and (\ref{u4v4-alpha-beta-04})
in Appendix~\ref{section-appendix-A}.
Then, the following relations hold between the probabilities and the eight functions:
\begin{eqnarray}
\sum_{i\in\{1,-1\}}\mbox{prob}(\sigma_{x}=i,\sigma_{\xi}=1,r_{1})
&\propto&
u_{1}(\alpha,\beta), \nonumber \\
\sum_{i\in\{1,-1\}}\mbox{prob}(\sigma_{x}=i,\sigma_{\xi}=-1,r_{1})
&\propto&
v_{1}(\alpha,\beta), \nonumber \\
\sum_{i\in\{1,-1\}}\mbox{prob}(\sigma_{z}=i,\sigma_{\xi}=1,r_{1})
&\propto&
u_{2}(\alpha,\beta), \nonumber \\
\sum_{i\in\{1,-1\}}\mbox{prob}(\sigma_{z}=i,\sigma_{\xi}=-1,r_{1})
&\propto&
v_{2}(\alpha,\beta), \nonumber \\
\sum_{i\in\{1,-1\}}\mbox{prob}(\sigma_{x}=i,\sigma_{\xi}=1,r_{4})
&\propto&
u_{3}(\alpha,\beta), \nonumber \\
\sum_{i\in\{1,-1\}}\mbox{prob}(\sigma_{x}=i,\sigma_{\xi}=-1,r_{4})
&\propto&
v_{3}(\alpha,\beta), \nonumber \\
\sum_{i\in\{1,-1\}}\mbox{prob}(\sigma_{z}=i,\sigma_{\xi}=1,r_{4})
&\propto&
u_{4}(\alpha,\beta), \nonumber \\
\sum_{i\in\{1,-1\}}\mbox{prob}(\sigma_{z}=i,\sigma_{\xi}=-1,r_{4})
&\propto&
v_{4}(\alpha,\beta).
\label{probabilities-uv}
\end{eqnarray}

We can derive the probability that Eve guesses right at the random secret bit Alice obtains as follows.
If Bob observes $\sigma_{x}$ and Alice detects $|r_{1}\rangle_{\mbox{\scriptsize AC}}$,
it is given by
\\
$u_{1}(\alpha,\beta)/[u_{1}(\alpha,\beta)+v_{1}(\alpha,\beta)]$.
If Bob observes $\sigma_{z}$ and Alice finds $|r_{1}\rangle_{\mbox{\scriptsize AC}}$,
it is given by
$u_{2}(\alpha,\beta)/[u_{2}(\alpha,\beta)+v_{2}(\alpha,\beta)]$.
If Bob measures $\sigma_{x}$ and Alice detects $|r_{4}\rangle_{\mbox{\scriptsize AC}}$,
it is given by
$v_{3}(\alpha,\beta)/[u_{3}(\alpha,\beta)+v_{3}(\alpha,\beta)]$.
If Bob measures $\sigma_{z}$ and Alice finds $|r_{4}\rangle_{\mbox{\scriptsize AC}}$,
it is given by
$v_{4}(\alpha,\beta)/[u_{4}(\alpha,\beta)+v_{4}(\alpha,\beta)]$.

We describe the probability that Alice detects $|r_{i}\rangle_{\mbox{\scriptsize AC}}$ for $i=1,2,3,4$ as $Q_{i}$.
Then, we obtain the following relation:
\begin{equation}
Q_{1}:Q_{2}:Q_{3}:Q_{4}
=
u_{1}+v_{1}+u_{2}+v_{2}
:
f_{1}+g_{1}+f_{2}+g_{2}
:
f_{3}+g_{3}+f_{4}+g_{4}
:
u_{3}+v_{3}+u_{4}+v_{4}.
\end{equation}

\section{\label{section-symmetric-strategy-Eve}Eve's strategies where Eq.~(\ref{Eve-strategy-symmetric-condition-01}) holds}
In this section,
we consider Eve's strategies where Eq.~(\ref{Eve-strategy-symmetric-condition-01}) holds.
Then, we obtain the following relation:
\begin{equation}
\cos^{2}\beta-\cos^{2}\alpha\sin^{2}\beta=0.
\end{equation}
Hence, the parameter $\beta$ is a function of the parameter $\alpha$,
\begin{equation}
\beta
=
\arctan
\Biggl(
\pm \frac{1}{\cos\alpha}
\Biggr).
\end{equation}

\subsection{\label{subsection-01}The case where $\beta=\arctan(1/\cos\alpha)$ holds}
In this subsection, we consider the case where $\beta=\arctan(1/\cos\alpha)$ holds.
Substituting $\beta=\arctan(1/\cos\alpha)$ into Eqs.~(\ref{f1g1-alpha-beta-01}),
(\ref{f2g2-alpha-beta-02}),
(\ref{f3g3-alpha-beta-03}),
(\ref{f4g4-alpha-beta-04}),
(\ref{u1v1-alpha-beta-01}),
(\ref{u2v2-alpha-beta-02}),
(\ref{u3v3-alpha-beta-03}),
and (\ref{u4v4-alpha-beta-04}),
we obtain the following functions, where we use the notation
$\left.
f_{1}(\alpha)=f_{1}(\alpha,\beta)
\right|_{\beta=\arctan(1/\cos\alpha)}$:
\begin{eqnarray}
f_{1}(\alpha)
&=&
\frac{1}{8[3+\cos(2\alpha)]}, \nonumber \\
g_{1}(\alpha)
&=&
\frac{4+\cos(2\alpha)+\sin(2\alpha)}{8[3+\cos(2\alpha)]},
\end{eqnarray}
\begin{eqnarray}
f_{2}(\alpha)
&=&
\frac{1}{32}
\Biggl[
4+\frac{4-2\sin(2\alpha)}{3+\cos(2\alpha)}
\Biggr], \nonumber \\
g_{2}(\alpha)
&=&
\frac{1}{8[3+\cos(2\alpha)]},
\end{eqnarray}
\begin{eqnarray}
f_{3}(\alpha)
&=&
\frac{4+\cos(2\alpha)-\sin(2\alpha)}{8[3+\cos(2\alpha)]}, \nonumber \\
g_{3}(\alpha)
&=&
\frac{1}{8[3+\cos(2\alpha)]},
\end{eqnarray}
\begin{eqnarray}
f_{4}(\alpha)
&=&
\frac{1}{8[3+\cos(2\alpha)]}, \nonumber \\
g_{4}(\alpha)
&=&
\frac{4+\cos(2\alpha)+\sin(2\alpha)}{8[3+\cos(2\alpha)]},
\end{eqnarray}
\begin{eqnarray}
u_{1}(\alpha)
&=&
\frac{1}{16}
\Biggl(
2
+
\frac{3+\tan\alpha}{\sqrt{1+\sec^{2}\alpha}}
+
\frac{1+\tan\alpha}{1+\sec^{2}\alpha}
\Biggr), \nonumber \\
v_{1}(\alpha)
&=&
\frac{1}{16}
\Biggl(
2
-
\frac{3+\tan\alpha}{\sqrt{1+\sec^{2}\alpha}}
+
\frac{1+\tan\alpha}{1+\sec^{2}\alpha}
\Biggr),
\end{eqnarray}
\begin{eqnarray}
u_{2}(\alpha)
&=&
\frac{1}{8}
\cos^{2}
\Biggl(
\frac{1}{2}\arctan(\sec\alpha)
\Biggr)
\Biggl(
2+\frac{1+\tan\alpha}{\sqrt{1+\sec^{2}\alpha}}
\Biggr), \nonumber \\
v_{2}(\alpha)
&=&
-
\frac{1}{8}
\sin^{2}
\Biggl(
\frac{1}{2}\arctan(\sec\alpha)
\Biggr)
\Biggl(
-2+\frac{1+\tan\alpha}{\sqrt{1+\sec^{2}\alpha}}
\Biggr),
\end{eqnarray}
\begin{eqnarray}
u_{3}(\alpha)
&=&
\frac{1}{16(1+\sec^{2}\alpha)^{3/2}}
[-(-1+\sqrt{1+\sec^{2}\alpha})(-3+\tan\alpha) \nonumber \\
&&
\quad
+
\sec^{2}\alpha(-3+2\sqrt{1+\sec^{2}\alpha}+\tan\alpha)], \nonumber \\
v_{3}(\alpha)
&=&
\frac{1}{16(1+\sec^{2}\alpha)^{3/2}}
[-(1+\sqrt{1+\sec^{2}\alpha})(-3+\tan\alpha) \nonumber \\
&&
\quad
+
\sec^{2}\alpha(3+2\sqrt{1+\sec^{2}\alpha}-\tan\alpha)],
\end{eqnarray}
\begin{eqnarray}
u_{4}(\alpha)
&=&
\frac{1}{8}
\sin^{2}
\Biggl(
\frac{1}{2}\arctan(\sec\alpha)
\Biggr)
\Biggl(
2+\frac{-1+\tan\alpha}{\sqrt{1+\sec^{2}\alpha}}
\Biggr), \nonumber \\
v_{4}(\alpha)
&=&
\frac{1}{8}
\cos^{2}
\Biggl(
\frac{1}{2}\arctan(\sec\alpha)
\Biggr)
\Biggl(
2+\frac{1-\tan\alpha}{\sqrt{1+\sec^{2}\alpha}}
\Biggr).
\end{eqnarray}

Here, we pay attention to a relation,
\begin{equation}
[f_{1}(\alpha)+g_{1}(\alpha)+f_{2}(\alpha)+g_{2}(\alpha)]
-
[f_{3}(\alpha)+g_{3}(\alpha)+f_{4}(\alpha)+g_{4}(\alpha)]
=0.
\end{equation}
Thus, the ratio of the probability that Alice detects $|r_{2}\rangle_{\mbox{\scriptsize AC}}$
to the probability that she finds $|r_{3}\rangle_{\mbox{\scriptsize AC}}$
is given by one to one.
Then, we obtain the probability $P_{\mbox{\scriptsize AB}}$ that Alice and Bob do not notice Eve's illegal acts as
\begin{equation}
P_{\mbox{\scriptsize AB}}(\alpha)
=
\frac{1}{4}
\Biggl[
\frac{g_{1}(\alpha)}{f_{1}(\alpha)+g_{1}(\alpha)}
+
\frac{f_{2}(\alpha)}{f_{2}(\alpha)+g_{2}(\alpha)}
+
\frac{f_{3}(\alpha)}{f_{3}(\alpha)+g_{3}(\alpha)}
+
\frac{g_{4}(\alpha)}{f_{4}(\alpha)+g_{4}(\alpha)}
\Biggr].
\label{P-1-alpha}
\end{equation}

By contrast, we pay attention to a relation,
\begin{equation}
[u_{1}(\alpha)+v_{1}(\alpha)+u_{2}(\alpha)+v_{2}(\alpha)]
-
[u_{3}(\alpha)+v_{3}(\alpha)+u_{4}(\alpha)+v_{4}(\alpha)]
=
\frac{\cos\alpha\sin\alpha}{3+\cos(2\alpha)}.
\end{equation}
Thus, in general,
the ratio of the probability that Alice detects $|r_{1}\rangle_{\mbox{\scriptsize AC}}$
to the probability that she finds $|r_{4}\rangle_{\mbox{\scriptsize AC}}$ is not always given by one to one.
Hence, we obtain the probability $P_{\mbox{\scriptsize E}}$ that Eve guesses right at the random secret bit Alice obtains as
\begin{eqnarray}
P_{\mbox{\scriptsize E}}(\alpha)
&=&
\frac{R_{1}(\alpha)}{2}
\Biggl[
\frac{u_{1}(\alpha)}{u_{1}(\alpha)+v_{1}(\alpha)}
+
\frac{u_{2}(\alpha)}{u_{2}(\alpha)+v_{2}(\alpha)}
\Biggr] \nonumber \\
&&\quad
+
\frac{R_{2}(\alpha)}{2}
\Biggl[
\frac{v_{3}(\alpha)}{u_{3}(\alpha)+v_{3}(\alpha)}
+
\frac{v_{4}(\alpha)}{u_{4}(\alpha)+v_{4}(\alpha)}
\Biggr],
\label{P-2-alpha}
\end{eqnarray}
\begin{equation}
R_{1}(\alpha)
=
\frac{1}{R(\alpha)}
[u_{1}(\alpha)+v_{1}(\alpha)+u_{2}(\alpha)+v_{2}(\alpha)],
\label{Q-1-alpha}
\end{equation}
\begin{equation}
R_{2}(\alpha)
=
\frac{1}{R(\alpha)}
[u_{3}(\alpha)+v_{3}(\alpha)+u_{4}(\alpha)+v_{4}(\alpha)],
\label{Q-2-alpha}
\end{equation}
\begin{equation}
R(\alpha)
=
u_{1}(\alpha)+v_{1}(\alpha)+u_{2}(\alpha)+v_{2}(\alpha)+u_{3}(\alpha)+v_{3}(\alpha)+u_{4}(\alpha)+v_{4}(\alpha).
\label{R-alpha}
\end{equation}

\begin{figure}
\begin{center}
\includegraphics{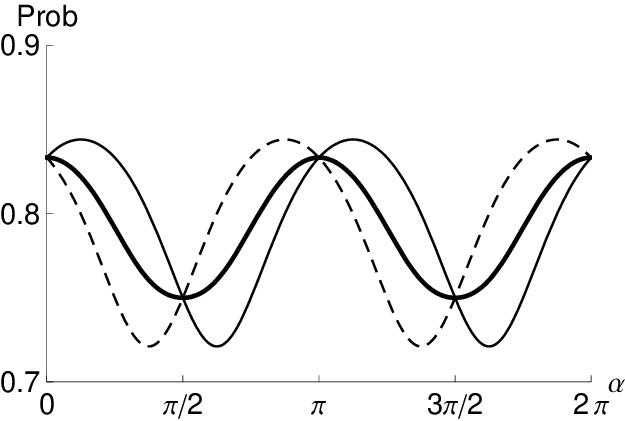}
\end{center}
\caption{Graphs of the probabilities that Alice and Bob do not notice Eve's eavesdropping as functions of the parameter $\alpha$.
The thick solid curve, thin solid curve, and thin dashed curve represent the functions
$P_{\mbox{\tiny AB}}(\alpha)$,
$g_{1}(\alpha)/[f_{1}(\alpha)+g_{1}(\alpha)]=g_{4}(\alpha)/[f_{4}(\alpha)+g_{4}(\alpha)]$,
and
$f_{2}(\alpha)/[f_{2}(\alpha)+g_{2}(\alpha)]=f_{3}(\alpha)/[f_{3}(\alpha)+g_{3}(\alpha)]$,
respectively.
The function $P_{\mbox{\tiny AB}}(\alpha)$ becomes maximum at $\alpha=0$ and $\alpha=\pi$.}
\label{Figure_01}
\end{figure}

\begin{figure}
\begin{center}
\includegraphics{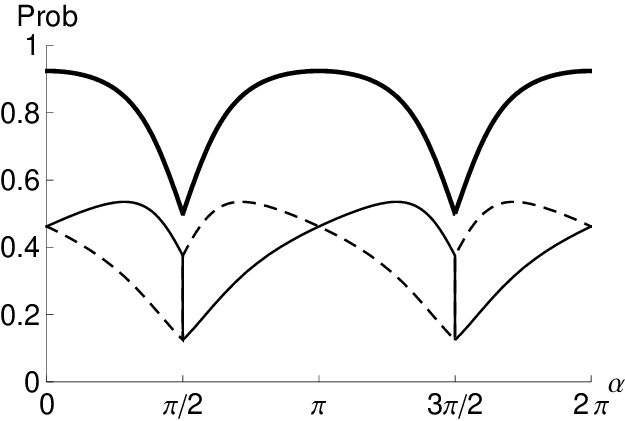}
\end{center}
\caption{Graphs of the probabilities that Eve guesses right at the random secret bit Alice obtains as functions of the parameter $\alpha$.
The thick solid curve, thin solid curve, and thin dashed curve represent the functions
$P_{\mbox{\tiny E}}(\alpha)$,
$R_{1}(\alpha)u_{1}(\alpha)/[u_{1}(\alpha)+v_{1}(\alpha)]=R_{1}(\alpha)u_{2}(\alpha)/[u_{2}(\alpha)+v_{2}(\alpha)]$,
and
$R_{2}(\alpha)v_{3}(\alpha)/[u_{3}(\alpha)+v_{3}(\alpha)]=R_{2}(\alpha)v_{4}(\alpha)/[u_{4}(\alpha)+v_{4}(\alpha)]$, respectively.
The function $P_{\mbox{\tiny E}}(\alpha)$ becomes maximum at $\alpha=0$ and $\alpha=\pi$.}
\label{Figure_02}
\end{figure}

We show graphs of $P_{\mbox{\scriptsize AB}}(\alpha)$ and $P_{\mbox{\scriptsize E}}(\alpha)$ in Figs.~\ref{Figure_01} and \ref{Figure_02}, respectively.
We put $0\leq\alpha<2\pi$ because it is enough for the range of $\alpha$.
In Fig.~\ref{Figure_01},
the graphs show the probabilities that Alice and Bob do not find a sign of Eve's malicious acts against the parameter $\alpha$.
The function $P_{\mbox{\scriptsize AB}}(\alpha)$ becomes maximum at $\alpha=0$ and $\alpha=\pi$.
In Fig.~\ref{Figure_02}, the graphs show the probabilities that Eve guesses right at the random secret bit Alice obtains
against the parameter $\alpha$.
The function $P_{\mbox{\scriptsize E}}(\alpha)$ becomes maximum at $\alpha=0$ and $\alpha=\pi$.
Thus, Eve's optimum strategies are given by
\begin{equation}
(\alpha,\beta)
=
(0,\pi/4),
(\pi,3\pi/4),
(0,5\pi/4),
(\pi,7\pi/4),
\end{equation}
where we use $\beta=\arctan(1/\cos\alpha)$.
In particular,
$(\alpha,\beta)=(0,\pi/4)$ represents the attack with the Breidbart basis \cite{Bennett1992b}.

\subsection{\label{subsection-02}The case where $(\alpha,\beta)=(0,\pi/4)$ holds: the Breidbart basis}
In this subsection, we consider the case where $(\alpha,\beta)=(0,\pi/4)$ holds.
Because
\begin{equation}
f_{1}(\alpha,\beta)
=
g_{2}(\alpha,\beta)
=
g_{3}(\alpha,\beta)
=
f_{4}(\alpha,\beta)
=
1/32,
\end{equation}
\begin{equation}
g_{1}(\alpha,\beta)
=
f_{2}(\alpha,\beta)
=
f_{3}(\alpha,\beta)
=
g_{4}(\alpha,\beta)
=
5/32,
\end{equation}
using Eq.~(\ref{P-1-alpha}),
the probability $P_{\mbox{\scriptsize AB}}$ that Alice and Bob cannot find Eve's malicious acts is equal to $5/6\simeq 0.833$.
Moreover, because
\begin{equation}
u_{1}(\alpha,\beta)
=
u_{2}(\alpha,\beta)
=
v_{3}(\alpha,\beta)
=
v_{4}(\alpha,\beta)
=
(1/32)(5+3\sqrt{2}),
\end{equation}
\begin{equation}
v_{1}(\alpha,\beta)
=
v_{2}(\alpha,\beta)
=
u_{3}(\alpha,\beta)
=
u_{4}(\alpha,\beta)
=
(1/32)(5-3\sqrt{2}),
\end{equation}
using Eqs.~(\ref{P-2-alpha}), (\ref{Q-1-alpha}), (\ref{Q-2-alpha}), and (\ref{R-alpha}),
the probability $P_{\mbox{\scriptsize E}}$ that Eve guesses right at a random secret bit Alice obtains is given by
$(1/10)(5+3\sqrt{2})
\simeq
0.924$.
Furthermore, ratios of the probabilities that Alice detects $|r_{i}\rangle_{\mbox{\scriptsize AC}}$ for $i=1,2,3,4$ are
\begin{equation}
Q_{1}:Q_{2}:Q_{3}:Q_{4}
=
1:1:1:1.
\end{equation}

\subsection{\label{subsection-03}The case where $\beta=\arctan(-1/\cos\alpha)$ holds}
In this subsection, we consider the case where $\beta=\arctan(-1/\cos\alpha)$ holds.
Substituting $\beta=\arctan(-1/\cos\alpha)$ into
Eqs.~(\ref{u1v1-alpha-beta-01}),
(\ref{u2v2-alpha-beta-02}),
(\ref{u3v3-alpha-beta-03}),
and
(\ref{u4v4-alpha-beta-04}),
we obtain the following functions,
where we use the notation,
$\left.
\tilde{u}_{1}(\alpha)=u_{1}(\alpha,\beta)
\right|_{\beta=\arctan(-1/\cos\alpha)}$:
\begin{eqnarray}
\tilde{u}_{1}(\alpha)
&=&
\frac{1}{16(1+\sec^{2}\alpha)^{3/2}}
[(-1+\sqrt{1+\sec^{2}\alpha})(1+\tan\alpha) \nonumber \\
&&
\quad
+
\sec^{2}\alpha(-1+2\sqrt{1+\sec^{2}\alpha}-\tan\alpha)], \nonumber \\
\tilde{v}_{1}(\alpha)
&=&
\frac{1}{16(1+\sec^{2}\alpha)^{3/2}}
[(1+\sqrt{1+\sec^{2}\alpha})(1+\tan\alpha) \nonumber \\
&&
\quad
+
\sec^{2}\alpha(1+2\sqrt{1+\sec^{2}\alpha}+\tan\alpha)],
\end{eqnarray}
\begin{eqnarray}
\tilde{u}_{2}(\alpha)
&=&
\frac{1}{8}
\cos^{2}
\Biggl(
\frac{1}{2}\arctan(\sec\alpha)
\Biggr)
\Biggl(
2-\frac{1+\tan\alpha}{\sqrt{1+\sec^{2}\alpha}}
\Biggr), \nonumber \\
\tilde{v}_{2}(\alpha)
&=&
\frac{1}{8}
\sin^{2}
\Biggl(
\frac{1}{2}\arctan(\sec\alpha)
\Biggr)
\Biggl(
2+\frac{1+\tan\alpha}{\sqrt{1+\sec^{2}\alpha}}
\Biggr),
\end{eqnarray}
\begin{eqnarray}
\tilde{u}_{3}(\alpha)
&=&
\frac{1}{16(1+\sec^{2}\alpha)^{3/2}}
[\sec^{2}\alpha(1+2\sqrt{1+\sec^{2}\alpha}-\tan\alpha) \nonumber \\
&&
\quad
-(1+\sqrt{1+\sec^{2}\alpha})(-1+\tan\alpha)], \nonumber \\
\tilde{v}_{3}(\alpha)
&=&
\frac{1}{16(1+\sec^{2}\alpha)^{3/2}}
[(1-\sqrt{1+\sec^{2}\alpha})(-1+\tan\alpha) \nonumber \\
&&
\quad
+
\sec^{2}\alpha(-1+2\sqrt{1+\sec^{2}\alpha}+\tan\alpha)],
\end{eqnarray}
\begin{eqnarray}
\tilde{u}_{4}(\alpha)
&=&
\frac{1}{8}
\sin^{2}
\Biggl(
\frac{1}{2}\arctan(\sec\alpha)
\Biggr)
\Biggl(
2+\frac{1-\tan\alpha}{\sqrt{1+\sec^{2}\alpha}}
\Biggr), \nonumber \\
\tilde{v}_{4}(\alpha)
&=&
\frac{1}{8}
\cos^{2}
\Biggl(
\frac{1}{2}\arctan(\sec\alpha)
\Biggr)
\Biggl(
2+\frac{-1+\tan\alpha}{\sqrt{1+\sec^{2}\alpha}}
\Biggr).
\end{eqnarray}

Here, we pay attention to a relation,
\begin{equation}
[\tilde{u}_{1}(\alpha)+\tilde{v}_{1}(\alpha)+\tilde{u}_{2}(\alpha)+\tilde{v}_{2}(\alpha)]
-
[\tilde{u}_{3}(\alpha)+\tilde{v}_{3}(\alpha)+\tilde{u}_{4}(\alpha)+\tilde{v}_{4}(\alpha)]
=0.
\end{equation}
Thus, the probability that Alice detects $|r_{1}\rangle_{\mbox{\scriptsize AC}}$
to the probability that she finds $|r_{4}\rangle_{\mbox{\scriptsize AC}}$ is one to one.
Hence, the probability that Eve guesses right at a random secret bit Alice obtains is given by
\begin{eqnarray}
\tilde{P}_{\mbox{\scriptsize E}}(\alpha)
&=&
\frac{1}{4}
\Biggl[
\frac{\tilde{u}_{1}(\alpha)}{\tilde{u}_{1}(\alpha)+\tilde{v}_{1}(\alpha)}
+
\frac{\tilde{u}_{2}(\alpha)}{\tilde{u}_{2}(\alpha)+\tilde{v}_{2}(\alpha)}
+
\frac{\tilde{v}_{3}(\alpha)}{\tilde{u}_{3}(\alpha)+\tilde{v}_{3}(\alpha)}
+
\frac{\tilde{v}_{4}(\alpha)}{\tilde{u}_{4}(\alpha)+\tilde{v}_{4}(\alpha)}
\Biggr] \nonumber \\
&=&
\frac{1}{2}.
\end{eqnarray}
This implies that Eve obtains a completely random bit,
and therefore there is no correlation between Eve and Alice's bits.
Hence, Eve cannot gain any information although she eavesdrops on the transmission from Alice to Bob.

\subsection{\label{subsection-04}The case where $(\alpha,\beta)=(\pi,\pi/4)$ holds}
In this subsection, we consider the case where the parameters are given by $(\alpha,\beta)=(\pi,\pi/4)$.
In this case, $\beta=\arctan(-1/\cos\alpha)$ holds.
Because
\begin{equation}
f_{1}(\alpha,\beta)
=
g_{2}(\alpha,\beta)
=
g_{3}(\alpha,\beta)
=
f_{4}(\alpha,\beta)
=
1/32,
\end{equation}
\begin{equation}
g_{1}(\alpha,\beta)
=
f_{2}(\alpha,\beta)
=
f_{3}(\alpha,\beta)
=
g_{4}(\alpha,\beta)
=
9/32,
\end{equation}
using Eq.~(\ref{P-1-alpha}),
the probability $P_{\mbox{\scriptsize AB}}$ that Alice and Bob cannot notice Eve's illegal acts is given by $9/10$.
Moreover, because
\begin{equation}
u_{1}(\alpha,\beta)
=
v_{2}(\alpha,\beta)
=
v_{3}(\alpha,\beta)
=
u_{4}(\alpha,\beta)
=
(3-\sqrt{2})/32,
\end{equation}
\begin{equation}
v_{1}(\alpha,\beta)
=
u_{2}(\alpha,\beta)
=
u_{3}(\alpha,\beta)
=
v_{4}(\alpha,\beta)
=
(3+\sqrt{2})/32,
\end{equation}
the probability $P_{\mbox{\scriptsize E}}$ that Eve guesses right at a random secret bit Alice obtains is given as follows.
If Bob observes $\sigma_{x}$ and Alice detects $|r_{1}\rangle_{\mbox{\scriptsize AC}}$,
it is equal to $(3-\sqrt{2})/6$.
If Bob observes $\sigma_{z}$ and Alice finds $|r_{1}\rangle_{\mbox{\scriptsize AC}}$,
it is equal to $(3+\sqrt{2})/6$.
If Bob measures $\sigma_{x}$ and Alice detects $|r_{4}\rangle_{\mbox{\scriptsize AC}}$,
it is given by $(3-\sqrt{2})/6$.
If Bob measures $\sigma_{z}$ and Alice finds $|r_{4}\rangle_{\mbox{\scriptsize AC}}$,
it is given by $(3+\sqrt{2})/6$.
Thus, the average of $P_{\mbox{\scriptsize E}}$ is equal to $1/2$.
Hence, Eve's eavesdropping is useless for this strategy.
Furthermore, ratios of the probabilities that Alice detects $|r_{i}\rangle_{\mbox{\scriptsize AC}}$ for $i=1,2,3,4$ are
\begin{equation}
Q_{1}:Q_{2}:Q_{3}:Q_{4}
=
3:5:5:3.
\end{equation}

\section{\label{section-two-way-intercept-resend-attack}The two-way intercept/resend attack}
In this section, we estimate the security against the two-way intercept/resend attack.
On the way from Alice to Bob, we assume that Eve measures $\sigma_{\xi}$ on the channel qubit.
Moreover, on the way from Bob to Alice, we assume that Eve observes $\sigma_{\mu}$ on the channel qubit.
The projection operators of $\sigma_{\xi}$ and $\sigma_{\mu}$ are given by Eqs.~(\ref{projection-sigma-plus}) and (\ref{projection-sigma-minus}),
where $\hat{P}(\sigma_{\mu}=\pm 1)$ are parametrized by $\gamma$ and $\delta$ instead of $\alpha$ and $\beta$. 

Referring to Eq.~(\ref{prob-sigma-x-r2-one-way-intercept-resend-attak}),
for example,
we describe the probability the Bob has $\sigma_{x}=i$ and Eve obtains $\sigma_{\xi}=j$ and $\sigma_{\mu}=k$
on condition that Alice detects $|r_{2}\rangle_{\mbox{\scriptsize AC}}$ as
\begin{eqnarray}
&&
\mbox{prob}(\sigma_{x}=i,\sigma_{\xi}=j,\sigma_{\mu}=k,r_{2}) \nonumber \\
&=&
\frac{
|_{\mbox{\scriptsize AC}}\langle r_{2}|\hat{P}(\sigma_{\mu}=k)\hat{P}(\sigma_{x}=i)\hat{P}(\sigma_{\xi}=j)|\psi\rangle_{\mbox{\scriptsize AC}}|^{2}
}
{
\sum_{l,m,n\in \{1,-1\}}|_{\mbox{\scriptsize AC}}\langle r_{2}|\hat{P}(\sigma_{\mu}=n)\hat{P}(\sigma_{x}=l)
\hat{P}(\sigma_{\xi}=m)|\psi\rangle_{\mbox{\scriptsize AC}}|^{2}
} \nonumber \\
&&
\quad
\mbox{for $i,j,k\in \{1,-1\}$}.
\end{eqnarray}

To evaluate the probabilities that Alice and Bob do not notice Eve's illegal acts in the subsequence $S_{23}$,
we prepare eight functions $f_{k}(\alpha,\beta,\gamma,\delta)$ and $g_{k}(\alpha,\beta,\gamma,\delta)$ for $k=1,2,3,4$
in Eqs.~(\ref{appendix-B-f1-g1}), (\ref{appendix-B-f2-g2}),
(\ref{appendix-B-f3-g3}), and (\ref{appendix-B-f4-g4}) in Appendix~\ref{section-appendix-B}.
Then, the following relations hold between the probabilities and the eight functions:
\begin{eqnarray}
\sum_{j,k\in\{1,-1\}}
\mbox{prob}(\sigma_{x}=1,\sigma_{\xi}=j,\sigma_{\mu}=k,r_{2})
& \propto &
f_{1}(\alpha,\beta,\gamma,\delta), \nonumber \\
\sum_{j,k\in\{1,-1\}}
\mbox{prob}(\sigma_{x}=-1,\sigma_{\xi}=j,\sigma_{\mu}=k,r_{2})
& \propto &
g_{1}(\alpha,\beta,\gamma,\delta), \nonumber \\
\sum_{j,k\in\{1,-1\}}
\mbox{prob}(\sigma_{z}=1,\sigma_{\xi}=j,\sigma_{\mu}=k,r_{2})
& \propto &
f_{2}(\alpha,\beta,\gamma,\delta), \nonumber \\
\sum_{j,k\in\{1,-1\}}
\mbox{prob}(\sigma_{z}=-1,\sigma_{\xi}=j,\sigma_{\mu}=k,r_{2})
& \propto &
g_{2}(\alpha,\beta,\gamma,\delta), \nonumber \\
\sum_{j,k\in\{1,-1\}}
\mbox{prob}(\sigma_{x}=1,\sigma_{\xi}=j,\sigma_{\mu}=k,r_{3})
& \propto &
f_{3}(\alpha,\beta,\gamma,\delta), \nonumber \\
\sum_{j,k\in\{1,-1\}}
\mbox{prob}(\sigma_{x}=-1,\sigma_{\xi}=j,\sigma_{\mu}=k,r_{3})
& \propto &
g_{3}(\alpha,\beta,\gamma,\delta), \nonumber \\
\sum_{j,k\in\{1,-1\}}
\mbox{prob}(\sigma_{z}=1,\sigma_{\xi}=j,\sigma_{\mu}=k,r_{3})
& \propto &
f_{4}(\alpha,\beta,\gamma,\delta), \nonumber \\
\sum_{j,k\in\{1,-1\}}
\mbox{prob}(\sigma_{z}=-1,\sigma_{\xi}=j,\sigma_{\mu}=k,r_{3})
& \propto &
g_{4}(\alpha,\beta,\gamma,\delta).
\label{prob-fg-two-way-intercept-resend}
\end{eqnarray}

Because of Eq.~(\ref{prob-fg-two-way-intercept-resend}),
we reach the final form of the probability $P_{\mbox{\scriptsize AB}}$ that Alice and Bob do not notice Eve's illegal acts as
\begin{equation}
P_{\mbox{\scriptsize AB}}(\alpha,\beta,\gamma,\delta)
=
\frac{R_{1}}{2}
\Biggl(
\frac{g_{1}}{f_{1}+g_{1}}
+
\frac{f_{2}}{f_{2}+g_{2}}
\Biggr)
+
\frac{R_{2}}{2}
\Biggl(
\frac{f_{3}}{f_{3}+g_{3}}
+
\frac{g_{4}}{f_{4}+g_{4}}
\Biggr),
\label{P-AB-two-way-intercept-resend}
\end{equation}
\begin{equation}
R_{1}
=
\frac{1}{R}
(f_{1}+g_{1}+f_{2}+g_{2}),
\label{Q1-two-way-intercept-resend}
\end{equation}
\begin{equation}
R_{2}
=
\frac{1}{R}
(f_{3}+g_{3}+f_{4}+g_{4}),
\label{Q2-two-way-intercept-resend}
\end{equation}
\begin{equation}
R=f_{1}+g_{1}+f_{2}+g_{2}+f_{3}+g_{3}+f_{4}+g_{4}.
\label{R-two-way-intercept-resend}
\end{equation}
In Eqs.~(\ref{P-AB-two-way-intercept-resend}), (\ref{Q1-two-way-intercept-resend}),
(\ref{Q2-two-way-intercept-resend}), and (\ref{R-two-way-intercept-resend}),
we omit symbols of variables $\alpha$, $\beta$, $\gamma$, and $\delta$
from functions
$f_{1}$, $g_{1}$, $f_{2}$, $g_{2}$, $f_{3}$, $g_{3}$, $f_{4}$, and $g_{4}$.

To estimate the probabilities that Eve guesses right at the random secret bit Alice obtains in the subsequence $S_{14}$,
we prepare eight functions
$u_{k}(\alpha,\beta,\gamma,\delta)$ and $v_{k}(\alpha,\beta,\gamma,\delta)$ for $k=1,2,3,4$
in Eqs.~(\ref{appendix-B-u1-v1}), (\ref{appendix-B-u2-v2}),
(\ref{appendix-B-u3-v3}), (\ref{appendix-B-u4-v4}) in Appendix~\ref{section-appendix-B}.
Then, the following relations hold between the probabilities and the eight functions:
\begin{eqnarray}
\sum_{j,k\in\{1,-1\}}
\mbox{prob}(\sigma_{x}=j,\sigma_{\xi}=k,\sigma_{\mu}=k,r_{1})
& \propto &
u_{1}(\alpha,\beta,\gamma,\delta), \nonumber \\
\sum_{j,k\in\{1,-1\}}
\mbox{prob}(\sigma_{x}=j,\sigma_{\xi}=k,\sigma_{\mu}=\bar{k},r_{1})
& \propto &
v_{1}(\alpha,\beta,\gamma,\delta), \nonumber \\
\sum_{j,k\in\{1,-1\}}
\mbox{prob}(\sigma_{z}=j,\sigma_{\xi}=k,\sigma_{\mu}=k,r_{1})
& \propto &
u_{2}(\alpha,\beta,\gamma,\delta), \nonumber \\
\sum_{j,k\in\{1,-1\}}
\mbox{prob}(\sigma_{z}=j,\sigma_{\xi}=k,\sigma_{\mu}=\bar{k},r_{1})
& \propto &
v_{2}(\alpha,\beta,\gamma,\delta), \nonumber \\
\sum_{j,k\in\{1,-1\}}
\mbox{prob}(\sigma_{x}=j,\sigma_{\xi}=k,\sigma_{\mu}=k,r_{4})
& \propto &
u_{3}(\alpha,\beta,\gamma,\delta), \nonumber \\
\sum_{j,k\in\{1,-1\}}
\mbox{prob}(\sigma_{x}=j,\sigma_{\xi}=k,\sigma_{\mu}=\bar{k},r_{4})
& \propto &
v_{3}(\alpha,\beta,\gamma,\delta), \nonumber \\
\sum_{j,k\in\{1,-1\}}
\mbox{prob}(\sigma_{z}=j,\sigma_{\xi}=k,\sigma_{\mu}=k,r_{4})
& \propto &
u_{4}(\alpha,\beta,\gamma,\delta), \nonumber \\
\sum_{j,k\in\{1,-1\}}
\mbox{prob}(\sigma_{z}=j,\sigma_{\xi}=k,\sigma_{\mu}=\bar{k},r_{4})
& \propto &
v_{4}(\alpha,\beta,\gamma,\delta),
\label{prob-uv-two-way-intercept-resend}
\end{eqnarray}
where $\bar{k}$ is defined as
\begin{equation}
\bar{k}
=
\left\{
\begin{array}{ll}
1  & \mbox{for $k=-1$} \\
-1 & \mbox{for $k=1$}
\end{array}
\right.
.
\label{def-bar-k}
\end{equation}

Because of Eq.~(\ref{prob-uv-two-way-intercept-resend}),
we attain the final form of the probability $P_{\mbox{\scriptsize E}}$ that Eve guesses right at the random secret bit Alice obtains as
\begin{equation}
P_{\mbox{\scriptsize E}}(\alpha,\beta,\gamma,\delta)
=
\frac{\tilde{R}_{1}}{2}
\Biggl(
\frac{u_{1}}{u_{1}+v_{1}}
+
\frac{u_{2}}{u_{2}+v_{2}}
\Biggr)
+
\frac{\tilde{R}_{2}}{2}
\Biggl(
\frac{v_{3}}{u_{3}+v_{3}}
+
\frac{v_{4}}{u_{4}+v_{4}}
\Biggr),
\label{P-E-two-way-intercept-resend}
\end{equation}
\begin{equation}
\tilde{R}_{1}
=
\frac{1}{\tilde{R}}
(u_{1}+v_{1}+u_{2}+v_{2}),
\label{tilde-Q1-two-way-intercept-resend}
\end{equation}
\begin{equation}
\tilde{R}_{2}
=
\frac{1}{\tilde{R}}
(u_{3}+v_{3}+u_{4}+v_{4}),
\label{tilde-Q2-two-way-intercept-resend}
\end{equation}
\begin{equation}
\tilde{R}=u_{1}+v_{1}+u_{2}+v_{2}+u_{3}+v_{3}+u_{4}+v_{4}.
\label{tilde-R-two-way-intercept-resend}
\end{equation}
In Eqs.~(\ref{P-E-two-way-intercept-resend}), (\ref{tilde-Q1-two-way-intercept-resend}),
(\ref{tilde-Q2-two-way-intercept-resend}), and (\ref{tilde-R-two-way-intercept-resend}),
we omit symbols of variables $\alpha$, $\beta$, $\gamma$, and $\delta$
from functions 
$u_{1}$, $v_{1}$, $u_{2}$, $v_{2}$, $u_{3}$, $v_{3}$, $u_{4}$, and $v_{4}$.

Calculating $P_{\mbox{\scriptsize AB}}(\alpha,\beta,\gamma,\delta)$ and $P_{\mbox{\scriptsize E}}(\alpha,\beta,\gamma,\delta)$
at points in a mesh
\\
$(\alpha,\beta,\gamma,\delta)\in\{(j/400)\pi:j=0,1,2,...,400\}^{\otimes 4}$,
we obtain the following results.
When $(\alpha,\beta,\gamma,\delta)=(0,3\pi/4,0,3\pi/4)$,
$P_{\mbox{\scriptsize AB}}$ takes the maximum value $9/10$
and $P_{\mbox{\scriptsize E}}$ is equal to $1/2$.
When $(\alpha,\beta,\gamma,\delta)=(0,\pi/4,\pi/2,\pi/2)$,
$P_{\mbox{\scriptsize E}}$ takes the maximum value $0.854$ and $P_{\mbox{\scriptsize AB}}$
is equal to $1/2$.
Therefore, we can conclude that the two-way intercept/resend attack is not preferable to the one-way intercept/resend attack for Eve.

\section{\label{section-one-way-translucent-attack}The one-way translucent attack}
We give a short review of the translucent attack as follows.
First, Eve keeps her own some auxiliary qubits close at hand as a probe.
The qubits are initialized in a particular state.
Second, Eve applies a unitary transformation to her probe and the single channel qubit
in order to generate entanglement between them.
Third, Eve leaves her probe untouched and sends the channel qubit to one of the legitimate users.
Fourth, after listening to the public discussion between Alice and Bob,
Eve makes a measurement on her probe depending on the classical information disclosed by Alice and Bob.
Fifth, Eve guesses at the secret bit Alice obtains according to the result of the observation on her probe.

In this section,
we consider the case where Eve makes the one-way translucent attack on the channel qubit flying from Bob to Alice.
Here, we trace Eve's attack step by step in concrete terms in the following paragraphs.

First, Alice prepares the sate $|\psi\rangle_{\mbox{\scriptsize AC}}$ given by Eq.~(\ref{initial-state}) as an initial state.
Second, we assume that Bob observes $\sigma_{z}$ on the channel qubit,
for example.
Then, the wave function of the whole system reduces to the following state:
\begin{equation}
|0\rangle_{\mbox{\scriptsize C}}{}_{\mbox{\scriptsize C}}\langle 0|
\otimes
\mbox{Tr}_{\mbox{\scriptsize C}}
[|0\rangle_{\mbox{\scriptsize C}}{}_{\mbox{\scriptsize C}}\langle 0|\psi\rangle_{\mbox{\scriptsize AC}}{}_{\mbox{\scriptsize AC}}\langle\psi|]
+
|1\rangle_{\mbox{\scriptsize C}}{}_{\mbox{\scriptsize C}}\langle 1|
\otimes
\mbox{Tr}_{\mbox{\scriptsize C}}
[|1\rangle_{\mbox{\scriptsize C}}{}_{\mbox{\scriptsize C}}\langle 1|\psi\rangle_{\mbox{\scriptsize AC}}{}_{\mbox{\scriptsize AC}}\langle\psi|].
\label{state-second-Bob}
\end{equation}
Third, Bob returns the channel qubit to Alice.

Fourth, in the middle of the channel qubit's travelling from Bob to Alice,
Eve lets her probe interact with it.
In general, a unitary transformation applied by Eve to her probe and the channel qubit is described as
\begin{equation}
U|0\rangle_{\mbox{\scriptsize C}}|X\rangle_{\mbox{\scriptsize E}}
=
\sqrt{F}|0\rangle_{\mbox{\scriptsize C}}|\alpha\rangle_{\mbox{\scriptsize E}}
+
\sqrt{1-F}|1\rangle_{\mbox{\scriptsize C}}|\beta\rangle_{\mbox{\scriptsize E}},
\label{state-fourth-Eve-1}
\end{equation}
\begin{equation}
U|1\rangle_{\mbox{\scriptsize C}}|X\rangle_{\mbox{\scriptsize E}}
=
\sqrt{1-F'}|0\rangle_{\mbox{\scriptsize C}}|\gamma\rangle_{\mbox{\scriptsize E}}
+
\sqrt{F'}|1\rangle_{\mbox{\scriptsize C}}|\delta\rangle_{\mbox{\scriptsize E}},
\label{state-fourth-Eve-2}
\end{equation}
where the index $\mbox{E}$ represents Eve's probe,
$|X\rangle_{\mbox{\scriptsize E}}$ denotes the initial state of the probe,
and
$|\alpha\rangle_{\mbox{\scriptsize E}}$, $|\beta\rangle_{\mbox{\scriptsize E}}$, $|\gamma\rangle_{\mbox{\scriptsize E}}$,
and $|\delta\rangle_{\mbox{\scriptsize E}}$
are arbitrary normalized states.
The dimension of a Hilbert space for Eve's probe is equal to four at the most.
After Eve makes the attack on the state given by Eq.~(\ref{state-second-Bob}),
it evolves into
\begin{eqnarray}
&&
(U|0\rangle_{\mbox{\scriptsize C}}|X\rangle_{\mbox{\scriptsize E}})
({}_{\mbox{\scriptsize E}}\langle X|{}_{\mbox{\scriptsize C}}\langle 0|U^{\dagger})
\otimes
\mbox{Tr}_{\mbox{\scriptsize C}}
[|0\rangle_{\mbox{\scriptsize C}}{}_{\mbox{\scriptsize C}}\langle 0|\psi\rangle_{\mbox{\scriptsize AC}}{}_{\mbox{\scriptsize AC}}\langle\psi|] \nonumber \\
&&
\quad
+
(U|1\rangle_{\mbox{\scriptsize C}}|X\rangle_{\mbox{\scriptsize E}})
({}_{\mbox{\scriptsize E}}\langle X|{}_{\mbox{\scriptsize C}}\langle 1|U^{\dagger})
\otimes
\mbox{Tr}_{\mbox{\scriptsize C}}
[|1\rangle_{\mbox{\scriptsize C}}{}_{\mbox{\scriptsize C}}\langle 1|\psi\rangle_{\mbox{\scriptsize AC}}{}_{\mbox{\scriptsize AC}}\langle\psi|].
\label{state-fourth-Eve-3}
\end{eqnarray}
Finally, Alice performs the orthogonal measurement $R$ upon the state of Eq.~(\ref{state-fourth-Eve-3})
with $|r_{j}\rangle_{\mbox{\scriptsize AC}}$ for $j\in\{1,2,3,4\}$.

Here, we let $K(\sigma_{t}=i,r_{j})$ denote a square of the amplitude of the wave function
where Bob has obtained $i\in\{1,-1\}$ with the measurement of $\sigma_{t}$ for $t\in\{x,z\}$
and Alice has detected $|r_{j}\rangle_{\mbox{\scriptsize AC}}$ for $j\in\{1,2,3,4\}$.
Then, $K(\sigma_{t}=i,r_{j})$ is given in the form,
\begin{eqnarray}
K(\sigma_{t}=i,r_{j})
&=&
{}_{\mbox{\scriptsize AC}}\langle r_{j}|({}_{\mbox{\scriptsize C}}\langle i_{t}|\psi\rangle_{\mbox{\scriptsize AC}})
\mbox{Tr}_{\mbox{\scriptsize E}}[
(U|i_{t}\rangle_{\mbox{\scriptsize C}}|X\rangle_{\mbox{\scriptsize E}})
({}_{\mbox{\scriptsize E}}\langle X|{}_{\mbox{\scriptsize C}}\langle i_{t}|U^{\dagger})
] \nonumber \\
&&\quad
\times
({}_{\mbox{\scriptsize AC}}\langle\psi|i_{t}\rangle_{\mbox{\scriptsize C}})|r_{j}\rangle_{\mbox{\scriptsize AC}},
\label{K-sigma-r-definition}
\end{eqnarray}
where
\begin{equation}
i_{z}
=
\left\{
\begin{array}{ll}
0 & \mbox{for $i=1$} \\
1 & \mbox{for $i=-1$} \\
\end{array}
\right.
,
\label{def-iz}
\end{equation}
\begin{equation}
i_{x}
=
\left\{
\begin{array}{ll}
+ & \mbox{for $i=1$} \\
- & \mbox{for $i=-1$} \\
\end{array}
\right.
.
\label{def-ix}
\end{equation}

It is very difficult for us to estimate the security of the protocol because the degrees of freedom
for Eve's attack given by Eqs.~(\ref{state-fourth-Eve-1}) and (\ref{state-fourth-Eve-2}) are very large.
Thus, we add some restrictions to Eqs.~(\ref{state-fourth-Eve-1}) and (\ref{state-fourth-Eve-2})
in order to make the number of parameters of Eve's strategy small.

First, for the symmetry, we assume $F=F'$ where $0\leq F\leq 1$.
Second, because $U|0\rangle_{\mbox{\scriptsize C}}|X\rangle_{\mbox{\scriptsize E}}$ and $U|1\rangle_{\mbox{\scriptsize C}}|X\rangle_{\mbox{\scriptsize E}}$
are orthogonal to each other,
we obtain
${}_{\mbox{\scriptsize E}}\langle\alpha|\gamma\rangle_{\mbox{\scriptsize E}}
+
{}_{\mbox{\scriptsize E}}\langle\beta|\delta\rangle_{\mbox{\scriptsize E}}
=0$.
Third, we assume that Eqs.~(\ref{state-fourth-Eve-1}) and (\ref{state-fourth-Eve-2}) are given in the form of the Schmidt decomposition \cite{Nielsen2000},
so that we obtain
${}_{\mbox{\scriptsize E}}\langle\alpha|\beta\rangle_{\mbox{\scriptsize E}}
=
{}_{\mbox{\scriptsize E}}\langle\gamma|\delta\rangle_{\mbox{\scriptsize E}}
=0$.
Fourth,
we put the following restraints on Eqs.~(\ref{state-fourth-Eve-1}) and (\ref{state-fourth-Eve-2}) in analogy with
Bechmann-Pasquinucci and Gisin's work \cite{Bechmann-Pasquinucci1999}:
${}_{\mbox{\scriptsize E}}\langle\alpha|\gamma\rangle_{\mbox{\scriptsize E}}
=
{}_{\mbox{\scriptsize E}}\langle\beta|\delta\rangle_{\mbox{\scriptsize E}}
=0$.
Imposing these constraints upon Eve's attack,
we can describe
$|\alpha\rangle_{\mbox{\scriptsize E}}$,
$|\beta\rangle_{\mbox{\scriptsize E}}$,
$|\gamma\rangle_{\mbox{\scriptsize E}}$,
and $|\delta\rangle_{\mbox{\scriptsize E}}$
with two real parameters.
In the following, we explain this fact.

First of all,
because of ${}_{\mbox{\scriptsize E}}\langle\alpha|\gamma\rangle_{\mbox{\scriptsize E}}=0$,
we can write down $|\alpha\rangle_{\mbox{\scriptsize E}}$ and $|\gamma\rangle_{\mbox{\scriptsize E}}$ as
\begin{equation}
|\alpha\rangle_{\mbox{\scriptsize E}}
=
(1,0,0,0)^{\mbox{\scriptsize T}},
\label{alpha-vector}
\end{equation}
\begin{equation}
|\gamma\rangle_{\mbox{\scriptsize E}}
=
(0,1,0,0)^{\mbox{\scriptsize T}},
\label{gamma-vector}
\end{equation}
where the symbol $\mbox{T}$ denotes the transpose of a vector.
Second, from ${}_{\mbox{\scriptsize E}}\langle\gamma|\delta\rangle_{\mbox{\scriptsize E}}=0$,
$|\delta\rangle_{\mbox{\scriptsize E}}$ is given by
\begin{equation}
|\delta\rangle_{\mbox{\scriptsize E}}
=
(r,0,s,t)^{\mbox{\scriptsize T}},
\label{delta-vector-1}
\end{equation}
where $r$, $s$, and $t$ are complex numbers.
Third, we describe $r$ as $r=e^{i\theta}\cos a$ where $\theta$ and $a$ are real numbers.
Because we can rewrite $e^{-i\theta}|\delta\rangle_{\mbox{\scriptsize E}}$ as $|\delta\rangle_{\mbox{\scriptsize E}}$
for deleting the phase $e^{i\theta}$,
we obtain
\begin{equation}
|\delta\rangle_{\mbox{\scriptsize E}}
=
(\cos a,0,s',t')^{\mbox{\scriptsize T}},
\label{delta-vector-2}
\end{equation}
where $s'$ and $t'$ are complex numbers.
Now, adjusting the basis vectors of the third and fourth components of the system $\mbox{E}$,
we can let $|\delta\rangle_{\mbox{\scriptsize E}}$ be given in the form,
\begin{equation}
|\delta\rangle_{\mbox{\scriptsize E}}
=
(\cos a,0,\sin a,0)^{\mbox{\scriptsize T}}.
\label{delta-vector-3}
\end{equation}

Fourth, because of ${}_{\mbox{\scriptsize E}}\langle\alpha|\beta\rangle_{\mbox{\scriptsize E}}=0$,
we can write $|\beta\rangle_{\mbox{\scriptsize E}}$ as
\begin{equation}
|\beta\rangle_{\mbox{\scriptsize E}}
=
(0,\tilde{r},\tilde{s},\tilde{t})^{\mbox{\scriptsize T}},
\label{beta-vector-1}
\end{equation}
where $\tilde{r}$, $\tilde{s}$, and $\tilde{t}$ are complex numbers.
Then, ${}_{\mbox{\scriptsize E}}\langle\beta|\delta\rangle_{\mbox{\scriptsize E}}=0$ leads to
\begin{equation}
|\beta\rangle_{\mbox{\scriptsize E}}
=
(0,\tilde{r},0,\tilde{t})^{\mbox{\scriptsize T}}.
\label{beta-vector-2}
\end{equation}
At this stage, from Eqs.~(\ref{alpha-vector}), (\ref{gamma-vector}), and (\ref{delta-vector-3}),
the fourth component of the three vectors,
$|\alpha\rangle_{\mbox{\scriptsize E}}$,
$|\gamma\rangle_{\mbox{\scriptsize E}}$,
and $|\delta\rangle_{\mbox{\scriptsize E}}$,
is equal to zero.
Thus, changing the fourth vector of the basis
in order to let the second and fourth components of $|\beta\rangle_{\mbox{\scriptsize E}}$ have a common phase $e^{i\tilde{\theta}}$,
we obtain
\begin{equation}
|\beta\rangle_{\mbox{\scriptsize E}}
=
(0,e^{i\tilde{\theta}}\cos b,0,e^{i\tilde{\theta}}\sin b)^{\mbox{\scriptsize T}}.
\label{beta-vector-3}
\end{equation}
Moreover, rewriting $e^{-i\tilde{\theta}}|\beta\rangle_{\mbox{\scriptsize E}}$ as $|\beta\rangle_{\mbox{\scriptsize E}}$,
we reach
\begin{equation}
|\beta\rangle_{\mbox{\scriptsize E}}
=
(0,\cos b,0,\sin b)^{\mbox{\scriptsize T}}.
\label{beta-vector-4}
\end{equation}
Hence, from Eqs.~(\ref{alpha-vector}), (\ref{gamma-vector}), (\ref{delta-vector-3}), and (\ref{beta-vector-4}),
we can describe
$|\alpha\rangle_{\mbox{\scriptsize E}}$,
$|\beta\rangle_{\mbox{\scriptsize E}}$,
$|\gamma\rangle_{\mbox{\scriptsize E}}$,
and $|\delta\rangle_{\mbox{\scriptsize E}}$
with the two real parameters, $a$ and $b$.

From slightly tedious calculations, we obtain
\begin{eqnarray}
K(\sigma_{z}=1,r_{2})
&=&
K(\sigma_{z}=-1,r_{3})
=
(1/8)(1+F), \nonumber \\
K(\sigma_{z}=-1,r_{2})
&=&
K(\sigma_{z}=1,r_{3})
=(1/8)(1-F), \nonumber \\
K(\sigma_{x}=1,r_{2})
&=&
K(\sigma_{x}=-1,r_{3})
=(1/16)[1-F\cos a-(1-F)\cos b], \nonumber \\
K(\sigma_{x}=-1,r_{2})
&=&
K(\sigma_{x}=1,r_{3})
=(1/16)[3+F\cos a+(1-F)\cos b].
\end{eqnarray}
Then, we request the following conditions:
\begin{eqnarray}
K(\sigma_{z}=1,r_{2})
&=&
K(\sigma_{z}=-1,r_{3})
=
K(\sigma_{x}=-1,r_{2})
=
K(\sigma_{x}=1,r_{3}) \nonumber \\
&=&
(1/8)(1+F), \nonumber \\
K(\sigma_{z}=-1,r_{2})
&=&
K(\sigma_{z}=1,r_{3})
=
K(\sigma_{x}=1,r_{2})
=
K(\sigma_{x}=-1,r_{3}) \nonumber \\
&=&
(1/8)(1-F).
\end{eqnarray}
From the above requirements,
we obtain
\begin{equation}
\cos b
=
\frac{-1+2F-F\cos a}{1-F},
\label{condition-a-b-F}
\end{equation}
so that we can specify Eve's unitary operator $U$ with the two real parameters $a$ and $b$ uniquely.
Assuming Eq.~(\ref{condition-a-b-F}),
we also obtain
\begin{eqnarray}
K(\sigma_{z}=1,r_{1})
&=&
K(\sigma_{z}=-1,r_{4})
=
K(\sigma_{x}=1,r_{1})
=
K(\sigma_{x}=-1,r_{4}) \nonumber \\
&=&
(1/8)(1+F), \nonumber \\
K(\sigma_{z}=-1,r_{1})
&=&
K(\sigma_{z}=1,r_{4})
=
K(\sigma_{x}=-1,r_{1})
=
K(\sigma_{x}=1,r_{4}) \nonumber \\
&=&
(1/8)(1-F).
\end{eqnarray}

Here, we remember the fact that Alice and Bob examine whether or not Eve interferes in the channel qubit during the subsequence $S_{23}$.
Hence, the probability $P_{\mbox{\scriptsize AB}}$ that Alice and Bob do not notice Eve's malicious acts is given by
\begin{eqnarray}
P_{\mbox{\scriptsize AB}}
&=&
K(\sigma_{x}=-1,r_{2})
+
K(\sigma_{x}=1,r_{3})
+
K(\sigma_{z}=1,r_{2})
+
K(\sigma_{z}=-1,r_{3}) \nonumber \\
&=&
(1/2)(1+F).
\label{P-AB-one-way-translucent}
\end{eqnarray}

Next, we consider states between which Eve has to discriminate for guessing right at Alice's secret bit.
If Alice detects $|r_{1}\rangle_{\mbox{\scriptsize AC}}$ or $|r_{4}\rangle_{\mbox{\scriptsize AC}}$,
she obtains the random secret bit.
Moreover, Alice and Bob disclose which channel qubit belongs to the subsequence $S_{14}$.

Now, we trace this process in concrete terms.
If Bob performs the observation of $\sigma_{z}$ upon the channel qubit and Eve lets her probe interact with it,
the wave function reduces to the state given by Eq.~(\ref{state-fourth-Eve-3}).
If Alice detects $|r_{1}\rangle_{\mbox{\scriptsize AC}}$ in the state of Eq.~(\ref{state-fourth-Eve-3}),
Eve obtains the following density operator:
\begin{equation}
|\phi(\sigma_{z}=1,r_{1})\rangle_{\mbox{\scriptsize E}}{}_{\mbox{\scriptsize E}}\langle\phi(\sigma_{z}=1,r_{1})|
+
|\phi(\sigma_{z}=-1,r_{1})\rangle_{\mbox{\scriptsize E}}{}_{\mbox{\scriptsize E}}\langle\phi(\sigma_{z}=-1,r_{1})|,
\end{equation}
where
\begin{equation}
|\phi(\sigma_{t}=i,r_{j})\rangle_{\mbox{\scriptsize E}}
=
{}_{\mbox{\scriptsize AC}}\langle r_{j}|({}_{\mbox{\scriptsize C}}\langle i_{t}|\psi\rangle_{\mbox{\scriptsize AC}})
(U|i_{t}\rangle_{\mbox{\scriptsize C}}|X\rangle_{\mbox{\scriptsize E}}),
\label{Eve-probe-state-0}
\end{equation}
$t\in\{x,z\}$, $i\in\{1,-1\}$, and $j\in\{1,4\}$.
In Appendix~\ref{section-appendix-C},
we give explicit forms of $|\phi(\sigma_{t}=i,r_{j})\rangle_{\mbox{\scriptsize E}}$.

Next, we define the following density operator:
\begin{equation}
\rho(\sigma_{t}=i,r_{j})_{\mbox{\scriptsize E}}
=
|\phi(\sigma_{t}=i,r_{j})\rangle_{\mbox{\scriptsize E}}{}_{\mbox{\scriptsize E}}\langle\phi(\sigma_{t}=i,r_{j})|.
\end{equation}
Then, the following relations hold:
\begin{eqnarray}
\mbox{Tr}\rho(\sigma_{z}=1,r_{1})_{\mbox{\scriptsize E}}
&=&
\mbox{Tr}\rho(\sigma_{x}=1,r_{1})_{\mbox{\scriptsize E}}
=
\mbox{Tr}\rho(\sigma_{x}=-1,r_{4})_{\mbox{\scriptsize E}}
=
\mbox{Tr}\rho(\sigma_{z}=-1,r_{4})_{\mbox{\scriptsize E}} \nonumber \\
&=&
(1/8)(1+F), \nonumber \\
\mbox{Tr}\rho(\sigma_{z}=-1,r_{1})_{\mbox{\scriptsize E}}
&=&
\mbox{Tr}\rho(\sigma_{x}=-1,r_{1})_{\mbox{\scriptsize E}}
=
\mbox{Tr}\rho(\sigma_{x}=1,r_{4})_{\mbox{\scriptsize E}}
=
\mbox{Tr}\rho(\sigma_{z}=1,r_{4})_{\mbox{\scriptsize E}} \nonumber \\
&=&
(1/8)(1-F).
\end{eqnarray}

Alice and Bob reveal to which subsequence the transmitted channel qubit belongs,
$S_{14}$ or $S_{23}$,
via the classical channel.
Eve observes her own probe only when the transmitted qubit is classified as $S_{14}$.
In order to guess right at Alice's secret bit,
Eve has to discriminate between the following two density operators:
\begin{eqnarray}
\rho_{0,\mbox{\scriptsize E}}
&=&
\rho(\sigma_{z}=1,r_{1})_{\mbox{\scriptsize E}}
+
\rho(\sigma_{x}=1,r_{1})_{\mbox{\scriptsize E}}
+
\rho(\sigma_{z}=-1,r_{1})_{\mbox{\scriptsize E}}
+
\rho(\sigma_{x}=-1,r_{1})_{\mbox{\scriptsize E}}, \nonumber \\
\rho_{1,\mbox{\scriptsize E}}
&=&
\rho(\sigma_{z}=-1,r_{4})_{\mbox{\scriptsize E}}
+
\rho(\sigma_{x}=-1,r_{4})_{\mbox{\scriptsize E}}
+
\rho(\sigma_{z}=1,r_{4})_{\mbox{\scriptsize E}}
+
\rho(\sigma_{x}=1,r_{4})_{\mbox{\scriptsize E}}.
\end{eqnarray}
We pay attention to the fact that $\mbox{Tr}\rho_{0,\mbox{\scriptsize E}}=\mbox{Tr}\rho_{1,\mbox{\scriptsize E}}=1/2$.
The probability $P_{\mbox{\scriptsize E}}$ that Eve recognizes the difference between $\rho_{0,\mbox{\scriptsize E}}$ and $\rho_{1,\mbox{\scriptsize E}}$
is given by
\begin{equation}
P_{\mbox{\scriptsize E}}
=
\frac{1}{2}
+
\frac{1}{2}
||\rho_{0,\mbox{\scriptsize E}}-\rho_{1,\mbox{\scriptsize E}}||_{\mbox{\scriptsize t}},
\label{PE-formula-0}
\end{equation}
where
$||X||_{\mbox{\scriptsize t}}=\mbox{Tr}|X|$
and
$|X|=\sqrt{X^{\dagger}X}$
for an arbitrary operator $X$ \cite{Nielsen2000,Fuchs1999}.

However, it is very difficult for us to express $||\rho_{0,\mbox{\scriptsize E}}-\rho_{1,\mbox{\scriptsize E}}||_{\mbox{\scriptsize t}}$
in a closed form in terms of elementary functions because we have to solve a quartic equation in an algebraic manner.
Eve's strategy of the translucent attack has the three real parameters $a$, $b$, and $F$
as shown in Eqs.~(\ref{state-fourth-Eve-1}), (\ref{state-fourth-Eve-2}),
(\ref{alpha-vector}), (\ref{gamma-vector}), (\ref{delta-vector-3}), and (\ref{beta-vector-4}).
We set the relation given by Eq.~(\ref{condition-a-b-F}) among them.
Thus, we can obtain $P_{\mbox{\scriptsize E}}$ and $F$ with numerical calculations in the following procedure.

\begin{figure}
\begin{center}
\includegraphics{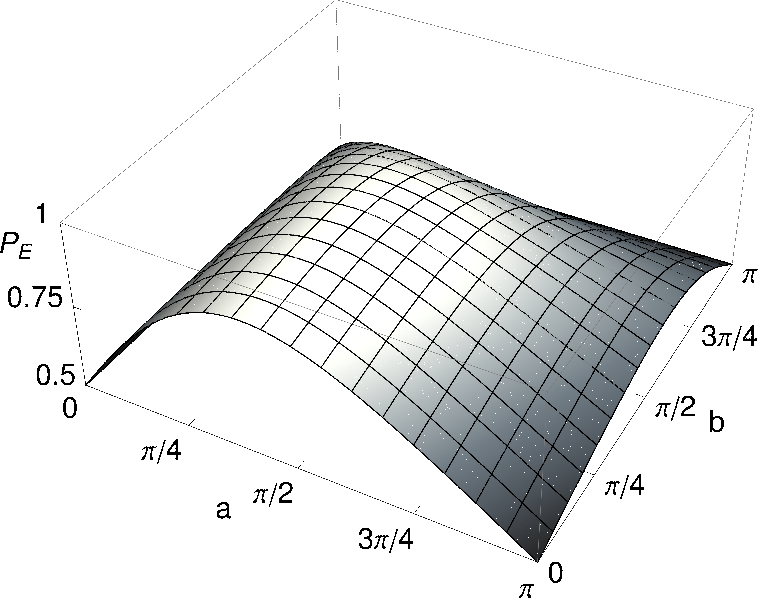}
\end{center}
\caption{A plot of $P_{\mbox{\tiny E}}$,
the probability that Eve makes a correct guess,
as a function of $a\in[0,\pi]$ and $b\in[0,\pi]$.
The probability $P_{\mbox{\tiny E}}$ has the maximum value $0.927$ at $a=1.30$ and $b=0.990$.}
\label{Figure_03}
\end{figure}

\begin{figure}
\begin{center}
\includegraphics{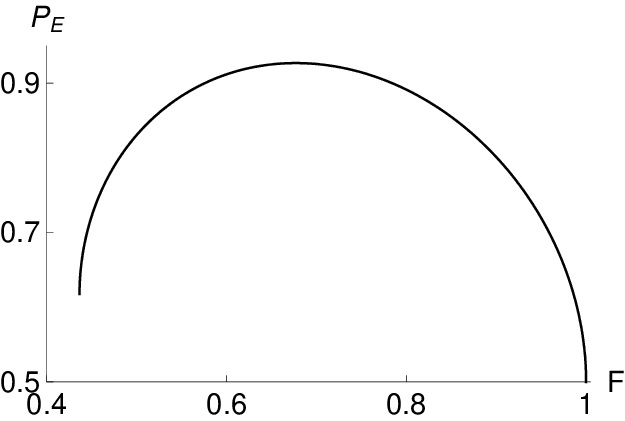}
\end{center}
\caption{A plot of $P_{\mbox{\tiny E}}$,
the probability that Eve makes a correct guess,
as a function of $F$ with fixing $b$ at $0.990$ and letting $a$ vary between zero and $\pi$.
The probability $P_{\mbox{\tiny E}}$ has the maximum value $0.927$ at $F=0.678$.}
\label{Figure_04}
\end{figure}

By feeding actual values into variables $a$ and $b$ as
\\
$(a,b)\in\{(j/200)\pi:j=0,1,2,...,200\}^{\otimes 2}$,
we can calculate $F$ and $P_{\mbox{\scriptsize E}}$ numerically with Eqs.~(\ref{condition-a-b-F}) and (\ref{PE-formula-0}).
In Fig.~\ref{Figure_03}, we plot $P_{\mbox{\scriptsize E}}$ as a function of $a$ and $b$.
The probability $P_{\mbox{\scriptsize E}}$ has the maximum value $0.927$ at $a=1.30$ and $b=0.990$.
Fixing $b$ at $0.990$ and letting $a$ vary between zero and $\pi$,
we plot $P_{\mbox{\scriptsize E}}$ as a function of $F$ in Fig.~\ref{Figure_04}.

Looking at Figs.~\ref{Figure_03} and \ref{Figure_04},
we become aware of the following.
When $a=1.30$ and $b=0.990$, $P_{\mbox{\scriptsize E}}$ takes the maximum value as $P_{\mbox{\scriptsize E}}=0.927$.
At this time, we obtain $F=0.678$ and $P_{\mbox{\scriptsize AB}}=0.839$.

Moreover, we notice that $P_{\mbox{\scriptsize E}}$ does not reach unity in Fig.~\ref{Figure_03}.
By contrast, in the BB84 scheme,
Eve's translucent attack allows $P_{\mbox{\scriptsize E}}$ to attain unity when the disturbance becomes maximum \cite{Cirac1997}.
This fact implies that the one-way translucent attack in Bub's protocol may not be optimum for Eve.
We confirm this suggestion from results obtained in the next section.

\section{\label{section-two-way-translucent-attack}The two-way translucent attack}
In this section,
we investigate the security against the two-way translucent attack.
Here, we pursue a series of Eve's acts step by step.

First, Alice prepares the initial entangled state $|\psi\rangle_{\mbox{\scriptsize AC}}$
given by Eq.~(\ref{initial-state}).
Second, Alice sends the channel qubit to Bob.
Third, in the middle of the quantum channel from Alice to Bob,
Eve lets her own probe $\mbox{E1}$ interact with the channel qubit using the unitary transformation described
in Eqs.~(\ref{state-fourth-Eve-1}) and (\ref{state-fourth-Eve-2}).
We assume that the initial state of the probe is given by $|X\rangle_{\mbox{\scriptsize E1}}$, $F=F'$,
and
$|\alpha\rangle_{\mbox{\scriptsize E1}}$,
$|\beta\rangle_{\mbox{\scriptsize E1}}$,
$|\gamma\rangle_{\mbox{\scriptsize E1}}$,
and $|\delta\rangle_{\mbox{\scriptsize E1}}$
are given by Eqs.~(\ref{alpha-vector}), (\ref{gamma-vector}), (\ref{delta-vector-3}), and (\ref{beta-vector-4}).
At this moment, the whole state is written down as
\begin{equation}
|\tilde{\psi}\rangle_{\mbox{\scriptsize ACE1}}
=
(1/\sqrt{2})
[
|0\rangle_{\mbox{\scriptsize A}}
(|0\rangle_{\mbox{\scriptsize C}}|A\rangle_{\mbox{\scriptsize E1}}+|1\rangle_{\mbox{\scriptsize C}}|B\rangle_{\mbox{\scriptsize E1}})
+
|1\rangle_{\mbox{\scriptsize A}}
(|0\rangle_{\mbox{\scriptsize C}}|C\rangle_{\mbox{\scriptsize E1}}+|1\rangle_{\mbox{\scriptsize C}}|D\rangle_{\mbox{\scriptsize E1}})
],
\end{equation}
where
\begin{eqnarray}
|A\rangle_{\mbox{\scriptsize E1}}
&=&
\sqrt{F}|\alpha\rangle_{\mbox{\scriptsize E1}}, \nonumber \\
|B\rangle_{\mbox{\scriptsize E1}}
&=&
\sqrt{1-F}|\beta\rangle_{\mbox{\scriptsize E1}}, \nonumber \\
|C\rangle_{\mbox{\scriptsize E1}}
&=&
\sqrt{1-F}|\gamma\rangle_{\mbox{\scriptsize E1}}, \nonumber \\
|D\rangle_{\mbox{\scriptsize E1}}
&=&
\sqrt{F}|\delta\rangle_{\mbox{\scriptsize E1}}.
\end{eqnarray}

Fourth, Bob performs a projective measurement with $\sigma_{x}$ or $\sigma_{z}$ on the channel qubit that belongs to $|\tilde{\psi}\rangle_{\mbox{\scriptsize ACE1}}$.
For example,
if Bob carries out the observation with $\sigma_{z}$,
the reduction of the wave packet occurs and the whole state changes into
\begin{eqnarray}
&&
|0\rangle_{\mbox{\scriptsize C}}{}_{\mbox{\scriptsize C}}\langle 0|
\otimes
\mbox{tr}_{\mbox{\scriptsize C}}
(
|0\rangle_{\mbox{\scriptsize C}}{}_{\mbox{\scriptsize C}}\langle 0|\tilde{\psi}\rangle_{\mbox{\scriptsize ACE1}}{}_{\mbox{\scriptsize ACE1}}\langle\tilde{\psi}|
)
+
|1\rangle_{\mbox{\scriptsize C}}{}_{\mbox{\scriptsize C}}\langle 1|
\otimes
\mbox{tr}_{\mbox{\scriptsize C}}
(
|1\rangle_{\mbox{\scriptsize C}}{}_{\mbox{\scriptsize C}}\langle 1|\tilde{\psi}\rangle_{\mbox{\scriptsize ACE1}}{}_{\mbox{\scriptsize ACE1}}\langle\tilde{\psi}|
) \nonumber \\
&=&
(1/2)
|0\rangle_{\mbox{\scriptsize C}}{}_{\mbox{\scriptsize C}}\langle 0|
\otimes
|\phi(z,0)\rangle_{\mbox{\scriptsize AE1}}{}_{\mbox{\scriptsize AE1}}\langle\phi(z,0)|
+
(1/2)
|1\rangle_{\mbox{\scriptsize C}}{}_{\mbox{\scriptsize C}}\langle 1|
\otimes
|\phi(z,1)\rangle_{\mbox{\scriptsize AE1}}{}_{\mbox{\scriptsize AE1}}\langle\phi(z,1)|, \nonumber \\
\end{eqnarray}
where
\begin{eqnarray}
|\phi(z,0)\rangle_{\mbox{\scriptsize AE1}}
&=&
|0\rangle_{\mbox{\scriptsize A}}|A\rangle_{\mbox{\scriptsize E1}}
+
|1\rangle_{\mbox{\scriptsize A}}|C\rangle_{\mbox{\scriptsize E1}}, \nonumber \\
|\phi(z,1)\rangle_{\mbox{\scriptsize AE1}}
&=&
|0\rangle_{\mbox{\scriptsize A}}|B\rangle_{\mbox{\scriptsize E1}}
+
|1\rangle_{\mbox{\scriptsize A}}|D\rangle_{\mbox{\scriptsize E1}}.
\end{eqnarray}
Then, Bob returns the channel qubit to Alice.

Fifth, halfway along the quantum channel from Bob to Alice,
Eve lets another probe $\mbox{E2}$ of hers interact with the channel qubit using the unitary transformation $U'$
given by Eqs.~(\ref{state-fourth-Eve-1}) and (\ref{state-fourth-Eve-2}),
where the fidelities $F$ and $F'$ are denoted by $F'$ together.
We assume that Eve puts the initial state of the probe $|X'\rangle_{\mbox{\scriptsize E2}}$.
Then, the whole state evolves into
\begin{eqnarray}
&&
(1/2)
|\varphi(z,0)\rangle_{\mbox{\scriptsize CE2}}{}_{\mbox{\scriptsize CE2}}\langle\varphi(z,0)|
\otimes
|\phi(z,0)\rangle_{\mbox{\scriptsize AE1}}{}_{\mbox{\scriptsize AE1}}\langle\phi(z,0)| \nonumber \\
&&
+
(1/2)
|\varphi(z,1)\rangle_{\mbox{\scriptsize CE2}}{}_{\mbox{\scriptsize CE2}}\langle\varphi(z,1)|
\otimes
|\phi(z,1)\rangle_{\mbox{\scriptsize AE1}}{}_{\mbox{\scriptsize AE1}}\langle\phi(z,1)|,
\end{eqnarray}
where
\begin{eqnarray}
|\varphi(z,0)\rangle_{\mbox{\scriptsize CE2}}
&=&
|0\rangle_{\mbox{\scriptsize C}}|A'\rangle_{\mbox{\scriptsize E2}}
+
|1\rangle_{\mbox{\scriptsize C}}|B'\rangle_{\mbox{\scriptsize E2}}, \nonumber \\
|\varphi(z,1)\rangle_{\mbox{\scriptsize CE2}}
&=&
|0\rangle_{\mbox{\scriptsize C}}|C'\rangle_{\mbox{\scriptsize E2}}
+
|1\rangle_{\mbox{\scriptsize C}}|D'\rangle_{\mbox{\scriptsize E2}},
\end{eqnarray}
\begin{eqnarray}
|A'\rangle_{\mbox{\scriptsize E2}}
&=&
\sqrt{F'}|\alpha'\rangle_{\mbox{\scriptsize E2}}, \nonumber \\
|B'\rangle_{\mbox{\scriptsize E2}}
&=&
\sqrt{1-F'}|\beta'\rangle_{\mbox{\scriptsize E2}}, \nonumber \\
|C'\rangle_{\mbox{\scriptsize E2}}
&=&
\sqrt{1-F'}|\gamma'\rangle_{\mbox{\scriptsize E2}}, \nonumber \\
|D'\rangle_{\mbox{\scriptsize E2}}
&=&
\sqrt{F'}|\delta'\rangle_{\mbox{\scriptsize E2}},
\end{eqnarray}
and
$|\alpha'\rangle_{\mbox{\scriptsize E2}}$,
$|\beta'\rangle_{\mbox{\scriptsize E2}}$,
$|\gamma'\rangle_{\mbox{\scriptsize E2}}$,
and $|\delta'\rangle_{\mbox{\scriptsize E2}}$
are given by Eqs.~(\ref{alpha-vector}), (\ref{gamma-vector}), (\ref{delta-vector-3}), and (\ref{beta-vector-4})
with parameters $a'$ and $b'$ instead of $a$ and $b$.

Sixth, Alice performs the orthogonal measurement with $\{|r_{j}\rangle_{\mbox{\scriptsize AC}}:j=1,2,3,4\}$.
A probability that Alice obtains $|r_{j}\rangle_{\mbox{\scriptsize AC}}$ is given by
\begin{eqnarray}
&&
{}_{\mbox{\scriptsize AC}}\langle r_{j}|
\mbox{tr}_{\mbox{\scriptsize E2}}
[
|\varphi(z,0)\rangle_{\mbox{\scriptsize CE2}}
{}_{\mbox{\scriptsize CE2}}\langle \varphi(z,0)|
]
\otimes
\mbox{tr}_{\mbox{\scriptsize E1}}
[
|\phi(z,0)\rangle_{\mbox{\scriptsize AE1}}
{}_{\mbox{\scriptsize AE1}}\langle \phi(z,0)|
]
|r_{j}\rangle_{\mbox{\scriptsize AC}}.
\end{eqnarray}
Here, we introduce the following notation to describe the probability that Bob observes $\sigma_{t}$
and obtains the output $i$ and Alice detects $|r_{j}\rangle_{\mbox{\scriptsize AC}}$ as
\begin{eqnarray}
K(\sigma_{t}=i,r_{j})
&=&
(1/2)
{}_{\mbox{\scriptsize AC}}\langle r_{j}|
\mbox{tr}_{\mbox{\scriptsize E2}}
[
|\varphi(t,i_{t})\rangle_{\mbox{\scriptsize CE2}}
{}_{\mbox{\scriptsize CE2}}\langle \varphi(t,i_{t})|
] \nonumber \\
&&
\otimes
\mbox{tr}_{\mbox{\scriptsize E1}}
[
|\phi(t,i_{t})\rangle_{\mbox{\scriptsize AE1}}
{}_{\mbox{\scriptsize AE1}}\langle \phi(t,i_{t})|
]
|r_{j}\rangle_{\mbox{\scriptsize AC}} \nonumber \\
&&
\mbox{for $t\in\{x,z\}, i\in\{1,-1\}, j\in\{1,2,3,4\}$,}
\end{eqnarray}
where $i_{t}$ is given by Eqs.~(\ref{def-iz}) and (\ref{def-ix}).

From some tedious calculations,
we obtain
\begin{eqnarray}
K(\sigma_{z}=1,r_{1})
&=&
K(\sigma_{z}=1,r_{2})
=
K(\sigma_{z}=-1,r_{3})
=
K(\sigma_{z}=-1,r_{4}) \nonumber \\
&=&
\frac{1}{8}(F+F'), \nonumber \\
K(\sigma_{z}=1,r_{3})
&=&
K(\sigma_{z}=1,r_{4})
=
K(\sigma_{z}=-1,r_{1})
=
K(\sigma_{z}=-1,r_{2}) \nonumber \\
&=&
\frac{1}{8}(2-F-F'), \nonumber \\
K(\sigma_{x}=1,r_{1})
&=&
K(\sigma_{x}=1,r_{3})
=
K(\sigma_{x}=-1,r_{2})
=
K(\sigma_{x}=-1,r_{4}) \nonumber \\
&=&
\frac{1}{16}
[2+F\cos a+F'\cos a'+(1-F)\cos b+(1-F')\cos b'], \nonumber \\
K(\sigma_{x}=1,r_{2})
&=&
K(\sigma_{x}=1,r_{4})
=
K(\sigma_{x}=-1,r_{1})
=
K(\sigma_{x}=-1,r_{3}) \nonumber \\
&=&
\frac{1}{16}
[2-F\cos a-F'\cos a'-(1-F)\cos b-(1-F')\cos b'].
\end{eqnarray}
Then, we request the following conditions:
\begin{eqnarray}
K(\sigma_{z}=1,r_{2})
&=&
K(\sigma_{z}=-1,r_{3})
=
K(\sigma_{x}=-1,r_{2})
=
K(\sigma_{x}=1,r_{3}), \nonumber \\
K(\sigma_{z}=-1,r_{2})
&=&
K(\sigma_{z}=1,r_{3})
=
K(\sigma_{x}=1,r_{2})
=
K(\sigma_{x}=-1,r_{3}), \nonumber \\
K(\sigma_{z}=1,r_{1})
&=&
K(\sigma_{z}=-1,r_{4})
=
K(\sigma_{x}=1,r_{1})
=
K(\sigma_{x}=-1,r_{4}), \nonumber \\
K(\sigma_{z}=-1,r_{1})
&=&
K(\sigma_{z}=1,r_{4})
=
K(\sigma_{x}=-1,r_{1})
=
K(\sigma_{x}=1,r_{4}).
\end{eqnarray}

From the above requirements,
we obtain Eq.~(\ref{condition-a-b-F}) and
\begin{equation}
\cos b'
=
\frac{-1+2F'-F'\cos a'}{1-F'}.
\label{condition-adash-bdash-Fdash}
\end{equation}
Hence, we can specify Eve's attack with four real parameters $a$, $b$, $a'$, and $b'$.

The probability $P_{\mbox{\scriptsize AB}}$ that Alice and Bob do not notice Eve's illegal acts is given by
\begin{eqnarray}
P_{\mbox{\scriptsize AB}}
&=&
K(\sigma_{x}=-1,r_{2})
+
K(\sigma_{x}=1,r_{3})
+
K(\sigma_{z}=1,r_{2})
+
K(\sigma_{z}=-1,r_{3}) \nonumber \\
&=&
\frac{1}{2}(F+F').
\end{eqnarray}

Next, we think about states between which Eve has to discriminate for guessing right at Alice's secret bit.
Here, we consider a concrete example.
If Bob obtains $\sigma_{z}=1$ and Alice detects $|r_{1}\rangle_{\mbox{\scriptsize AC}}$,
the state of Eve's probes suffers from reduction and becomes
\begin{equation}
\frac{1}{\sqrt{2}}
|\Phi(\sigma_{z}=1,r_{1})\rangle_{\mbox{\scriptsize E1E2}}
=
\frac{1}{\sqrt{2}}
{}_{\mbox{\scriptsize AC}}\langle r_{1}|
[
|\phi(z,0)\rangle_{\mbox{\scriptsize AE1}}
\otimes
|\varphi(z,0)\rangle_{\mbox{\scriptsize CE2}}
].
\end{equation}
In Appendix~\ref{section-appendix-D},
we list explicit forms of
\begin{equation}
|\Phi(\sigma_{t}=i,r_{j})\rangle_{\mbox{\scriptsize E1E2}}
=
{}_{\mbox{\scriptsize AC}}\langle r_{j}|
[
|\phi(t,i_{t})\rangle_{\mbox{\scriptsize AE1}}
\otimes
|\varphi(t,i_{t})\rangle_{\mbox{\scriptsize CE2}}
]
\label{def-Phi-E1E2}
\end{equation}
for $t\in\{x,z\}$, $i\in\{1,-1\}$, and $j\in\{1,4\}$.

Next, we define the following density operator:
\begin{equation}
\rho(\sigma_{t}=i,r_{j})_{\mbox{\scriptsize E1E2}}
=
\frac{1}{2}
|\Phi(\sigma_{t}=i,r_{j})\rangle_{\mbox{\scriptsize E1E2}}
{}_{\mbox{\scriptsize E1E2}}\langle\Phi(\sigma_{t}=i,r_{j})|.
\end{equation}
Then, the following relations hold:
\begin{eqnarray}
\mbox{tr}
[
\rho(\sigma_{z}=1,r_{1})_{\mbox{\scriptsize E1E2}}
]
&=&
\mbox{tr}
[
\rho(\sigma_{z}=-1,r_{4})_{\mbox{\scriptsize E1E2}}
]
=
\frac{1}{8}(F+F'), \nonumber \\
\mbox{tr}
[
\rho(\sigma_{z}=1,r_{4})_{\mbox{\scriptsize E1E2}}
]
&=&
\mbox{tr}
[
\rho(\sigma_{z}=-1,r_{1})_{\mbox{\scriptsize E1E2}}
]
=
\frac{1}{8}(2-F-F'), \nonumber \\
\mbox{tr}
[
\rho(\sigma_{x}=1,r_{1})_{\mbox{\scriptsize E1E2}}
]
&=&
\mbox{tr}
[
\rho(\sigma_{x}=-1,r_{4})_{\mbox{\scriptsize E1E2}}
] \nonumber \\
&=&
\frac{1}{16}
[
2+F\cos a+F'\cos a'+(1-F)\cos b+(1-F')\cos b'
] \nonumber \\
&=&
\frac{1}{8}
\Biggl(
2-\frac{1-\cos a}{2-\cos a+\cos b}-\frac{1-\cos a'}{2-\cos a'+\cos b'}
\Biggr), \nonumber \\
\mbox{tr}
[
\rho(\sigma_{x}=1,r_{4})_{\mbox{\scriptsize E1E2}}
]
&=&
\mbox{tr}
[
\rho(\sigma_{x}=-1,r_{1})_{\mbox{\scriptsize E1E2}}
] \nonumber \\
&=&
\frac{1}{16}
[
2-F\cos a-F'\cos a'-(1-F)\cos b-(1-F')\cos b'
] \nonumber \\
&=&
\frac{1}{8}
\Biggl(
\frac{1-\cos a}{2-\cos a+\cos b}+\frac{1-\cos a'}{2-\cos a'+\cos b'}
\Biggr).
\end{eqnarray}

In order to guess right at Alice's secret bit,
Eve has to discriminate between the following two density operators:
\begin{eqnarray}
\rho_{0,\mbox{\scriptsize E1E2}}
&=&
\rho(\sigma_{z}=1,r_{1})_{\mbox{\scriptsize E1E2}}
+
\rho(\sigma_{x}=1,r_{1})_{\mbox{\scriptsize E1E2}} \nonumber \\
&&
+
\rho(\sigma_{z}=-1,r_{1})_{\mbox{\scriptsize E1E2}}
+
\rho(\sigma_{x}=-1,r_{1})_{\mbox{\scriptsize E1E2}}, \nonumber \\
\rho_{1,\mbox{\scriptsize E1E2}}
&=&
\rho(\sigma_{z}=-1,r_{4})_{\mbox{\scriptsize E1E2}}
+
\rho(\sigma_{x}=-1,r_{4})_{\mbox{\scriptsize E1E2}} \nonumber \\
&&
+
\rho(\sigma_{z}=1,r_{4})_{\mbox{\scriptsize E1E2}}
+
\rho(\sigma_{x}=1,r_{4})_{\mbox{\scriptsize E1E2}}.
\end{eqnarray}
The probability $P_{\mbox{\scriptsize E}}$ that Eve recognizes the difference
between $\rho_{0,\mbox{\scriptsize E1E2}}$ and $\rho_{1,\mbox{\scriptsize E1E2}}$
is given by
\begin{equation}
P_{\mbox{\scriptsize E}}
=
\frac{1}{2}
+
\frac{1}{2}
||\rho_{0,\mbox{\scriptsize E1E2}}-\rho_{1,\mbox{\scriptsize E1E2}}||_{\mbox{\scriptsize t}}.
\label{PE-formula-1}
\end{equation}
Because $\rho_{0,\mbox{\scriptsize E1E2}}$ and $\rho_{1,\mbox{\scriptsize E1E2}}$ are $16\times 16$ matrices,
it is impossible to derive a closed analytical form of $P_{\mbox{\scriptsize E}}$.
Thus, we calculate $P_{\mbox{\scriptsize E}}$ numerically.

\begin{figure}
\begin{center}
\includegraphics{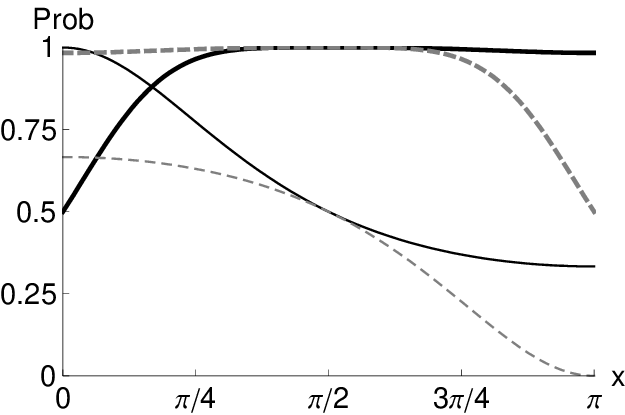}
\end{center}
\caption{The thick and thin black curves represent $P_{\mbox{\tiny E}}$ and $P_{\mbox{\tiny AB}}$ respectively
as functions of $x$,
where $a=a'=x$ and $b=b'=\pi/2$.
The thick and thin grey dashed curves represent $P_{\mbox{\tiny E}}$ and $P_{\mbox{\tiny AB}}$ respectively
as functions of $x$,
where $a=a'=\pi/2$ and $b=b'=x$.
Putting $a=a'=b=b'=\pi/2$,
$P_{\mbox{\tiny E}}$ attains unity and $P_{\mbox{\tiny AB}}$ is equal to $1/2$.}
\label{Figure_05}
\end{figure}

We can represent the probability $P_{\mbox{\scriptsize E}}$ as a function of four real parameters $a$, $b$, $a'$, and $b'$,
so that we describe it as $P_{\mbox{\scriptsize E}}(a,b,a',b')$.
Here, we examine two types of arrangements of the parameters.
In Fig.~\ref{Figure_05}, we plot $P_{\mbox{\scriptsize E}}(x,\pi/2,x,\pi/2)$ and $P_{\mbox{\scriptsize E}}(\pi/2,x,\pi/2,x)$
as functions of $x\in[0,\pi]$.
We also plot $P_{\mbox{\scriptsize AB}}(x,\pi/2,x,\pi/2)$ and $P_{\mbox{\scriptsize AB}}(\pi/2,x,\pi/2,x)$ in Fig.~\ref{Figure_05}.

Putting $x=\arccos(2/3)\simeq 0.841$, we obtain $P_{\mbox{\scriptsize E}}(x,\pi/2,x,\pi/2)\simeq 0.975$ and
\\
$P_{\mbox{\scriptsize AB}}(x,\pi/2,x,\pi/2)=3/4$.
Putting $x=0$, we obtain $P_{\mbox{\scriptsize E}}(\pi/2,x,\pi/2,x)\simeq 0.984$ and $P_{\mbox{\scriptsize AB}}(\pi/2,x,\pi/2,x)=2/3$.
Eve's optimum case for one-way translucent attack obtained in Sec.~\ref{section-one-way-translucent-attack} corresponds to
$(a,b,a',b')=(1.30,0.990,0,\pi/2)$.

Because the two-way translucent attack includes the one-way translucent attack as a special case,
the former one is better than the latter one for Eve obviously.
However, the two-way translucent attack does not give an overwhelming advantage to Eve.
For example, we can consider the following simple case.
If we adjust the parameters as $a=b=a'=b'=\pi/2$,
we can obtain $P_{\mbox{\scriptsize E}}=1$.
However, this choice of the parameters causes $P_{\mbox{\scriptsize AB}}=1/2$.

In the BB84 scheme, if Eve makes a translucent attack,
Eve's best strategy leads to $P_{\mbox{\scriptsize E}}\simeq 0.933$ with setting $P_{\mbox{\scriptsize AB}}=3/4$.
Thus, Bub's protocol is weaker than the BB84 scheme with respect to the security against Eve's translucent attack.

\section{\label{section-Eve-careful-attack}Eve's most careful one-way translucent attack}
In this section, we consider security against Eve's most careful one-way translucent attack.
Here, we assume that Eve is very cautious and thus she wants to let Alice and Bob not notice her illegal acts at all.
For the circumstances of this situation,
Eve prefers the one-way translucent attack to the two-way translucent attack
because the former one is gentler than the latter one.

Hence, Eve must choose a strategy that makes $P_{\mbox{\scriptsize AB}}$ defined in Eq.~(\ref{P-AB-one-way-translucent}) be unity.
However, we can show that Eve cannot learn anything about a secret bit shared by Alice and Bob if she selects this strategy.
We prove this fact in the following.

In the one-way translucent attack discussed in Sec.~\ref{section-one-way-translucent-attack},
Eve applies a unitary transformation to her probe and the channel qubit.
This transformation is given by Eqs.~(\ref{state-fourth-Eve-1}) and (\ref{state-fourth-Eve-2}).
However, in order not to leave evidence of her malicious acts,
Eve must let $F$ and $F'$ be unity,
so that her unitary transformation is rewritten down as
\begin{eqnarray}
U|0\rangle_{\mbox{\scriptsize C}}|X\rangle_{\mbox{\scriptsize E}}
&=&
|0\rangle_{\mbox{\scriptsize C}}|\alpha\rangle_{\mbox{\scriptsize E}}, \nonumber \\
U|1\rangle_{\mbox{\scriptsize C}}|X\rangle_{\mbox{\scriptsize E}}
&=&
|1\rangle_{\mbox{\scriptsize C}}|\delta\rangle_{\mbox{\scriptsize E}}.
\end{eqnarray}

Hence, the dimension of a Hilbert space for Eve's probe is equal to two at most.
Thus, we can describe $|\alpha\rangle_{\mbox{\scriptsize E}}$ and $|\delta\rangle_{\mbox{\scriptsize E}}$ as follows in general:
\begin{eqnarray}
|\alpha\rangle_{\mbox{\scriptsize E}}
&=&
(1,0)^{\mbox{\scriptsize T}}, \nonumber \\
|\delta\rangle_{\mbox{\scriptsize E}}
&=&
(r,\sqrt{1-r^{2}})^{\mbox{\scriptsize T}}.
\end{eqnarray}
(Although an arbitrary two-dimensional vector is given by $(r,\sqrt{1-r^{2}}e^{i\theta})^{\mbox{\scriptsize T}}$,
we can omit the factor $e^{i\theta}$ with adjusting the second vector of the basis.
This treatment is allowed because the second component of $|\alpha\rangle_{\mbox{\scriptsize E}}$ is equal to zero.)

Then, calculating $K(\sigma_{t}=i,r_{j})$ defined in Eq.~(\ref{K-sigma-r-definition}),
we obtain
\begin{eqnarray}
K(\sigma_{x}=-1,r_{2})
&=&
K(\sigma_{x}=1,r_{3})
=
\frac{1}{16}(3+r), \nonumber \\
K(\sigma_{z}=1,r_{2})
&=&
K(\sigma_{z}=-1,r_{3})
=
\frac{1}{4}.
\end{eqnarray}
Hence, from Eq.~(\ref{P-AB-one-way-translucent}), we obtain
\begin{equation}
P_{\mbox{\scriptsize AB}}
=
\frac{1}{8}(7+r).
\end{equation}

Therefore, if Eve wants to completely avert Alice and Bob's detection of her illegal acts,
she has to put $r=1$.
This arrangement implies that
$|\alpha\rangle_{\mbox{\scriptsize E}}
=
|\delta\rangle_{\mbox{\scriptsize E}}
=
(1,0)^{\mbox{\scriptsize T}}$.
Hence, Eve's probe and the channel qubit are perfectly disentangled and she cannot gain any information.
Thus, we can conclude that Eve's translucent attack gives her exactly zero information if she is restricted to make no noise.

\section{\label{section-discussion}Discussion}
In the current paper, we obtain two facts about Bub's quantum key distribution protocol.
The first one is the following.
If Eve makes the intercept/resend attack on the transmissions between Alice and Bob under the condition of Eq.~(\ref{Eve-strategy-symmetric-condition-01}),
Eve's best strategy is performing the measurement with the Breidbart basis
on the middle of the quantum channel from Alice to Bob.
The one-way intercept/resend attack is more favourable to Eve than the two-way one.
For this attack on the single transmission, the probability $P_{\mbox{\scriptsize AB}}$ that Alice and Bob do not notice Eve's illegal acts is equal to $5/6\simeq 0.833$
and the probability $P_{\mbox{\scriptsize E}}$ that Eve guesses right at the secret bit Alice obtains is given by
$(5+3\sqrt{2})/10\simeq 0.924$.
If Eve makes the intercept/resend attack on $n$ qubits Alice sends,
the probability that Eve's malicious acts are not revealed is given by $(5/6)^{n}$.
This probability decreases exponentially as $n$ becomes larger.
Thus, Bub's protocol is safe from the intercept/resend attack.
To specify Eve's optimum strategy,
we assume that the condition given by Eq.~(\ref{Eve-strategy-symmetric-condition-01}) holds.
We think that this constraint imposed upon Eve's strategy is natural and reasonable,
so that the Breidbart basis is best for Eve in general.

If Eve makes the intercept/resend attack on the BB84 scheme,
her best strategy is also the measurement with the Breidbart basis \cite{Bennett1992b}.
For this attack on the single transmission, the probability $P_{\mbox{\scriptsize AB}}$ that Alice and Bob do not find signs of Eve's illegal acts is equal to $3/4$
and the probability $P_{\mbox{\scriptsize E}}$ that Eve guesses right at the random secret bit Alice obtains is given by
$(1/4)(2+\sqrt{2})\simeq 0.854$.
Thus, we can conclude that the BB84 scheme is safer than Bub's protocol concerning to the intercept/resend attack.

The second fact that the current paper shows is the following.
If Eve mounts the one-way translucent attack on Bub's protocol,
she can let the probability $P_{\mbox{\scriptsize E}}$ that she guesses right at the secret bit Alice obtains be equal to $0.927$ at least.
When Eve choose this strategy,
the probability $P_{\mbox{\scriptsize AB}}$ that Alice and Bob do not notice Eve's illegal acts is given by $0.839$.
By contrast, if Eve makes the two-way translucent attack,
she can let $P_{\mbox{\scriptsize E}}$ be equal to unity.
However, in this case, $P_{\mbox{\scriptsize AB}}$ is equal to $1/2$.
In the two-way translucent attack, a trade-off between $P_{\mbox{\scriptsize AB}}$ and $P_{\mbox{\scriptsize E}}$ makes Eve think carefully
which values she must choose for parameters specifying the strategy.
However, one thing is certain,
the translucent attack is more dangerous than the intercept/resend attack for Bub's protocol.

If Eve makes the translucent attack on a single transmission of the BB84 scheme,
we can estimate $P_{\mbox{\scriptsize E}}$ at $(1/2)[1+(\sqrt{3}/2)]\simeq 0.933$ at least
with setting $P_{\mbox{\scriptsize AB}}$ at $3/4$ \cite{Cirac1997}.
This implies that the translucent attack is more dangerous than the intercept/resend attack for the BB84 scheme.
In the present paper,
we illustrate the fact that the BB84 scheme is robuster than Bub's protocol
against the translucent attack.

In the current paper, we do not intend to tell which protocol is better, Bub's one or other quantum key distribution scheme,
for example, the BB84 scheme.
In the present paper, we aim at clarifying characteristic properties of Bub's protocol from a neutral viewpoint.
In Refs.~\cite{Lo1999,Mayers2001,Biham2006}, and \cite{Shor2000},
the BB84 scheme was proven secure.
In other words, it was rigorously indicated that the BB84 scheme is secure against an enemy
who is able to perform any physical operation permitted by quantum mechanics.
Contrastingly, the security of Bub's protocol has not been studied well yet.
We have to admit that it is not full-grown theoretically or experimentally.

One of the most serious faults Bub's quantum key distribution protocol has is that Alice has to perform measurements of two-qubit states for detection of
$\{|r_{i}\rangle_{\mbox{\scriptsize AC}}:i=1,2,3,4\}$.
Because $\{|r_{i}\rangle_{\mbox{\scriptsize AC}}\}$ have entanglement,
Alice needs to prepare a quantum circuit, which was examined in Refs.~\cite{Bub2001} and \cite{Metzger2000}.

As mentioned above, experimental realization of Bub's protocol owns some difficulties.
Because the protocol uses a two-way quantum channel,
it is vulnerable to the channel loss and a noise source,
compared with the BB84 and E91 schemes that make use of a one-way quantum channel.
Moreover, in Bub's protocol,
Bob has to carry out the projective measurement,
so that he must not destroy the channel qubit.
In contrast, for the BB84 and E91 schemes,
Alice and Bob only need to perform an ordinary strong quantum measurement.

The investigation of the BB84 scheme has a long tradition and its practical use has been studied in many papers,
for example, Refs.~\cite{Lutkenhaus1999,Lutkenhaus2000,Gottesman2004}, and \cite{Scarani2009}.
Noises of the experimental setup lead to increase of the quantum bit error rate.
However, in the current paper, we do not evaluate the maximum quantum bit error rate that the legitimate users can accept
because we focus on an ideal case where the channel losses and noise sources are not assumed,
the photodetectors work perfectly, and so on.
In the present paper, we do not argue those experimental aspects and practical uses of Bub's protocol.
These problems remain to be examined in the future.

However, Bub's protocol is a natural application of the ABL-rule.
Thus, we can consider Bub's one to be a typical example of the strange nature of quantum mechanics.
Moreover, in the current paper, we show that we can analyse the security of Bub's protocol against some specified strategies of eavesdropping
in analytical and numerical manners.
These points give sound reasons why we study the quantum key distribution protocol based on the pre- and post-selection effect.

\appendix

\section{\label{section-appendix-A}Some useful functions for Secs.~\ref{section-one-way-intercept-resend-attack} and \ref{section-symmetric-strategy-Eve}}
For the sake of convenience in Secs.~\ref{section-one-way-intercept-resend-attack} and \ref{section-symmetric-strategy-Eve},
we calculate the following equations
from $|\psi\rangle_{\mbox{\scriptsize AC}}$, $|r_{i}\rangle_{\mbox{\scriptsize AC}}$ for $i=1, 2, 3, 4$, and $\hat{P}(\sigma_{\xi}=\pm 1)$
defined in Eqs.~(\ref{initial-state}), (\ref{Alice-measurement-basis}), (\ref{projection-sigma-plus}), and (\ref{projection-sigma-minus}).
We pay attention to the fact that $\hat{P}(\sigma_{x}=\pm 1)$, $\hat{P}(\sigma_{z}=\pm 1)$, and $\hat{P}(\sigma_{\xi}=\pm 1)$
act on the channel qubit.
\begin{eqnarray}
{}_{\mbox{\scriptsize AC}}\langle r_{1}|\hat{P}(\sigma_{x}=1)\hat{P}(\sigma_{\xi}=1)|\psi\rangle_{\mbox{\scriptsize AC}}
&=&
(1/8)[2+(1-i)\cos\beta+2\cos\alpha\sin\beta \nonumber \\
&&
\quad
+(1+i)\sin\alpha\sin\beta], \nonumber \\
{}_{\mbox{\scriptsize AC}}\langle r_{1}|\hat{P}(\sigma_{x}=1)\hat{P}(\sigma_{\xi}=-1)|\psi\rangle_{\mbox{\scriptsize AC}}
&=&
(1/8)[2-(1-i)\cos\beta-2\cos\alpha\sin\beta \nonumber \\
&&
\quad
-(1+i)\sin\alpha\sin\beta], \nonumber \\
{}_{\mbox{\scriptsize AC}}\langle r_{1}|\hat{P}(\sigma_{x}=-1)\hat{P}(\sigma_{\xi}=1)|\psi\rangle_{\mbox{\scriptsize AC}}
&=&
(1/8)[(1+i)\cos\beta+(1-i)\sin\alpha\sin\beta], \nonumber \\
{}_{\mbox{\scriptsize AC}}\langle r_{1}|\hat{P}(\sigma_{x}=-1)\hat{P}(\sigma_{\xi}=-1)|\psi\rangle_{\mbox{\scriptsize AC}}
&=&
-(1/8)(1-i)(i\cos\beta+\sin\alpha\sin\beta),
\label{probability-set-1}
\end{eqnarray}
\begin{eqnarray}
{}_{\mbox{\scriptsize AC}}\langle r_{1}|\hat{P}(\sigma_{z}=1)\hat{P}(\sigma_{\xi}=1)|\psi\rangle_{\mbox{\scriptsize AC}}
&=&
(1/2)\cos(\beta/2)
[\cos(\beta/2) \nonumber \\
&&
\quad
+(1/2)(1+i)e^{-i\alpha}\sin(\beta/2)], \nonumber \\
{}_{\mbox{\scriptsize AC}}\langle r_{1}|\hat{P}(\sigma_{z}=1)\hat{P}(\sigma_{\xi}=-1)|\psi\rangle_{\mbox{\scriptsize AC}}
&=&
(1/2)\sin(\beta/2)
[\sin(\beta/2) \nonumber \\
&&
\quad
-(1/2)(1+i)e^{-i\alpha}\cos(\beta/2)], \nonumber \\
{}_{\mbox{\scriptsize AC}}\langle r_{1}|\hat{P}(\sigma_{z}=-1)\hat{P}(\sigma_{\xi}=1)|\psi\rangle_{\mbox{\scriptsize AC}}
&=&
(1/4)(1-i)e^{i\alpha}\sin(\beta/2)\cos(\beta/2), \nonumber \\
{}_{\mbox{\scriptsize AC}}\langle r_{1}|\hat{P}(\sigma_{z}=-1)\hat{P}(\sigma_{\xi}=-1)|\psi\rangle_{\mbox{\scriptsize AC}}
&=&
-(1/4)(1-i)e^{i\alpha}\sin(\beta/2)\cos(\beta/2),
\label{probability-set-2}
\end{eqnarray}
\begin{eqnarray}
{}_{\mbox{\scriptsize AC}}\langle r_{2}|\hat{P}(\sigma_{x}=1)\hat{P}(\sigma_{\xi}=1)|\psi\rangle_{\mbox{\scriptsize AC}}
&=&
(1/8)(1+i)(\cos\beta+i\sin\alpha\sin\beta), \nonumber \\
{}_{\mbox{\scriptsize AC}}\langle r_{2}|\hat{P}(\sigma_{x}=1)\hat{P}(\sigma_{\xi}=-1)|\psi\rangle_{\mbox{\scriptsize AC}}
&=&
-(1/8)(1+i)(\cos\beta+i\sin\alpha\sin\beta), \nonumber \\
{}_{\mbox{\scriptsize AC}}\langle r_{2}|\hat{P}(\sigma_{x}=-1)\hat{P}(\sigma_{\xi}=1)|\psi\rangle_{\mbox{\scriptsize AC}}
&=&
(1/8)[2+(1-i)\cos\beta-2\cos\alpha\sin\beta \nonumber \\
&&
\quad
-(1+i)\sin\alpha\sin\beta], \nonumber \\
{}_{\mbox{\scriptsize AC}}\langle r_{2}|\hat{P}(\sigma_{x}=-1)\hat{P}(\sigma_{\xi}=-1)|\psi\rangle_{\mbox{\scriptsize AC}}
&=&
(1/8)[2-(1-i)\cos\beta+2\cos\alpha\sin\beta \nonumber \\
&&
\quad
+(1+i)\sin\alpha\sin\beta],
\label{probability-set-3}
\end{eqnarray}
\begin{eqnarray}
{}_{\mbox{\scriptsize AC}}\langle r_{2}|\hat{P}(\sigma_{z}=1)\hat{P}(\sigma_{\xi}=1)|\psi\rangle_{\mbox{\scriptsize AC}}
&=&
(1/2)\cos(\beta/2)
[\cos(\beta/2) \nonumber \\
&&
\quad
-(1/2)(1+i)e^{-i\alpha}\sin(\beta/2)], \nonumber \\
{}_{\mbox{\scriptsize AC}}\langle r_{2}|\hat{P}(\sigma_{z}=1)\hat{P}(\sigma_{\xi}=-1)|\psi\rangle_{\mbox{\scriptsize AC}}
&=&
(1/2)\sin(\beta/2)
[\sin(\beta/2) \nonumber \\
&&
\quad
+(1/2)(1+i)e^{-i\alpha}\cos(\beta/2)], \nonumber \\
{}_{\mbox{\scriptsize AC}}\langle r_{2}|\hat{P}(\sigma_{z}=-1)\hat{P}(\sigma_{\xi}=1)|\psi\rangle_{\mbox{\scriptsize AC}}
&=&
-(1/4)(1-i)e^{i\alpha}\sin(\beta/2)\cos(\beta/2), \nonumber \\
{}_{\mbox{\scriptsize AC}}\langle r_{2}|\hat{P}(\sigma_{z}=-1)\hat{P}(\sigma_{\xi}=-1)|\psi\rangle_{\mbox{\scriptsize AC}}
&=&
(1/4)(1-i)e^{i\alpha}\sin(\beta/2)\cos(\beta/2),
\label{probability-set-4}
\end{eqnarray}
\begin{eqnarray}
{}_{\mbox{\scriptsize AC}}\langle r_{3}|\hat{P}(\sigma_{x}=1)\hat{P}(\sigma_{\xi}=1)|\psi\rangle_{\mbox{\scriptsize AC}}
&=&
(1/8)[2-(1-i)\cos\beta+2\cos\alpha\sin\beta \nonumber \\
&&
\quad
-(1+i)\sin\alpha\sin\beta], \nonumber \\
{}_{\mbox{\scriptsize AC}}\langle r_{3}|\hat{P}(\sigma_{x}=1)\hat{P}(\sigma_{\xi}=-1)|\psi\rangle_{\mbox{\scriptsize AC}}
&=&
(1/8)[2+(1-i)\cos\beta-2\cos\alpha\sin\beta \nonumber \\
&&
\quad
+(1+i)\sin\alpha\sin\beta], \nonumber \\
{}_{\mbox{\scriptsize AC}}\langle r_{3}|\hat{P}(\sigma_{x}=-1)\hat{P}(\sigma_{\xi}=1)|\psi\rangle_{\mbox{\scriptsize AC}}
&=&
-(1/8)(1-i)(i\cos\beta+\sin\alpha\sin\beta), \nonumber \\
{}_{\mbox{\scriptsize AC}}\langle r_{3}|\hat{P}(\sigma_{x}=-1)\hat{P}(\sigma_{\xi}=-1)|\psi\rangle_{\mbox{\scriptsize AC}}
&=&
(1/8)[(1+i)\cos\beta+(1-i)\sin\alpha\sin\beta], \nonumber \\
\label{probability-set-5}
\end{eqnarray}
\begin{eqnarray}
{}_{\mbox{\scriptsize AC}}\langle r_{3}|\hat{P}(\sigma_{z}=1)\hat{P}(\sigma_{\xi}=1)|\psi\rangle_{\mbox{\scriptsize AC}}
&=&
(1/4)(1-i)e^{-i\alpha}\sin(\beta/2)\cos(\beta/2), \nonumber \\
{}_{\mbox{\scriptsize AC}}\langle r_{3}|\hat{P}(\sigma_{z}=1)\hat{P}(\sigma_{\xi}=-1)|\psi\rangle_{\mbox{\scriptsize AC}}
&=&
-(1/4)(1-i)e^{-i\alpha}\sin(\beta/2)\cos(\beta/2), \nonumber \\
{}_{\mbox{\scriptsize AC}}\langle r_{3}|\hat{P}(\sigma_{z}=-1)\hat{P}(\sigma_{\xi}=1)|\psi\rangle_{\mbox{\scriptsize AC}}
&=&
(1/2)\sin(\beta/2)[\sin(\beta/2) \nonumber \\
&&
\quad
+(1/2)(1+i)e^{i\alpha}\cos(\beta/2)], \nonumber \\
{}_{\mbox{\scriptsize AC}}\langle r_{3}|\hat{P}(\sigma_{z}=-1)\hat{P}(\sigma_{\xi}=-1)|\psi\rangle_{\mbox{\scriptsize AC}}
&=&
(1/2)\cos(\beta/2)[\cos(\beta/2) \nonumber \\
&&
\quad
-(1/2)(1+i)e^{i\alpha}\sin(\beta/2)],
\label{probability-set-6}
\end{eqnarray}
\begin{eqnarray}
{}_{\mbox{\scriptsize AC}}\langle r_{4}|\hat{P}(\sigma_{x}=1)\hat{P}(\sigma_{\xi}=1)|\psi\rangle_{\mbox{\scriptsize AC}}
&=&
-(1/8)(1+i)(\cos\beta+i\sin\alpha\sin\beta), \nonumber \\
{}_{\mbox{\scriptsize AC}}\langle r_{4}|\hat{P}(\sigma_{x}=1)\hat{P}(\sigma_{\xi}=-1)|\psi\rangle_{\mbox{\scriptsize AC}}
&=&
(1/8)(1+i)(\cos\beta+i\sin\alpha\sin\beta), \nonumber \\
{}_{\mbox{\scriptsize AC}}\langle r_{4}|\hat{P}(\sigma_{x}=-1)\hat{P}(\sigma_{\xi}=1)|\psi\rangle_{\mbox{\scriptsize AC}}
&=&
(1/8)[2-(1-i)\cos\beta-2\cos\alpha\sin\beta \nonumber \\
&&
\quad
+(1+i)\sin\alpha\sin\beta], \nonumber \\
{}_{\mbox{\scriptsize AC}}\langle r_{4}|\hat{P}(\sigma_{x}=-1)\hat{P}(\sigma_{\xi}=-1)|\psi\rangle_{\mbox{\scriptsize AC}}
&=&
(1/8)[2+(1-i)\cos\beta+2\cos\alpha\sin\beta \nonumber \\
&&
\quad
-(1+i)\sin\alpha\sin\beta],
\label{probability-set-7}
\end{eqnarray}
\begin{eqnarray}
{}_{\mbox{\scriptsize AC}}\langle r_{4}|\hat{P}(\sigma_{z}=1)\hat{P}(\sigma_{\xi}=1)|\psi\rangle_{\mbox{\scriptsize AC}}
&=&
-(1/4)(1-i)e^{-i\alpha}\sin(\beta/2)\cos(\beta/2), \nonumber \\
{}_{\mbox{\scriptsize AC}}\langle r_{4}|\hat{P}(\sigma_{z}=1)\hat{P}(\sigma_{\xi}=-1)|\psi\rangle_{\mbox{\scriptsize AC}}
&=&
(1/4)(1-i)e^{-i\alpha}\sin(\beta/2)\cos(\beta/2), \nonumber \\
{}_{\mbox{\scriptsize AC}}\langle r_{4}|\hat{P}(\sigma_{z}=-1)\hat{P}(\sigma_{\xi}=1)|\psi\rangle_{\mbox{\scriptsize AC}}
&=&
(1/2)\sin(\beta/2)[\sin(\beta/2) \nonumber \\
&&
\quad
-(1/2)(1+i)e^{i\alpha}\cos(\beta/2)], \nonumber \\
{}_{\mbox{\scriptsize AC}}\langle r_{4}|\hat{P}(\sigma_{z}=-1)\hat{P}(\sigma_{\xi}=-1)|\psi\rangle_{\mbox{\scriptsize AC}}
&=&
(1/2)\cos(\beta/2)[\cos(\beta/2) \nonumber \\
&&
\quad
+(1/2)(1+i)e^{i\alpha}\sin(\beta/2)].
\label{probability-set-8}
\end{eqnarray}

To evaluate the probabilities given in Eq.~(\ref{probabilities-fg}),
we prepare the following eight functions:
\begin{eqnarray}
f_{1}(\alpha,\beta)
&=&
\sum_{j\in\{1,-1\}}|_{\mbox{\scriptsize AC}}\langle r_{2}|\hat{P}(\sigma_{x}=1)\hat{P}(\sigma_{\xi}=j)|\psi\rangle_{\mbox{\scriptsize AC}}|^{2} \nonumber \\
&=&
(1/16)(\cos^{2}\beta+\sin^{2}\alpha\sin^{2}\beta), \nonumber \\
g_{1}(\alpha,\beta)
&=&
\sum_{j\in\{1,-1\}}|_{\mbox{\scriptsize AC}}\langle r_{2}|\hat{P}(\sigma_{x}=-1)\hat{P}(\sigma_{\xi}=j)|\psi\rangle_{\mbox{\scriptsize AC}}|^{2} \nonumber \\
&=&
(1/32)[4+2\cos^{2}\beta-4\cos\alpha\cos\beta\sin\beta \nonumber \\
&&
\quad
+(3+\cos(2\alpha)+2\sin(2\alpha))\sin^{2}\beta],
\label{f1g1-alpha-beta-01}
\end{eqnarray}
\begin{eqnarray}
f_{2}(\alpha,\beta)
&=&
\sum_{j\in\{1,-1\}}|_{\mbox{\scriptsize AC}}\langle r_{2}|\hat{P}(\sigma_{z}=1)\hat{P}(\sigma_{\xi}=j)|\psi\rangle_{\mbox{\scriptsize AC}}|^{2} \nonumber \\
&=&
(1/32)[7+\cos(2\beta)-2(\cos\alpha+\sin\alpha)\sin(2\beta)], \nonumber \\
g_{2}(\alpha,\beta)
&=&
\sum_{j\in\{1,-1\}}|_{\mbox{\scriptsize AC}}\langle r_{2}|\hat{P}(\sigma_{z}=-1)\hat{P}(\sigma_{\xi}=j)|\psi\rangle_{\mbox{\scriptsize AC}}|^{2} \nonumber \\
&=&
(1/16)\sin^{2}\beta,
\label{f2g2-alpha-beta-02}
\end{eqnarray}
\begin{eqnarray}
f_{3}(\alpha,\beta)
&=&
\sum_{j\in\{1,-1\}}|_{\mbox{\scriptsize AC}}\langle r_{3}|\hat{P}(\sigma_{x}=1)\hat{P}(\sigma_{\xi}=j)|\psi\rangle_{\mbox{\scriptsize AC}}|^{2} \nonumber \\
&=&
(1/32)[4+2\cos^{2}\beta-4\cos\alpha\cos\beta\sin\beta+(3+\cos(2\alpha)-2\sin(2\alpha))\sin^{2}\beta], \nonumber \\
g_{3}(\alpha,\beta)
&=&
\sum_{j\in\{1,-1\}}|_{\mbox{\scriptsize AC}}\langle r_{3}|\hat{P}(\sigma_{x}=-1)\hat{P}(\sigma_{\xi}=j)|\psi\rangle_{\mbox{\scriptsize AC}}|^{2} \nonumber \\
&=&
(1/16)(\cos^{2}\beta+\sin^{2}\alpha\sin^{2}\beta),
\label{f3g3-alpha-beta-03}
\end{eqnarray}
\begin{eqnarray}
f_{4}(\alpha,\beta)
&=&
\sum_{j\in\{1,-1\}}|_{\mbox{\scriptsize AC}}\langle r_{3}|\hat{P}(\sigma_{z}=1)\hat{P}(\sigma_{\xi}=j)|\psi\rangle_{\mbox{\scriptsize AC}}|^{2} \nonumber \\
&=&
(1/16)\sin^{2}\beta, \nonumber \\
g_{4}(\alpha,\beta)
&=&
\sum_{j\in\{1,-1\}}|_{\mbox{\scriptsize AC}}\langle r_{3}|\hat{P}(\sigma_{z}=-1)\hat{P}(\sigma_{\xi}=j)|\psi\rangle_{\mbox{\scriptsize AC}}|^{2} \nonumber \\
&=&
(1/32)[7+\cos(2\beta)+2(-\cos\alpha+\sin\alpha)\sin(2\beta)],
\label{f4g4-alpha-beta-04}
\end{eqnarray}
which are calculated using Eqs.~(\ref{probability-set-1}),
(\ref{probability-set-2}),
(\ref{probability-set-3}),
(\ref{probability-set-4}),
(\ref{probability-set-5}),
(\ref{probability-set-6}),
(\ref{probability-set-7}),
and
(\ref{probability-set-8}).

To evaluate the probabilities given in Eq.~(\ref{probabilities-uv}),
we prepare the following eight functions:
\begin{eqnarray}
u_{1}(\alpha,\beta)
&=&
\sum_{i\in\{1,-1\}}|_{\mbox{\scriptsize AC}}\langle r_{1}|\hat{P}(\sigma_{x}=i)\hat{P}(\sigma_{\xi}=1)|\psi\rangle_{\mbox{\scriptsize AC}}|^{2} \nonumber \\
&=&
(1/32)[2\cos\beta+2(1+\cos\alpha\sin\beta)(2+\sin\alpha\sin\beta)+\cos\alpha\sin(2\beta)], \nonumber \\
v_{1}(\alpha,\beta)
&=&
\sum_{i\in\{1,-1\}}|_{\mbox{\scriptsize AC}}\langle r_{1}|\hat{P}(\sigma_{x}=i)\hat{P}(\sigma_{\xi}=-1)|\psi\rangle_{\mbox{\scriptsize AC}}|^{2} \nonumber \\
&=&
(1/32)[-2\cos\beta+2(1-\cos\alpha\sin\beta)(2-\sin\alpha\sin\beta)+\cos\alpha\sin(2\beta)], \nonumber \\
\label{u1v1-alpha-beta-01}
\end{eqnarray}
\begin{eqnarray}
u_{2}(\alpha,\beta)
&=&
\sum_{i\in\{1,-1\}}|_{\mbox{\scriptsize AC}}\langle r_{1}|\hat{P}(\sigma_{z}=i)\hat{P}(\sigma_{\xi}=1)|\psi\rangle_{\mbox{\scriptsize AC}}|^{2} \nonumber \\
&=&
(1/8)\cos^{2}(\beta/2)[2+(\cos\alpha+\sin\alpha)\sin\beta], \nonumber \\
v_{2}(\alpha,\beta)
&=&
\sum_{i\in\{1,-1\}}|_{\mbox{\scriptsize AC}}\langle r_{1}|\hat{P}(\sigma_{z}=i)\hat{P}(\sigma_{\xi}=-1)|\psi\rangle_{\mbox{\scriptsize AC}}|^{2} \nonumber \\
&=&
-(1/8)\sin^{2}(\beta/2)[-2+(\cos\alpha+\sin\alpha)\sin\beta],
\label{u2v2-alpha-beta-02}
\end{eqnarray}
\begin{eqnarray}
u_{3}(\alpha,\beta)
&=&
\sum_{i\in\{1,-1\}}|_{\mbox{\scriptsize AC}}\langle r_{4}|\hat{P}(\sigma_{x}=i)\hat{P}(\sigma_{\xi}=1)|\psi\rangle_{\mbox{\scriptsize AC}}|^{2} \nonumber \\
&=&
(1/32)[-2\cos\beta-2(-1+\cos\alpha\sin\beta)(2+\sin\alpha\sin\beta)+\cos\alpha\sin(2\beta)], \nonumber \\
v_{3}(\alpha,\beta)
&=&
\sum_{i\in\{1,-1\}}|_{\mbox{\scriptsize AC}}\langle r_{4}|\hat{P}(\sigma_{x}=i)\hat{P}(\sigma_{\xi}=-1)|\psi\rangle_{\mbox{\scriptsize AC}}|^{2} \nonumber \\
&=&
(1/32)[2\cos\beta-2(1+\cos\alpha\sin\beta)(-2+\sin\alpha\sin\beta)+\cos\alpha\sin(2\beta)], \nonumber \\
\label{u3v3-alpha-beta-03}
\end{eqnarray}
\begin{eqnarray}
u_{4}(\alpha,\beta)
&=&
\sum_{i\in\{1,-1\}}|_{\mbox{\scriptsize AC}}\langle r_{4}|\hat{P}(\sigma_{z}=i)\hat{P}(\sigma_{\xi}=1)|\psi\rangle_{\mbox{\scriptsize AC}}|^{2} \nonumber \\
&=&
(1/8)\sin^{2}(\beta/2)[2+(-\cos\alpha+\sin\alpha)\sin\beta], \nonumber \\
v_{4}(\alpha,\beta)
&=&
\sum_{i\in\{1,-1\}}|_{\mbox{\scriptsize AC}}\langle r_{4}|\hat{P}(\sigma_{z}=i)\hat{P}(\sigma_{\xi}=-1)|\psi\rangle_{\mbox{\scriptsize AC}}|^{2} \nonumber \\
&=&
(1/8)\cos^{2}(\beta/2)[2+(\cos\alpha-\sin\alpha)\sin\beta],
\label{u4v4-alpha-beta-04}
\end{eqnarray}
which are calculated using Eqs.~(\ref{probability-set-1}),
(\ref{probability-set-2}),
(\ref{probability-set-3}),
(\ref{probability-set-4}),
(\ref{probability-set-5}),
(\ref{probability-set-6}),
(\ref{probability-set-7}),
and
(\ref{probability-set-8}).

\section{\label{section-appendix-B}Some useful functions for Sec.~\ref{section-two-way-intercept-resend-attack}}
To evaluate the probabilities given in Eq.~(\ref{prob-fg-two-way-intercept-resend}),
we prepare the following eight functions:
\begin{eqnarray}
&&
f_{1}(\alpha,\beta,\gamma,\delta) \nonumber \\
&=&
\sum_{j,k\in\{1,-1\}}
|_{\mbox{\scriptsize AC}}\langle r_{2}|\hat{P}(\sigma_{\mu}=k)\hat{P}(\sigma_{x}=1)\hat{P}(\sigma_{\xi}=j)|\psi\rangle_{\mbox{\scriptsize AC}}|^{2}
\nonumber \\
&=&
\frac{1}{512} \Biggl{(}4 \cos (2 (\alpha - \beta) )+4 \cos (2 (\alpha +\beta ))-\cos (\alpha +2 (\beta -
   \gamma) )-\cos (\alpha -2 (\beta - \gamma) ) \nonumber \\
&&
+\cos (\alpha -2 (\beta +\gamma ))+\cos (\alpha +2 (\beta
   +\gamma ))+8 \cos (2 \delta )+12 \cos \alpha  \sin (2 \beta )-6 \sin (2 \gamma ) \nonumber \\
&&
+4 \cos (2 \gamma
   ) \Bigl{[}\sin (2 \alpha ) \sin ^2\beta -4 \cos \alpha \sin
   \beta \cos \beta  \sin ^2\delta  
 +2 \cos (2 \delta )-2\Bigr{]} \nonumber \\
&&
+12 \cos \gamma  \sin (2 \delta ) \nonumber \\
&&
+4 \cos (2 \beta )
   \Bigl{[}-\sin (2 \gamma ) \cos ^2\alpha +2 \cos \alpha \cos \gamma  \cos (2 \delta ) \sin (\alpha +\gamma )
    +\cos \gamma  \sin (2 \delta )+2\Bigr{]} \nonumber \\
&&
+2 \cos (2 \alpha ) \Bigl{[}-4 \cos \gamma 
   \sin (2 \delta ) \sin ^2\beta +\sin (2 \gamma )-4\Bigr{]} \nonumber \\
&&
-2 \Bigl{\{}2 \sin (2 \alpha )\sin ^2\beta \Bigl{[}2 \cos
   \gamma  \sin (2 \delta )+3\Bigr{]}   \nonumber \\
&&
+\cos (2 \delta ) \Bigl{[}\sin (2 \alpha )-3 \sin (2 \gamma
   )
-2 \cos \alpha  \sin (2 \beta ) (\sin (2 \gamma )+1)+\sin (2 (\alpha +\gamma
   ))\Bigr{]}\Bigr{\}} \nonumber \\
&&
+48\Biggr{)}, \nonumber \\
&&
g_{1}(\alpha,\beta,\gamma,\delta) \nonumber \\
&=&
\sum_{j,k\in\{1,-1\}}
|_{\mbox{\scriptsize AC}}\langle r_{2}|\hat{P}(\sigma_{\mu}=k)\hat{P}(\sigma_{x}=-1)\hat{P}(\sigma_{\xi}=j)|\psi\rangle_{\mbox{\scriptsize AC}}|^{2}
\nonumber \\
&=&
\frac{1}{512} \Biggl{(}-4 \cos (2 (\alpha - \beta) )-4 \cos (2 (\alpha +\beta ))-\cos (\alpha +2 (\beta -
   \gamma) )-\cos (\alpha -2( \beta - \gamma) ) \nonumber \\
&&
+\cos (\alpha -2 (\beta +\gamma ))+\cos (\alpha +2 (\beta
   +\gamma ))-8 \cos (2 \delta ) \nonumber \\
&&
+10 \Bigl{[}-2 \cos \alpha  \sin (2 \beta )+\sin (2 \gamma )-2 \cos \gamma
    \sin (2 \delta )+8\Bigl{]} \nonumber \\
&&
+4 \cos (2 \beta ) \Bigl{[}-\sin (2 \gamma ) \cos ^2\alpha +2 \cos \alpha\cos \gamma 
   \cos (2 \delta ) \sin (\alpha +\gamma )  +\cos \gamma  \sin (2 \delta )-2\Bigr{]} \nonumber \\
&&
+2
   \cos (2 \alpha ) \Bigl{[}-4 \cos \gamma  \sin (2 \delta ) \sin ^2\beta +\sin (2 \gamma
   )+4\Bigr{]} \nonumber \\
&&
-2 \Bigl{\{}2 \sin (2 \alpha ) \sin ^2\beta \Bigl{[}2 \cos \gamma  \sin (2 \delta )-5\Bigr{]}  \nonumber \\
&&
+\cos (2
   \delta ) \Bigr{[}\sin (2 \alpha )+5 \sin (2 \gamma )-2 \cos \alpha  \sin (2 \beta ) (\sin (2 \gamma
   )+1)+\sin (2 (\alpha +\gamma ))\Bigr{]} \nonumber \\
&&
+\cos (2 \gamma ) \Bigl{[}-2 \sin (2 \alpha ) \sin ^2\beta +4 \cos
   \alpha  \sin (2 \beta ) \sin ^2\delta +4 \cos (2 \delta )-4\Bigr{]}\Bigr{\}}\Biggl{)},
\label{appendix-B-f1-g1}
\end{eqnarray}
\begin{eqnarray}
&&
f_{2}(\alpha,\beta,\gamma,\delta) \nonumber \\
&=&
\sum_{j,k\in\{1,-1\}}
|_{\mbox{\scriptsize AC}}\langle r_{2}|\hat{P}(\sigma_{\mu}=k)\hat{P}(\sigma_{z}=1)\hat{P}(\sigma_{\xi}=j)|\psi\rangle_{\mbox{\scriptsize AC}}|^{2}
\nonumber \\
&=&
\frac{1}{128} \Biggl{(}4 \Bigl{[}1-\cos \delta  (\cos \gamma +\sin \gamma ) \sin \delta \Bigr{]} \cos ^2\beta
    \nonumber \\
&&
+2 \Bigl{\{}\cos (2 \beta )+2 \cos (2 \delta )-\Bigl{[}\cos (2 \delta )+3\Bigr{]} (\cos \alpha +\sin \alpha ) \sin
   (2 \beta )+11\Bigr{\}} \nonumber \\
&&
-\Bigl{\{}\Bigl{[}\cos (2 \beta )+5\Bigr{]} (\cos \gamma +\sin \gamma )-2 \sin (2 \beta ) \sin (\alpha
   +\gamma )\Bigr{\}} \sin (2 \delta )\Biggr{)}, \nonumber \\
&&
g_{2}(\alpha,\beta,\gamma,\delta) \nonumber \\
&=&
\sum_{j,k\in\{1,-1\}}
|_{\mbox{\scriptsize AC}}\langle r_{2}|\hat{P}(\sigma_{\mu}=k)\hat{P}(\sigma_{z}=-1)\hat{P}(\sigma_{\xi}=j)|\psi\rangle_{\mbox{\scriptsize AC}}|^{2}
\nonumber \\
&=&
\frac{1}{128} \Bigl{\{}-4 \Bigl{[}\cos \delta \sin \delta (\cos \gamma +\sin \gamma )  +1\Bigr{]} \cos ^2\beta \nonumber \\
&&
   +8 \sin \beta \cos \beta \sin \delta  \Bigl{[}\cos \delta  \sin (\alpha +\gamma )+(\cos \alpha +\sin
   \alpha ) \sin \delta \Bigr{]}  \nonumber \\
&&
-2 \Bigl{[}\cos (2 \beta )+2 \cos (2 \delta )-5\Bigl{]}-\Bigl{[}\cos (2 \beta
   )-3\Bigr{]} (\cos \gamma +\sin \gamma ) \sin (2 \delta )\Bigr{\}},
\label{appendix-B-f2-g2}
\end{eqnarray}
\begin{eqnarray}
&&
f_{3}(\alpha,\beta,\gamma,\delta) \nonumber \\
&=&
\sum_{j,k\in\{1,-1\}}
|_{\mbox{\scriptsize AC}}\langle r_{3}|\hat{P}(\sigma_{\mu}=k)\hat{P}(\sigma_{x}=1)\hat{P}(\sigma_{\xi}=j)|\psi\rangle_{\mbox{\scriptsize AC}}|^{2}
\nonumber \\
&=&
\frac{1}{512} \Biggl{[}-4 \cos (2 (\alpha - \beta) )-4 \cos (2 (\alpha +\beta ))+\cos (\alpha +2 (\beta -
   \gamma) )+\cos (\alpha -2 (\beta - \gamma) ) \nonumber \\
&&
-\cos (\alpha -2 (\beta +\gamma ))-\cos (\alpha +2 (\beta
   +\gamma )) \nonumber \\
&&
+4 \cos (2 \beta ) \Bigl{[}\sin (2 \gamma ) \cos ^2\alpha -2 \cos \alpha \cos \gamma  \cos (2
   \delta ) \sin (\alpha +\gamma )  +\cos \gamma  \sin (2 \delta )-2\Bigr{]} \nonumber \\
&&
-2 \cos (2
   \alpha ) \Bigl{[}4 \cos \gamma  \sin (2 \delta ) \sin ^2\beta +\sin (2 \gamma )-4\Bigr{]} \nonumber \\
&&
+2
   \Biggl{(}\cos (2 \delta ) \Bigl{[}\sin (2 \alpha )-2 \cos \alpha  \sin (2 \beta ) (\sin (2 \gamma )-1)+5
   \sin (2 \gamma )+\sin (2 (\alpha +\gamma ))-4\Bigr{]} \nonumber \\
&&
-5 \Bigl{[}2 \cos \alpha  \sin (2 \beta )+\sin (2 \gamma
   )+2 \cos \gamma  \sin (2 \delta )-8\Bigr{]} \nonumber \\
&&
-2 \Bigl{\{}\sin (2 \alpha ) \sin ^2\beta \Bigl{[}\cos (2 \gamma )-2 \cos \gamma 
   \sin (2 \delta )+5\Bigr{]}   \nonumber \\
&&
+2 \cos (2 \gamma ) \sin
   ^2\delta \Bigl{[}\cos \alpha  \sin (2 \beta )-2\Bigr{]}  \Bigr{\}}\Biggr{)}\Biggr{]}, \nonumber \\
&&
g_{3}(\alpha,\beta,\gamma,\delta) \nonumber \\
&=&
\sum_{j,k\in\{1,-1\}}
|_{\mbox{\scriptsize AC}}\langle r_{3}|\hat{P}(\sigma_{\mu}=k)\hat{P}(\sigma_{x}=-1)\hat{P}(\sigma_{\xi}=j)|\psi\rangle_{\mbox{\scriptsize AC}}|^{2}
\nonumber \\
&=&
\frac{1}{512} \Biggl{[}4 \cos (2 (\alpha - \beta) )+4 \cos (2 (\alpha +\beta ))+\cos (\alpha +2 (\beta -
   \gamma) )+\cos (\alpha -2 (\beta - \gamma) ) \nonumber \\
&&
-\cos (\alpha -2 (\beta +\gamma ))-\cos (\alpha +2 (\beta
   +\gamma ))+8 \cos (2 \delta )+12 \cos \alpha  \sin (2 \beta )+6 \sin (2 \gamma ) \nonumber \\
&&
+12 \cos \gamma
    \sin (2 \delta ) \nonumber \\
&&
+4 \cos (2 \beta ) \Bigl{[}\sin (2 \gamma ) \cos ^2\alpha -2 \cos \alpha \cos \gamma  \cos
   (2 \delta ) \sin (\alpha +\gamma )  +\cos \gamma  \sin (2 \delta )+2\Bigr{]} \nonumber \\
&&
-2 \cos
   (2 \alpha ) \Bigl{[}4 \cos \gamma  \sin (2 \delta ) \sin ^2\beta +\sin (2 \gamma )+4\Bigr{]} \nonumber \\
&&
+2
   \Biggl{(}2 \sin (2 \alpha )  \sin ^2\beta  \Bigl{[}2 \cos \gamma  \sin (2 \delta )+3\Bigr{]} \nonumber \\
&&
+\cos (2 \delta )
   \Bigl{[}\sin (2 \alpha )-2 \cos \alpha  \sin (2 \beta ) (\sin (2 \gamma )-1)-3 \sin (2 \gamma )+\sin (2
   (\alpha +\gamma ))\Bigr{]} \nonumber \\
&&
+\cos (2 \gamma ) \Bigl{\{}4 \cos (2 \delta )-2 \Bigl{[}\sin (2 \alpha ) \sin
   ^2\beta +4 \cos \alpha  \cos \beta \sin \beta \sin ^2\delta  
   +2\Bigr{]}\Bigr{\}}\Biggr{)} \nonumber \\
&&
+48\Biggr{]},
\label{appendix-B-f3-g3}
\end{eqnarray}
\begin{eqnarray}
&&
f_{4}(\alpha,\beta,\gamma,\delta) \nonumber \\
&=&
\sum_{j,k\in\{1,-1\}}
|_{\mbox{\scriptsize AC}}\langle r_{3}|\hat{P}(\sigma_{\mu}=k)\hat{P}(\sigma_{z}=1)\hat{P}(\sigma_{\xi}=j)|\psi\rangle_{\mbox{\scriptsize AC}}|^{2}
\nonumber \\
&=&
\frac{1}{128} \Bigl{\{}-4 \Bigl{[}\cos \delta  \sin \delta (\cos \gamma -\sin \gamma ) +1\Bigr{]} \cos ^2\beta
    \nonumber \\
&&
-8 \sin \beta \cos \beta \sin \delta  \Bigl{[}\cos \delta  \sin (\alpha +\gamma )+(\sin \alpha -\cos
   \alpha ) \sin \delta \Bigr{]}  \nonumber \\
&&
 -2 \Bigl{[}\cos (2 \beta )+2 \cos (2 \delta )-5\Bigr{]}-\Bigl{[}\cos (2 \beta
   )-3\Bigr{]} (\cos \gamma -\sin \gamma ) \sin (2 \delta )\Bigr{\}}, \nonumber \\
&&
g_{4}(\alpha,\beta,\gamma,\delta) \nonumber \\
&=&
\sum_{j,k\in\{1,-1\}}
|_{\mbox{\scriptsize AC}}\langle r_{3}|\hat{P}(\sigma_{\mu}=k)\hat{P}(\sigma_{z}=-1)\hat{P}(\sigma_{\xi}=j)|\psi\rangle_{\mbox{\scriptsize AC}}|^{2}
\nonumber \\
&=&
\frac{1}{128} \Biggl{(}4 \Bigl{[}\cos \delta  (\sin \gamma -\cos \gamma ) \sin \delta +1\Bigr{]} \cos ^2\beta
    \nonumber \\
&&
+2 \Bigl{\{}\cos (2 \beta )+2 \cos (2 \delta )-\Bigl{[}\cos (2 \delta )+3\Bigr{]} (\cos \alpha -\sin \alpha ) \sin
   (2 \beta )+11\Bigr{\}} \nonumber \\
&&
-\Bigl{\{}\Bigl{[}\cos (2 \beta )+5\Bigr{]} (\cos \gamma -\sin \gamma )+2 \sin (2 \beta ) \sin (\alpha
   +\gamma )\Bigr{\}} \sin (2 \delta )\Biggr{)}.
\label{appendix-B-f4-g4}
\end{eqnarray}

To estimate the probabilities given in Eq.~(\ref{prob-uv-two-way-intercept-resend}),
we prepare the following eight functions:
\begin{eqnarray}
&&
u_{1}(\alpha,\beta,\gamma,\delta) \nonumber \\
&=&
\sum_{j,k\in\{1,-1\}}
|_{\mbox{\scriptsize AC}}\langle r_{1}|\hat{P}(\sigma_{\mu}=k)\hat{P}(\sigma_{x}=j)\hat{P}(\sigma_{\xi}=k)|\psi\rangle_{\mbox{\scriptsize AC}}|^{2}
\nonumber \\
&=&
\frac{1}{256} \Biggl{[}4 \cos ^2\alpha \sin ^2\beta  \cos \gamma  \Bigl{[}4 \sin \gamma  \sin ^2\delta +\sin (2
   \delta )\Bigr{]}  \nonumber \\
&&
-\cos \alpha  \Biggl{(}2 \sin (2 \beta )\cos ^2\delta\Bigl{[}\cos (2 \gamma )+\sin (2
   \gamma )+1\Bigr{]}  \nonumber \\
&&
+8 \cos \delta\Bigl{[} \sin \alpha \cos ^2\beta  \cos \gamma  \sin \delta -2
   \sin \beta \Bigr{]}   \nonumber \\
&&
+ \sin (2 \beta ) \Bigl{[}\cos (2 \gamma )+\sin (2
   \gamma )+1\Bigr{]}\Bigl{[}\cos (2 \delta )-3\Bigr{]} \nonumber \\
&&
-16 \sin \beta  \sin \delta  \Bigl{\{}\sin \gamma +\cos \gamma  \Bigl{[}  \sin
   \alpha  \sin \beta \cos \gamma \sin \delta +2\Bigr{]}\Bigr{\}}\Biggr{)} \nonumber \\
&&
+2 \Biggl{(}\sin ^2\alpha\cos ^2\beta  \cos \gamma  \sin (2
   \delta )  +8 \cos \beta  (\cos \gamma +\sin \gamma ) \sin \delta \nonumber \\
&&
+\cos
   \delta  \Bigl{\{}-2 \cos \gamma  \sin \delta  \cos ^2\beta +8 \sin \alpha  \sin \beta
   +\Bigl{[}\cos (2 \beta )-3\Bigr{]} \cos \gamma \sin \delta \sin ^2\alpha   \Bigr{\}} \nonumber \\
&&
+\cos \gamma  \Bigl{\{}8
   \sin \alpha  \sin \beta  \sin \delta +\Bigl{[}\sin (2 \alpha )+1\Bigr{]} (\sin ^2\beta +1)
   \sin (2 \delta )\Bigr{\}}+16\Biggr{)}\Biggr{]}, \nonumber \\
&&
v_{1}(\alpha,\beta,\gamma,\delta) \nonumber \\
&=&
\sum_{j,k\in\{1,-1\}}
|_{\mbox{\scriptsize AC}}\langle r_{1}|\hat{P}(\sigma_{\mu}=k)\hat{P}(\sigma_{x}=j)\hat{P}(\sigma_{\xi}=\bar{k})|\psi\rangle_{\mbox{\scriptsize AC}}|^{2}
\nonumber \\
&=&
\frac{1}{256} \Biggl{(}8 \cos ^2\alpha \sin ^2\beta \cos \gamma  \sin (2 \delta )  -16 
    \sin \alpha  \sin \beta \cos \delta +\sin (2 \gamma )+\sin (2 (\alpha +\gamma )) \nonumber \\
&&
-16 \sin \delta \Bigl{\{}\cos \beta  (\cos
   \gamma +\sin \gamma )+\sin \beta  \Bigl{[}\cos (\alpha -\gamma )+\cos (\alpha +\gamma )+\sin (\alpha
   +\gamma )\Bigr{]}\Bigr{\}}  \nonumber \\
&&
 +4 \cos \alpha  \Bigl{\{}\sin (2 \beta ) \Bigl{[}\cos (2 \gamma )+\sin (2 \gamma
   )+1\Bigr{]} \sin ^2 \delta -4 \cos \delta  \sin \beta \nonumber \\
&&
-\cos \gamma \sin (\alpha +\gamma ) \Bigl{[}2 \cos (2 \delta ) \sin
   ^2 \beta +\cos (2 \beta )\Bigr{]} \Bigr{\}} \nonumber \\
&&
+\sin (2 \alpha ) \Bigl{[}4 \cos
   \gamma  \sin (2 \delta ) \sin ^2\beta +1\Bigr{]}+32\Biggr{)},
\label{appendix-B-u1-v1}
\end{eqnarray}
\begin{eqnarray}
&&
u_{2}(\alpha,\beta,\gamma,\delta) \nonumber \\
&=&
\sum_{j,k\in\{1,-1\}}
|_{\mbox{\scriptsize AC}}\langle r_{1}|\hat{P}(\sigma_{\mu}=k)\hat{P}(\sigma_{z}=j)\hat{P}(\sigma_{\xi}=k)|\psi\rangle_{\mbox{\scriptsize AC}}|^{2}
\nonumber \\
&=&
\frac{1}{16} (\cos \beta  \cos \delta +1) \Bigl{\{}\cos \beta  (\cos \gamma +\sin \gamma ) \sin
   \delta \nonumber \\
&&
+\sin \beta  \Bigl{[}\cos \delta  (\cos \alpha +\sin \alpha )+\sin (\alpha +\gamma ) \sin
   \delta \Bigr{]}+2\Bigr{\}}, \nonumber \\
&&
v_{2}(\alpha,\beta,\gamma,\delta) \nonumber \\
&=&
\sum_{j,k\in\{1,-1\}}
|_{\mbox{\scriptsize AC}}\langle r_{1}|\hat{P}(\sigma_{\mu}=k)\hat{P}(\sigma_{z}=j)\hat{P}(\sigma_{\xi}=\bar{k})|\psi\rangle_{\mbox{\scriptsize AC}}|^{2}
\nonumber \\
&=&
\frac{1}{16} (\cos \beta  \cos \delta -1) \Bigl{\{}\cos \beta  (\cos \gamma +\sin \gamma ) \sin
   \delta  \nonumber \\
&&
+\sin \beta  \Bigl{[}\cos \delta  (\cos \alpha +\sin \alpha )+\sin (\alpha +\gamma ) \sin
   \delta \Bigr{]}-2\Bigr{\}},
\label{appendix-B-u2-v2}
\end{eqnarray}
\begin{eqnarray}
&&
u_{3}(\alpha,\beta,\gamma,\delta) \nonumber \\
&=&
\sum_{j,k\in\{1,-1\}}
|_{\mbox{\scriptsize AC}}\langle r_{4}|\hat{P}(\sigma_{\mu}=k)\hat{P}(\sigma_{x}=j)\hat{P}(\sigma_{\xi}=k)|\psi\rangle_{\mbox{\scriptsize AC}}|^{2}
\nonumber \\
&=&
\frac{1}{256} \Biggl{[}8 \cos ^2\alpha \sin ^2\beta \cos \gamma  \sin \delta  (\cos \delta -2 \sin \gamma 
   \sin \delta )  \nonumber \\
&&
-\cos \alpha  \Bigl{\{}2 \sin (2 \beta ) \Bigl{[}\cos (2 \gamma )-\sin (2
   \gamma )+1\Bigr{]} \cos ^2\delta -16 \sin \beta  \cos \delta \nonumber \\
&&
+ \sin (2 \beta )
   \Bigr{[}\cos (2 \gamma )-\sin (2 \gamma )+1\Bigr{]}\Bigl{[}\cos (2 \delta )-3\Bigr{]} \nonumber \\
&&
+16 \sin \beta  \sin \delta  \Bigl{[}\sin \gamma +\cos \gamma
    (  \sin \alpha  \sin \beta \cos \gamma \sin \delta -2)\Bigr{]} \nonumber \\
&&
-2 \sin \alpha\Bigl{[}\cos (2 \beta )-3\Bigr{]} \cos
   \gamma    \sin (2 \delta )\Bigr{\}} \nonumber \\
&&
+2 \Biggl{(}-8 \Bigl{[}  \sin \alpha  \sin
   \beta\cos \gamma +\cos \beta  (\sin \gamma -\cos \gamma )\Bigr{]} \sin \delta \nonumber \\
&&
+\cos \delta  \Bigl{\{}-2 \cos
   \gamma  \sin \delta  \cos ^2\beta -8 \sin \alpha  \sin \beta +\Bigl{[}\cos (2 \beta )-3\Bigr{]} \cos
   \gamma    \sin \delta \sin ^2\alpha \Bigr{\}} \nonumber \\
&&
+\cos \gamma  \Bigl{\{}\Bigl{[}\sin ^2\alpha +\sin
   (2 \alpha )\Bigr{]} \cos ^2\beta +\sin ^2\beta +1\Bigr{\}} \sin (2 \delta )+16\Biggr{)}\Biggr{]}, \nonumber \\
&&
v_{3}(\alpha,\beta,\gamma,\delta) \nonumber \\
&=&
\sum_{j,k\in\{1,-1\}}
|_{\mbox{\scriptsize AC}}\langle r_{4}|\hat{P}(\sigma_{\mu}=k)\hat{P}(\sigma_{x}=j)\hat{P}(\sigma_{\xi}=\bar{k})|\psi\rangle_{\mbox{\scriptsize AC}}|^{2}
\nonumber \\
&=&
\frac{1}{256} \Biggl{[}8 \cos ^2\alpha  \cos \gamma  \sin \delta  (\cos \delta -2 \sin \gamma 
   \sin \delta ) \sin ^2\beta  \nonumber \\
&&
-\cos \alpha  \Bigl{\{}2 \sin (2 \beta ) \Bigl{[}\cos (2 \gamma )-\sin (2
   \gamma )+1\Bigr{]} \cos ^2 \delta +16 \sin \beta  \cos \delta  \nonumber \\
&&
+ \sin (2 \beta )
   \Bigl{[}\cos (2 \gamma )-\sin (2 \gamma )+1\Bigr{]}\Bigl{[}\cos (2 \delta )-3\Bigr{]} \nonumber \\
&&
+16 \sin \beta  \sin \delta  \Bigl{[}\cos \gamma  (
    \sin \alpha  \sin \beta \cos \gamma \sin \delta +2)-\sin \gamma \Bigr{]} \nonumber \\
&&
-2 \Bigl{[}\cos (2 \beta )-3\Bigr{]}\sin \alpha \cos \gamma
      \sin (2 \delta )\Bigr{\}} \nonumber \\
&&
+2 \Biggl{(}-8 \cos \beta  \cos \gamma  \sin \delta +8
   (  \sin \alpha  \sin \beta\cos \gamma +\cos \beta  \sin \gamma ) \sin \delta \nonumber \\
&&
+\cos
   \delta  \Bigl{\{}-2 \cos \gamma  \sin \delta  \cos ^2\beta +8 \sin \alpha  \sin \beta
+\Bigl{[}\cos (2 \beta )-3\Bigr{]} \cos \gamma  \sin ^2\alpha  \sin \delta \Bigr{\}} \nonumber \\
&&
+\cos \gamma 
   \Bigl{\{}\Bigl{[}\sin ^2 \alpha +\sin (2 \alpha )\Bigr{]} \cos ^2\beta +\sin ^2\beta +1\Bigr{\}} \sin
   (2 \delta )+16\Biggr{)}\Biggr{]},
\label{appendix-B-u3-v3}
\end{eqnarray}
\begin{eqnarray}
&&
u_{4}(\alpha,\beta,\gamma,\delta) \nonumber \\
&=&
\sum_{j,k\in\{1,-1\}}
|_{\mbox{\scriptsize AC}}\langle r_{4}|\hat{P}(\sigma_{\mu}=k)\hat{P}(\sigma_{z}=j)\hat{P}(\sigma_{\xi}=k)|\psi\rangle_{\mbox{\scriptsize AC}}|^{2}
\nonumber \\
&=&
\frac{1}{16} (\cos \beta  \cos \delta +1) \Bigl{\{}\cos \beta  (\cos \gamma -\sin \gamma ) \sin
   \delta \nonumber \\
&&
+\sin \beta  \Bigl{[}\cos \delta  (\cos \alpha -\sin \alpha )-\sin (\alpha +\gamma ) \sin
   \delta \Bigr{]}+2\Bigr{\}}, \nonumber \\
&&
v_{4}(\alpha,\beta,\gamma,\delta) \nonumber \\
&=&
\sum_{j,k\in\{1,-1\}}
|_{\mbox{\scriptsize AC}}\langle r_{4}|\hat{P}(\sigma_{\mu}=k)\hat{P}(\sigma_{z}=j)\hat{P}(\sigma_{\xi}=\bar{k})|\psi\rangle_{\mbox{\scriptsize AC}}|^{2}
\nonumber \\
&=&
\frac{1}{16} (\cos \beta  \cos \delta -1) \Bigl{\{}\cos \beta  (\cos \gamma -\sin \gamma ) \sin
   \delta  \nonumber \\
&&
+\sin \beta  \Bigl{[}\cos \delta  (\cos \alpha -\sin \alpha )-\sin (\alpha +\gamma ) \sin
   \delta \Bigr{]}-2\Bigr{\}},
\label{appendix-B-u4-v4}
\end{eqnarray}
where $\bar{k}$ is given by Eq.~(\ref{def-bar-k}).

\section{\label{section-appendix-C}Explicit forms of $|\phi(\sigma_{t}=i,r_{j})\rangle_{\mbox{\scriptsize E}}$}
In this section, we give explicit forms of $|\phi(\sigma_{t}=i,r_{j})\rangle_{\mbox{\scriptsize E}}$
for $t\in\{x,z\}$, $i\in\{1,-1\}$, and $j\in\{1,4\}$
defined in Eq.~(\ref{Eve-probe-state-0}),
\begin{eqnarray}
|\phi(\sigma_{z}=1,r_{1})\rangle_{\mbox{\scriptsize E}}
&=&
(\sqrt{F}/2)|\alpha\rangle_{\mbox{\scriptsize E}}
+
[(1-i)/4]\sqrt{1-F}|\beta\rangle_{\mbox{\scriptsize E}}, \nonumber \\
|\phi(\sigma_{z}=-1,r_{1})\rangle_{\mbox{\scriptsize E}}
&=&
[(1+i)/4]\sqrt{1-F}|\gamma\rangle_{\mbox{\scriptsize E}}, \nonumber \\
|\phi(\sigma_{x}=1,r_{1})\rangle_{\mbox{\scriptsize E}}
&=&
(1/8)
[(3+i)\sqrt{F}|\alpha\rangle_{\mbox{\scriptsize E}}
+
(1-i)\sqrt{1-F}|\beta\rangle_{\mbox{\scriptsize E}} \nonumber \\
&&
\quad
+
(3+i)\sqrt{1-F}|\gamma\rangle_{\mbox{\scriptsize E}}
+
(1-i)\sqrt{F}|\delta\rangle_{\mbox{\scriptsize E}}], \nonumber \\
|\phi(\sigma_{x}=-1,r_{1})\rangle_{\mbox{\scriptsize E}}
&=&
(1/8)(1-i)
[\sqrt{F}|\alpha\rangle_{\mbox{\scriptsize E}}
+
\sqrt{1-F}|\beta\rangle_{\mbox{\scriptsize E}} \nonumber \\
&&
\quad
-
\sqrt{1-F}|\gamma\rangle_{\mbox{\scriptsize E}}
-
\sqrt{F}|\delta\rangle_{\mbox{\scriptsize E}}],
\end{eqnarray}
\begin{eqnarray}
|\phi(\sigma_{z}=1,r_{4})\rangle_{\mbox{\scriptsize E}}
&=&
-[(1+i)/4]\sqrt{1-F}|\beta\rangle_{\mbox{\scriptsize E}}, \nonumber \\
|\phi(\sigma_{z}=-1,r_{4})\rangle_{\mbox{\scriptsize E}}
&=&
-[(1-i)/4]\sqrt{1-F}|\gamma\rangle_{\mbox{\scriptsize E}}
+
(\sqrt{F}/2)|\delta\rangle_{\mbox{\scriptsize E}}, \nonumber \\
|\phi(\sigma_{x}=1,r_{4})\rangle_{\mbox{\scriptsize E}}
&=&
(1/8)(1-i)
[-\sqrt{F}|\alpha\rangle_{\mbox{\scriptsize E}}
+
\sqrt{1-F}|\beta\rangle_{\mbox{\scriptsize E}}
-
\sqrt{1-F}|\gamma\rangle_{\mbox{\scriptsize E}}
+
\sqrt{F}|\delta\rangle_{\mbox{\scriptsize E}}], \nonumber \\
|\phi(\sigma_{x}=-1,r_{4})\rangle_{\mbox{\scriptsize E}}
&=&
(1/8)
[(1-i)\sqrt{F}|\alpha\rangle_{\mbox{\scriptsize E}}
-
(3+i)\sqrt{1-F}|\beta\rangle_{\mbox{\scriptsize E}} \nonumber \\
&&
\quad
-
(1-i)\sqrt{1-F}|\gamma\rangle_{\mbox{\scriptsize E}}
+
(3+i)\sqrt{F}|\delta\rangle_{\mbox{\scriptsize E}}].
\end{eqnarray}

\section{\label{section-appendix-D}Explicit forms of $|\Phi(\sigma_{t}=i,r_{j})\rangle_{\mbox{\scriptsize E1E2}}$}
In this section,
we give explicit forms of $|\Phi(\sigma_{t}=i,r_{j})\rangle_{\mbox{\scriptsize E1E2}}$
for $t\in\{x,z\}$, $i\in\{1,-1\}$, and $j\in\{1,4\}$
defined in Eq.~(\ref{def-Phi-E1E2}),
\begin{eqnarray}
&&
|\Phi(\sigma_{z}=1,r_{1})\rangle_{\mbox{\scriptsize E1E2}} \nonumber \\
&=&
\frac{1}{\sqrt{2}}\sqrt{FF'}|\alpha\rangle_{\mbox{\scriptsize E1}}|\alpha'\rangle_{\mbox{\scriptsize E2}} \nonumber \\
&&
+
\frac{1}{2}e^{-i\pi/4}\sqrt{F(1-F')}|\alpha\rangle_{\mbox{\scriptsize E1}}|\beta'\rangle_{\mbox{\scriptsize E2}}
+
\frac{1}{2}e^{i\pi/4}\sqrt{(1-F)F'}|\gamma\rangle_{\mbox{\scriptsize E1}}|\alpha'\rangle_{\mbox{\scriptsize E2}}, \nonumber \\
&&
|\Phi(\sigma_{z}=1,r_{4})\rangle_{\mbox{\scriptsize E1E2}} \nonumber \\
&=&
\frac{1}{\sqrt{2}}\sqrt{(1-F)(1-F')}|\gamma\rangle_{\mbox{\scriptsize E1}}|\beta'\rangle_{\mbox{\scriptsize E2}} \nonumber \\
&&
-
\frac{1}{2}e^{i\pi/4}\sqrt{F(1-F')}|\alpha\rangle_{\mbox{\scriptsize E1}}|\beta'\rangle_{\mbox{\scriptsize E2}}
-
\frac{1}{2}e^{-i\pi/4}\sqrt{(1-F)F'}|\gamma\rangle_{\mbox{\scriptsize E1}}|\alpha'\rangle_{\mbox{\scriptsize E2}}, \nonumber \\
&&
|\Phi(\sigma_{z}=-1,r_{1})\rangle_{\mbox{\scriptsize E1E2}} \nonumber \\
&=&
\frac{1}{\sqrt{2}}\sqrt{(1-F)(1-F')}|\beta\rangle_{\mbox{\scriptsize E1}}|\gamma'\rangle_{\mbox{\scriptsize E2}} \nonumber \\
&&
+
\frac{1}{2}e^{-i\pi/4}\sqrt{(1-F)F'}|\beta\rangle_{\mbox{\scriptsize E1}}|\delta'\rangle_{\mbox{\scriptsize E2}}
+
\frac{1}{2}e^{i\pi/4}\sqrt{F(1-F')}|\delta\rangle_{\mbox{\scriptsize E1}}|\gamma'\rangle_{\mbox{\scriptsize E2}}, \nonumber \\
&&
|\Phi(\sigma_{z}=-1,r_{4})\rangle_{\mbox{\scriptsize E1E2}} \nonumber \\
&=&
\frac{1}{\sqrt{2}}\sqrt{FF'}|\delta\rangle_{\mbox{\scriptsize E1}}|\delta'\rangle_{\mbox{\scriptsize E2}} \nonumber \\
&&
-
\frac{1}{2}e^{i\pi/4}\sqrt{(1-F)F'}|\beta\rangle_{\mbox{\scriptsize E1}}|\delta'\rangle_{\mbox{\scriptsize E2}}
-
\frac{1}{2}e^{-i\pi/4}\sqrt{F(1-F')}|\delta\rangle_{\mbox{\scriptsize E1}}|\gamma'\rangle_{\mbox{\scriptsize E2}},
\end{eqnarray}
\begin{eqnarray}
&&
|\Phi(\sigma_{x}=1,r_{1})\rangle_{\mbox{\scriptsize E1E2}} \nonumber \\
&=&
\frac{1}{2\sqrt{2}}
(\sqrt{F}|\alpha\rangle_{\mbox{\scriptsize E1}}+\sqrt{1-F}|\beta\rangle_{\mbox{\scriptsize E1}})
(\sqrt{F'}|\alpha'\rangle_{\mbox{\scriptsize E2}}+\sqrt{1-F'}|\gamma'\rangle_{\mbox{\scriptsize E2}}) \nonumber \\
&&
+
\frac{1}{4}e^{-i\pi/4}
(\sqrt{F}|\alpha\rangle_{\mbox{\scriptsize E1}}+\sqrt{1-F}|\beta\rangle_{\mbox{\scriptsize E1}})
(\sqrt{1-F'}|\beta'\rangle_{\mbox{\scriptsize E2}}+\sqrt{F'}|\delta'\rangle_{\mbox{\scriptsize E2}}) \nonumber \\
&&
+
\frac{1}{4}e^{i\pi/4}
(\sqrt{1-F}|\gamma\rangle_{\mbox{\scriptsize E1}}+\sqrt{F}|\delta\rangle_{\mbox{\scriptsize E1}})
(\sqrt{F'}|\alpha'\rangle_{\mbox{\scriptsize E2}}+\sqrt{1-F'}|\gamma'\rangle_{\mbox{\scriptsize E2}}), \nonumber \\
&&
|\Phi(\sigma_{x}=1,r_{4})\rangle_{\mbox{\scriptsize E1E2}} \nonumber \\
&=&
\frac{1}{2\sqrt{2}}
(\sqrt{1-F}|\gamma\rangle_{\mbox{\scriptsize E1}}+\sqrt{F}|\delta\rangle_{\mbox{\scriptsize E1}})
(\sqrt{1-F'}|\beta'\rangle_{\mbox{\scriptsize E2}}+\sqrt{F'}|\delta'\rangle_{\mbox{\scriptsize E2}}) \nonumber \\
&&
-
\frac{1}{4}e^{i\pi/4}
(\sqrt{F}|\alpha\rangle_{\mbox{\scriptsize E1}}+\sqrt{1-F}|\beta\rangle_{\mbox{\scriptsize E1}})
(\sqrt{1-F'}|\beta'\rangle_{\mbox{\scriptsize E2}}+\sqrt{F'}|\delta'\rangle_{\mbox{\scriptsize E2}}) \nonumber \\
&&
-
\frac{1}{4}e^{-i\pi/4}
(\sqrt{1-F}|\gamma\rangle_{\mbox{\scriptsize E1}}+\sqrt{F}|\delta\rangle_{\mbox{\scriptsize E1}})
(\sqrt{F'}|\alpha'\rangle_{\mbox{\scriptsize E2}}+\sqrt{1-F'}|\gamma'\rangle_{\mbox{\scriptsize E2}}), \nonumber \\
&&
|\Phi(\sigma_{x}=-1,r_{1})\rangle_{\mbox{\scriptsize E1E2}} \nonumber \\
&=&
\frac{1}{2\sqrt{2}}
(\sqrt{F}|\alpha\rangle_{\mbox{\scriptsize E1}}-\sqrt{1-F}|\beta\rangle_{\mbox{\scriptsize E1}})
(\sqrt{F'}|\alpha'\rangle_{\mbox{\scriptsize E2}}-\sqrt{1-F'}|\gamma'\rangle_{\mbox{\scriptsize E2}}) \nonumber \\
&&
+
\frac{1}{4}e^{-i\pi/4}
(\sqrt{F}|\alpha\rangle_{\mbox{\scriptsize E1}}-\sqrt{1-F}|\beta\rangle_{\mbox{\scriptsize E1}})
(\sqrt{1-F'}|\beta'\rangle_{\mbox{\scriptsize E2}}-\sqrt{F'}|\delta'\rangle_{\mbox{\scriptsize E2}}) \nonumber \\
&&
+
\frac{1}{4}e^{i\pi/4}
(\sqrt{1-F}|\gamma\rangle_{\mbox{\scriptsize E1}}-\sqrt{F}|\delta\rangle_{\mbox{\scriptsize E1}})
(\sqrt{F'}|\alpha'\rangle_{\mbox{\scriptsize E2}}-\sqrt{1-F'}|\gamma'\rangle_{\mbox{\scriptsize E2}}), \nonumber \\
&&
|\Phi(\sigma_{x}=-1,r_{4})\rangle_{\mbox{\scriptsize E1E2}} \nonumber \\
&=&
\frac{1}{2\sqrt{2}}
(\sqrt{1-F}|\gamma\rangle_{\mbox{\scriptsize E1}}-\sqrt{F}|\delta\rangle_{\mbox{\scriptsize E1}})
(\sqrt{1-F'}|\beta'\rangle_{\mbox{\scriptsize E2}}-\sqrt{F'}|\delta'\rangle_{\mbox{\scriptsize E2}}) \nonumber \\
&&
-
\frac{1}{4}e^{i\pi/4}
(\sqrt{F}|\alpha\rangle_{\mbox{\scriptsize E1}}-\sqrt{1-F}|\beta\rangle_{\mbox{\scriptsize E1}})
(\sqrt{1-F'}|\beta'\rangle_{\mbox{\scriptsize E2}}-\sqrt{F'}|\delta'\rangle_{\mbox{\scriptsize E2}}) \nonumber \\
&&
-
\frac{1}{4}e^{-i\pi/4}
(\sqrt{1-F}|\gamma\rangle_{\mbox{\scriptsize E1}}-\sqrt{F}|\delta\rangle_{\mbox{\scriptsize E1}})
(\sqrt{F'}|\alpha'\rangle_{\mbox{\scriptsize E2}}-\sqrt{1-F'}|\gamma'\rangle_{\mbox{\scriptsize E2}}).
\end{eqnarray}


\begin{thebibliography}{99}
%
\bibitem{Bennett1984}
C.~H.~Bennett and G.~Brassard,
Quantum cryptography: public key distribution and coin tossing,
in {\it Proceedings of the IEEE International Conference on Computers, Systems, and Signal Processing},
Bangalore, India (IEEE, New York, 1984), pp. 175--179;
{\it Theor. Comput. Sci.} {\bf 560}(Part 1), 7--11 (2014).
doi:10.1016/j.tcs.2014.05.025
%
\bibitem{Ekert1991}
A.~K.~Ekert,
Quantum cryptography based on Bell's theorem,
{\it Phys. Rev. Lett.} {\bf 67}(6), 661--663 (1991).
doi:10.1103/PhysRevLett.67.661
%
\bibitem{Bennett1992a}
C.~H.~Bennett, G.~Brassard, and A.~K.~Ekert,
Quantum cryptography,
{\it Sci. Am.} {\bf 267}(4), 50--57 (October 1992).
doi:10.1038/scientificamerican1092-50
%
\bibitem{Bennett1992b}
C.~H.~Bennett, F.~Bessette, G.~Brassard, L.~Salvail, and J.~Smolin,
Experimental quantum cryptography,
{\it J. Cryptology} {\bf 5}(1), 3--28 (1992).
doi:10.1007/BF00191318
%
\bibitem{Ekert1994}
A.~K.~Ekert, B.~Huttner, G.~M.~Palma, and A.~Peres,
Eavesdropping on quantum-cryptographical systems,
{\it Phys. Rev. A} {\bf 50}(2), 1047--1056 (1994).
\\
doi:10.1103/PhysRevA.50.1047
%
\bibitem{Biham1997}
E.~Biham and T.~Mor,
Security of quantum cryptography against collective attacks,
{\it Phys. Rev. Lett.} {\bf 78}(11), 2256--2259 (1997).
doi:10.1103/PhysRevLett.78.2256
%
\bibitem{Enzer2002}
D.~G.~Enzer, P.~G.~Hadley, R.~J.~Hughes, C.~G.~Peterson, and P.~G.~Kwiat,
Entangled-photon six-state quantum cryptography,
{\it New J. Phys.} {\bf 4}, 45.1--45.8 (2002).
doi:10.1088/1367-2630/4/1/345
%
\bibitem{Cirac1997}
J.~I.~Cirac and N.~Gisin,
Coherent eavesdropping strategies for the four state quantum cryptography protocol,
{\it Phys. Lett. A} {\bf 229}(1), 1--7 (1997).
doi:10.1016/S0375-9601(97)00176-X
%
\bibitem{Bub2001}
J.~Bub,
Secure key distribution via pre- and postselected quantum states,
{\it Phys. Rev. A} {\bf 63}(3), 032309 (2001).
doi:10.1103/PhysRevA.63.032309
%
\bibitem{Aharonov1964}
Y.~Aharonov, P.~G.~Bergmann, and J.~L.~Lebowitz,
Time symmetry in the quantum process of measurement,
{\it Phys. Rev.} {\bf 134}(6B), B1410--B1416 (1964).
doi:10.1103/PhysRev.134.B1410
%
\bibitem{Aharonov2008}
Y.~Aharonov and L.~Vaidman,
The two-state vector formalism: an updated review,
in {\it Time in Quantum Mechanics}, vol.~1, 2nd ed., Lect. Notes Phys. {\bf 734}, 
eds. J.~G.~Muga, R.~S.~Mayato, and {\'I}.~L.~Egusquiza
(Springer-Verlag, Berlin Heidelberg, 2008),
pp. 399--447.
doi:10.1007/978-3-540-73473-4{\_}13
%
\bibitem{Aharonov1991}
Y.~Aharonov and L.~Vaidman,
Complete description of a quantum system at a given time,
{\it J. Phys. A: Math. Gen.} {\bf 24}(10), 2315--2328 (1991).
doi:10.1088/0305-4470/24/10/018
%
\bibitem{Vaidman1987}
L.~Vaidman, Y.~Aharonov, and D.~Z.~Albert,
How to ascertain the values of $\sigma_{x}$, $\sigma_{y}$, and $\sigma_{z}$ of a spin-$1/2$ particle,
{\it Phys. Rev. Lett.} {\bf 58}(14), 1385--1387 (1987).
doi:10.1103/PhysRevLett.58.1385
%
\bibitem{Aharonov2001}
Y.~Aharonov and B.-G.~Englert,
The mean king's problem: Spin $1$,
{\it Z. Naturforsch.} {\bf 56}a(1-2), 16--19 (2001).
doi:10.1515/zna-2001-0104
%
\bibitem{Yoshida2010}
M.~Yoshida, T.~Miyadera, and H.~Imai,
On the security of the quantum key distribution using the mean king problem,
in {\it International Symposium on Information Theory {\&} Its Application} (ISITA), IEEE, Taichung, Taiwan, 2010,
pp. 917--921.
\\
doi:10.1109/ISITA.2010.5649556 
%
\bibitem{Yoshida2012}
M.~Yoshida, T.~Miyadera, and H.~Imai,
Quantum key distribution using mean king problem with modified measurement scheme,
in {\it International Symposium on Information Theory {\&} Its Application} (ISITA), IEEE, Honolulu, Hawaii, USA, 2012,
pp. 317--321.
%
\bibitem{Werner2009}
A.~H.~Werner, T.~Franz, and R.~F.~Werner,
Quantum cryptography as a retrodiction problem,
{\it Phys. Rev. A} {\bf 103}(22), 220504 (2009).
\\
doi:10.1103/PhysRevLett.103.220504
%
\bibitem{Reimpell2007}
M.~Reimpell and R.~F.~Werner,
Meaner king uses biased bases,
{\it Phys. Rev. A} {\bf 75}(6), 062334 (2007).
doi:10.1103/PhysRevA.75.062334
%
\bibitem{Sakurai}
J.~J.~Sakurai,
{\it Modern Quantum Mechanics},
revised ed.
(Addison-Wesley Publishing Company, Reading, Massachusetts, 1994),
Eq.~(3.3.21).
%
\bibitem{Nielsen2000}
M.~A.~Nielsen and I.~L.~Chuang,
{\it Quantum Computation and Quantum Information}
(Cambridge University Press, Cambridge, United Kingdom, 2000).
%
\bibitem{Bechmann-Pasquinucci1999}
H.~Bechmann-Pasquinucci and N.~Gisin,
Incoherent and coherent eavesdropping in the six-state protocol of quantum cryptography,
{\it Phys. Rev. A} {\bf 59}(6), 4238--4248 (1999).
doi:10.1103/PhysRevA.59.4238
%
\bibitem{Fuchs1999}
C.~A.~Fuchs and J.~van~de~Graaf,
Cryptographic distinguishability measures for quantum-mechanical states,
{\it IEEE Transactions and Information Theory} {\bf 45}(4), 1216--1227 (1999).
doi:10.1109/18.761271
%
\bibitem{Lo1999}
H.-K.~Lo and H.~F.~Chau,
Unconditional security of quantum key distribution over arbitrarily long distances,
{\it Science} {\bf 283}(5410), 2050--2056 (1999).
\\
doi:10.1126/science.283.5410.2050
%
\bibitem{Mayers2001}
D.~Mayers,
Unconditional security in quantum cryptography,
{\it J. ACM} {\bf 48}(3), 351--406 (2001).
doi:10.1145/382780.382781
%
\bibitem{Biham2006}
E.~Biham, M.~Boyer, P.~O.~Boykin, T.~Mor, and V.~Roychowdhury,
A proof of the security of quantum key distribution,
{\it J. Cryptology} {\bf 19}(4), 381--439 (2006).
doi:10.1007/s00145-005-0011-3
%
\bibitem{Shor2000}
P.~W.~Shor and J.~Preskill,
Simple proof of security of the BB84 quantum key distribution protocol,
{\it Phys. Rev. Lett.} {\bf 85}(2), 441--444 (2000).
\\
doi:10.1103/PhysRevLett.85.441
%
\bibitem{Metzger2000}
S.~Metzger,
Spin-measurement retrodiction revisited,
arXiv:quant-ph/0006115.
%
\bibitem{Lutkenhaus1999}
N.~L{\"u}tkenhaus,
Estimates for practical quantum cryptography,
{\it Phys. Rev. A} {\bf 59}(5), 3301--3319 (1999).
doi:10.1103/PhysRevA.59.3301
%
\bibitem{Lutkenhaus2000}
N.~L{\"u}tkenhaus,
Security against individual attacks for realistic quantum key distribution,
{\it Phys. Rev. A} {\bf 61}(5), 052304 (2000).
doi:10.1103/PhysRevA.61.052304
%
\bibitem{Gottesman2004}
D.~Gottesman, H.-K.~Lo, N.~L{\"u}tkenhaus, and J.~Preskill,
Security of quantum key distribution with imperfect devices,
in {\it Proceedings of 2004 IEEE International Symposium on Information Theory}, ISIT 2004,
Chicago, USA (IEEE, Piscataway, 2004), 136.
doi:10.1109/ISIT.2004.1365172
%
\bibitem{Scarani2009}
V.~Scarani, H.~Bechmann-Pasquinucci, N.~J.~Cerf,
M.~Du{\u{s}}ek, N.~L{\"u}tkenhaus, and M.~Peev,
The security of practical quantum key distribution,
{\it Rev. Mod. Phys.} {\bf 81}(3), 1301--1350 (2009).
doi:10.1103/RevModPhys.81.1301
%
\end{thebibliography}
\end{document}